\documentclass[final,3p,times]{elsarticle}

\usepackage{amssymb}
\usepackage{amsthm}

\usepackage{amsmath, enumerate, enumitem, subcaption}

\usepackage{lineno}

\journal{Progress in Nuclear Energy}

\begin{document}

\begin{frontmatter}

\title{A Digital Twin Framework for Generation-IV Reactors with Reinforcement Learning-Enabled Health-Aware Supervisory Control}
\author[inst1]{Jasmin Y. Lim\corref{cor1}}
\ead{jaslim@umich.edu}
\author[inst2]{Dimitrios Pylorof}
\author[inst2]{Humberto E. Garcia}
\author[inst1,inst3]{Karthik Duraisamy}

\cortext[cor1]{Corresponding Author}

\affiliation[inst1]{organization={Department of Aerospace Engineering, University of Michigan},
            addressline={1320 Beal Ave}, 
            city={Ann Arbor},
            postcode={48109}, 
            state={MI},
            country={USA}}

\affiliation[inst2]{organization={Systems Science and Engineering Division, U.S. DOE Idaho National Laboratory},
            addressline={1955 N. Fremont Ave.}, 
            city={Idaho Falls},
            postcode={83415}, 
            state={ID},
            country={USA}}

\affiliation[inst3]{organization={Michigan Institute for Computational Discovery \& Engineering (MICDE), University of Michigan},
            addressline={915 E Washington St, Suite 3530}, 
            city={Ann Arbor},
            postcode={48109}, 
            state={MI},
            country={USA}}

\begin{abstract}
Generation IV (Gen-IV) nuclear power plants are envisioned to replace the current reactor fleet, bringing improvements in performance, safety, reliability, and sustainability.
However, large cost investments currently inhibit the deployment of these advanced reactor concepts. 
Digital twins bridge real-world systems with digital tools to reduce costs, enhance decision-making, and boost operational efficiency. 
In this work, a digital twin framework is designed to operate the Gen-IV Fluoride-salt-cooled High-temperature Reactor, utilizing data-enhanced methods to optimize operational and maintenance policies while adhering to system constraints.
The closed-loop framework integrates surrogate modeling, reinforcement learning, and Bayesian inference to streamline end-to-end communication for online regulation and self-adjustment. 
Reinforcement learning is used to consider component health and degradation to drive the target power generations, with constraints enforced through a Reference Governor control algorithm that ensures compliance with pump flow rate and temperature limits. These input driving modules benefit from detailed online simulations that are assimilated to measurement data with Bayesian filtering. 
The digital twin is demonstrated in three  case studies: a one-year long-term operational period showcasing maintenance planning capabilities, short-term accuracy refinement with high-frequency measurements, and system shock capturing that demonstrates real-time recalibration capabilities when change in boundary conditions. 
These demonstrations validate robustness for health-aware and constraint-informed nuclear plant operation, with general applicability to other advanced reactor concepts and complex engineering systems.
\end{abstract}



\end{frontmatter}

\section{Introduction}



The large-scale deployment of nuclear energy faces several challenges, including high development costs, managing nuclear waste, and stringent safety requirements.
Generation-IV (Gen-IV) Nuclear Reactors, the successors of the current reactor fleet, aim to address these issues with designs that enable greater operational efficiency, have inherent safety features, have lower radioactive waste, and contain anti-proliferation measures~\cite{GIF:02}.
While these Gen-IV reactors offer promising solutions, there are still barriers hindering their full-scale implementation.
Compared to other non-renewable and renewable energy sources, nuclear reactors have higher construction costs~\cite{mit:18,Thellufusen:24} and development periods~\cite{IEA:25}.
This is partially attributed to the lack of information and data, leaving gaps in knowledge that are necessary for cost analysis, energy output projections, and development of regulations.
These uncertainties increase the likelihood of unforeseen delays in the construction of testing facilities and plants, which will increase investment costs and extending project timelines. 
Major nuclear accidents have also instilled that these advanced reactor must continue to advance in safety analysis measures for secure and resilient operations. 

The Generation-IV International Forum~(GIF) stated in the \textit{R\&D Outlook for Gen-IV Nuclear Energy Systems}~\cite{GIF:18} report that digital Product Lifecycle Management (PLM) tools are to be investigated for reducing the time to market of Gen-IV reactors.
These PLM tools are also useful for system health tracking, providing analyses of components for maintenance diagnosis and prognosis~\cite{Meyer:02}. 
\textit{Digital twins} are an emerging technology that connect a physical asset with a virtual counterpart, enabling global benefits through real-time data acquisition, adaptive computational models, and intelligently determined inputs.
The modern digital twin was originally conceptualized by Michael Grieves in 2003 as a PLM tool~\cite{Grieves:05}. 
The dynamic data-informed framework aims to drive and optimize a real-world asset throughout its individualized existence; enabling the detection and response to both predictable and unpredictable events~\cite{Grieves:17}.
Movement towards digital twins research and deployment is fueled by innovations in both the instrumentation and computational sides~\cite{Jones:20}. 
This includes development of facilities with autonomous sensors, communication technologies, and robotic automation; and the success of simulation and prediction advancements like Artificial Intelligence (AI) and computing hardware~\cite{Rasheed:20}. 
Digital twins are prevalent in engineering applications such as manufacturing~\cite{KRITZINGER:18}, urban planning~\cite{Batty:24,Ge:25}, fleet management~\cite{Kraft:17}, operations \& maintenance~\cite{Zohdi:22,Torzoni:24}, and supply-chain networks~\cite{Shrivastava:22}.
They have also experienced a widespread appearance in the medical field for drug development, surgical planning, and personalized healthcare~\cite{Erol:20,Lal:20}.

Within the nuclear energy field, the promise of digital twin technologies is applicable to all stages of a reactor's lifecycle, from initial design and construction to waste management and decommissioning~\cite{Yadav:21}. 
There have been several works in developing digital twin frameworks specifically for nuclear reactors, with focuses on the generalization of software architectures~\cite{Patterson:16,Bowman:22} and the incorporation of uncertainty analysis~\cite{Kochunas:21,Lin:21}. 
The AGN-201 digital twin~\cite{stewart:25} was the first digital twin to pilot real-time anomaly detection of a research reactor.
The framework consisted of several surrogate modeling techniques --- including Gaussian process regression and machine learning --- coupled with reactor physical models.
The near real-time monitoring demonstrated the capabilities of digital twins and provided insights for future developments in remote monitoring and securing measures. 
In the work done by Ndum et al.~\cite{Ndum:24}, a proof-of-concept digital twin was developed for a Lead-cooled Fast Reactor (LFR), focused on coupled neutronic-thermal-hydraulics models and front-end real-time visualization tools.
A Dynamic Operation and maintenance Optimization (DyOMO) Bayesian network framework was developed by Rivas et al.~\cite{Rivas:25} for predicting maintenance of Pebble-Bed High-Temperature Gas-cooled Reactors (PB-HTGRs), featuring a neural network that quantifies component remaining useful life.

Nuclear Power Plants (NPPs) are complex systems, containing dynamic sub-systems that couple various phenomena (e.g., nuclear physics, fluids, heat transfer, chemistry) to meet safety and economic requirements. 
All of which are assessable with digital twin technologies that contain modules to optimize, assist, and automate operational decisions.
What separates a digital twin from conventional modeling is their individualized approach to create unique instances of a product, rather than using a generalized approach formed from average population characteristics~\cite{thelen:22}. 
In order to support these beneficial capabilities, digital twins are largely dependent on information and computing power for their continued progress~\cite{Grieves:17}.
Advancements in physics-based simulations have improved the sophistication of computational modeling for accurate prediction of complex systems. 
Additionally, data-driven methods such as machine learning, Reduced-Order Models, and Bayesian inference are effective and scalable interpreters of diverse data formats.

Reinforcement Learning (RL) is a field in AI that combines both perceptual and decision-making abilities with high-potential for managing complex systems to meet multi-objective goals in a robust and adaptable manner~\cite{Liu:24}. 
The sequential action-reaction approach that RL algorithms use to learn and drive a system is compatible with the digital twin idea; making their use here apparent. 
In the nuclear field, RL research has been concentrated in a few areas such as autonomous controllers~\cite{Nguyen:24,Tunkle:25}, design optimization~\cite{Kim:24}, and safety diagnostics~\cite{Zhong:23,Park:20}.

Operating a NPP is a complicated optimization problem that involves the management of components, economic profitability and safety guidelines; and RL has been applied for such problems.
Zhao and Smidts uses a model-based RL approach to learn the degradation model starting from an imperfect model and then determining the optimal maintenance policy for reactor pumps with limited data observability~\cite{ZhaoSmidts:22}.
In the paper by Hao et al.~\cite{HAO:24}, a Monte-Carlo tree search RL algorithm is applied to scheduling pump maintenance in a Lead-cooled Fast Reactor, using a multi-objective policy that maximizes profit while keeping various variables within their safety thresholds. 
Bae et al. also address multi-objective operations but with multiple manipulative devices, training an agent to control the reactor coolant status during a NPP start up by activating and deactivating reactor coolant pumps and steam generators, respectively~\cite{Bae:23}. 

Digital twins have the potential to aid Gen-IV reactor development and eventually assisting in their general adoption.
Coupled with advanced data-driven methods, these digital twin technologies support researchers and operators with around-the-clock monitoring and predictive capabilities that enhance safety and reliability.
RL is promising prospect for supervisory plant management, with multi-component control informed by comprehensive information from diverse knowledge sources.
In this work, a digital twin framework is developed for the operation of a Gen-IV reactor concept --- the Fluoride-salt-cooled High Temperature Reactor (FHR) --- with RL-based methods. 
Supported by a reduced-complexity surrogate model, the framework uses a RL powered supervisory controller to drive the reactor power output and manage maintenance of the system pumps.  
Target power levels are evaluated with a Reference Governor (RG) constraint enforcement agent before sending operable set-points to the physical plant. 
Information about the real-life system is collected through data and is integrated into the surrogate model representation using the Ensemble Kalman Filter data assimilation algorithm; enabling real-time digital twin calibration for accuracy refinement. 
All combined, these elements build the health-aware digital twin framework with an online adaptation capabilities for informed operational decisions and maintenance scheduling. 

The remainder of this work is organized by first introducing the nuclear reactor application in Section~\ref{sec:FHR} and then presenting the digital twin framework in Section~\ref{sec:dt-frame}.
The digital twin is demonstrated on three cases presented in Section~\ref{sec:demo}, which include a long-term health analysis case, a short-term accuracy refinement case, and a system shock capturing case. 
Finally, concluding thoughts and future work are discussed in Section~\ref{sec:conc}.

\subsection*{Notation}
We use $\mathbb{N}$, $\mathbb{Z}$, $\mathbb{Z}_{\geq}$, $\mathbb{R}$, $\mathbb{R}^n$, $\mathbb{R}^n_{\geq}$ to denote the sets, respectively, of natural numbers, integers, non-negative integers, reals, $n$-dimensional real vectors, and $n$-dimensional real vectors with non-negative elements. 
Regular font Latin or Greek letters (e.g., $c,\gamma$) are scalars; lowercase bold font Latin or Greek letters (e.g., $\mathbf{x}, \boldsymbol{\theta}$) are column vectors; and upper case bold font Latin or Greek letters (e.g., $\mathbf{A}, \boldsymbol{\Gamma}$) are matrices. 
A superscript with parentheses (e.g., $x^(k)$) denotes the variable value at the discrete time-step $k\in \mathbb{Z}_{>}$; and the superscript with parentheses and colon (e.g., $\mathcal{A}^{(k:\ell)}$) denotes a set of values from the discrete time-step $k\in\mathbb{Z}_{\geq}$ to $\ell\in\mathbb{Z}_{\geq}$ where $k<\ell$.
A letter followed by parenthesis closing arguments is a function.

A \textit{random variable} is denoted by uppercase Latin letters or lowercase Greek letters (e.g., $X, \xi$). 
A \textit{multivariate random variable} is a column vector denoted by bold uppercase Latin letters or bold lowercase Greek letters (e.g., $\mathbf{X}$, $\boldsymbol{\xi}$).
A particular \textit{realization} of a random variable is written in corresponding lowercase letters. 
For instance, $x\in\mathcal{X}$ is a realization of the random variable $X$ distributed as $p(x)$, where $\mathcal{X}$ is the set of all possible realizations of the random variable. 
The realization of a multivariate random variable is written in corresponding bold font lowercase letters or denoted with a subscript, $\mathbf{x}$ is a realization of random variable $\mathbf{X}$ and $\boldsymbol{\xi}_j$ is a realization of random variable $\boldsymbol{\xi}$.
The use of random variables and their respective realizations will be made apparent from context. 
The \textit{expectation} of a random variable $X$ is $\mathbb{E}{[X]}$. 
The \textit{variance} of a random variable $X$ is $\mathbb{V}\text{ar}{[X]}$.
The \textit{covariance} between the two random variables $X$ and $Y$ is $\mathbb{C}\text{ov}{[X,Y]}$.
A function $p(\cdot)$ is a \textit{probability density function} (PDF); for instance, $X \sim p(\cdot)$, where ``$\sim$" means ``is distributed as".

\section{Advanced Reactor System}\label{sec:FHR}

The six Generation-IV~(Gen-IV) reactors types were identified by GIF~\cite{GIF:02}, including the Molten Salt Reactor (MSR), a thermal reactor that uses a molten salt as coolant or as a fuel salt.
In this work the FHR, a type of MSR, is studied.
The FHR is characterized by its use of solid fuel pebbles, a graphite moderator, and a molten salt coolant; yielding a design that has efficient power generation with inherent safety features. 
This high-temperature reactor facilitates efficient thermal output while keeping operations near atmospheric pressure by using the high boiling point liquid fluoride-salt-coolant (Li$_2$BeF$_4$, FLiBe).
Additionally, the nuclear fission is contained within thermally-stable tristructural isotropic~(TRISO) fuel pebbles with failure temperatures well beyond the FHR's operating temperature~\cite{Jiang:22}.
With these qualities, the FHR has attracted R\&D attention from both research~\cite{Zhang:23, Andreades:16} and commercial~\cite{Zhao:23} institutions.

\subsection{Physical Model}\label{subsec:samfhr}

\begin{figure}
    \centering
    \includegraphics[width=\textwidth]{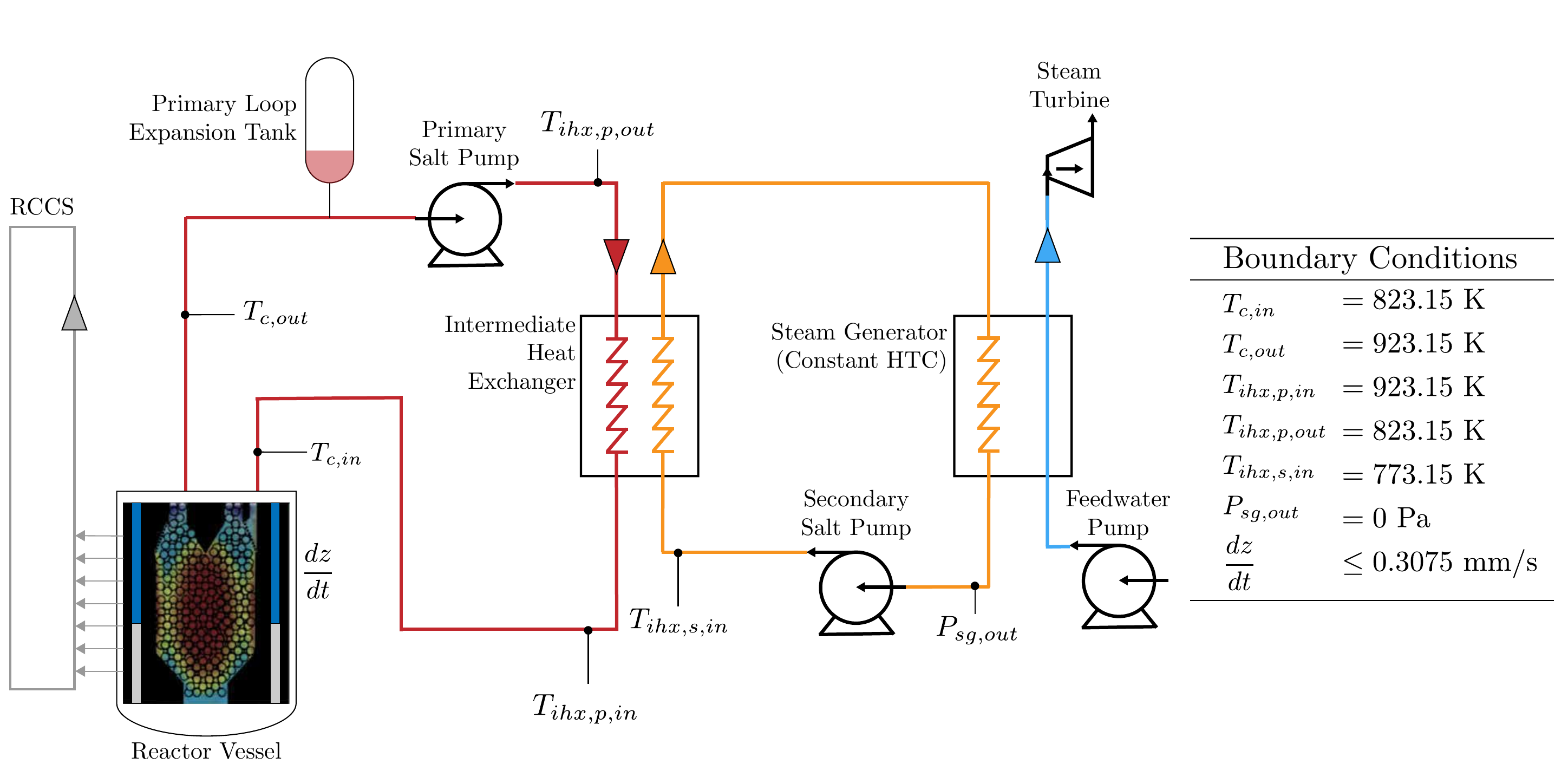}
    \caption{SAM gFHR reactor configuration. This schematic is from~\cite{Lim:25} and is adapted with permission from~\cite{Dave:2023}.}
    \label{fig:gFHR_system}
\end{figure}

The digital twin framework is designed to digitally model, control, and enhance an FHR configuration.
Physics-based, higher-fidelity transient simulations are provided by the System Analysis Module (SAM)~\cite{Hu:21}, a computational system analysis tool developed by Argonne National Laboratory (ANL) for system-level analysis of advanced reactor concepts --- with verification for FHRs~\cite{OGrady:21}. 
The SAM gFHR model~\cite{Li:24}, shown in Figure~\ref{fig:gFHR_system}, is a conceptual model based on the generic FHR (gFHR)~\cite{Satvat:21} core-only benchmark and the thermal-fluid design of the Kairos Power FHR (KP-FHR) by Kairos Power LLC.
The hybrid SAM gFHR model combines high-fidelity stochastic and deterministic core modeling codes with the intermediate-fidelity SAM code for a coupled neutronics and thermal-hydraulics system analysis. 

The SAM gFHR model consists of a two-loop heat transport system connected through an Intermediate Heat eXchanger (IHX) to keep the system's velocity less than 2 m/s. 
Each loop contains a pump, both of which are dynamic centrifugal pumps~\cite{Lee:23} modeled using the homologous pump theory~\cite{Kennedy:1980}. 
The neutronics of the gFHR core are implemented through the SAM Point Kinetics Model (PKE)~\cite{Fanning:17, Hu:19}, providing the reactor core transient capability through reactivity feedback effects from the coolant, fuel, moderator, and xenon poisoning. 
The core temperatures are supported by a water-based Reactor Cavity Cooling System (RCCS) that models core decay heat removal~\cite{Hu:20}.

A 3D-1D flow coupling scheme is applied, where the core is modeled in the three dimensions and the primary and intermediate (also called secondary) loops are modeled in one dimension. 
Since the fuel pebbles within the core are randomly packed, which cause irregular flows and variability in the cross-sectional areas, the two-dimensional porous media model is used.
The porous media equations are utilized for their permeability and porosity parameters to model the flow resistance and the void fraction present in pebble-bed phenomenon~\cite{Li:24}.
In the SAM gFHR model, both the core neutronics analysis and SAM use the porous-medium governing equations to perform a thermal-hydraulics analysis as follows:
\begin{align}
    \varphi \frac{\partial \rho_f}{\partial t} + \nabla \cdot (\rho_f \mathbf{v}) = 0, \label{eq:SAM_mass}\\
    \rho_f \frac{\partial \mathbf{v}}{\partial t} + \frac{\rho_f}{\varphi} (\mathbf{v} \cdot \nabla)\mathbf{v} + \varphi \nabla p - \varphi \rho_f \mathbf{g} + \beta \mathbf{v} + \alpha | \mathbf{v} | \mathbf{v} = 0, \label{eq:SAM_mom}\\
    \varphi \rho_f c_{p,f} \frac{\partial T_f}{\partial t} + \rho c_{p,f} \mathbf{v} \cdot \nabla T_f - \nabla \cdot (\varphi k_f \nabla T_f) - q_f''' + a_w h (T_f-T_s) = 0, \label{eq:SAM_energyF}\\
    (1-\varphi) \rho_s c_{p,s} \frac{\partial T_s}{\partial t} - \nabla \cdot (k_{s,eff} \nabla T_s) + q_s ''' + a_w h(T_s -T_f) = 0,\label{eq:SAM_energyS}
\end{align}
where Equation~\ref{eq:SAM_mass} is the mass balance equation; Equation~\ref{eq:SAM_mom} is the momentum equation; Equation~\ref{eq:SAM_energyF} is the fluid energy equation; and Equation~\ref{eq:SAM_energyS} is the solid energy equation. 
The subscripts $f$ and $s$ denote the fluid and solid phase variables, and $\varphi$ is the porosity of the pebble bed. 
$\rho, \mathbf{g}, \beta,\alpha,c,T,k,a_w,h, q'''$ are density, gravity, the linear dynamic fluid viscosity constant, the quadratic dynamic fluid viscosity constant, specific heat capacity at constant pressure, temperature, thermal conductivity, the heating surface area density per unit volume and the heat production per unit volume. 
$\mathbf{v}$ is a superficial velocity that is related to the intrinsic velocity $\mathbf{V}$ by $\mathbf{v} = \varphi\mathbf{V}$. 

Full reactor power is set to $280$ megawatts (MW), and design limitations restrict the operation of the plant to stay within 50\% to 100\% of full power.
For operation, a user provides the SAM gFHR model with two inputs, a time-dependent target power distribution and the pump degradation in both pumps. 
These are presented with a discrete target power set points $\dot{Q}_{RX,T}^{(k)}$ and a pump degradation defined with a loss coefficient $K^{(k)}$.

In order for the SAM gFHR model to meet a user-inputted target power, a simple control system with three PID controllers are used to meet energy requirements while maintaining boundary conditions. 
The first controller aims to meet the requested reactor thermal power by changing the control rod position, $z_{cr}$ at a maximum rate of $\frac{d z_{cr}}{dt} \le 0.3075$ mm/s, which translate to 5\% of full power per minute. The second controller maintains the constant core outlet temperature $T_{c,out}=923.15$~K by manipulating the primary pump power. The third controller maintains the core inlet temperature to a constant $T_{c,in}=823.15$~K by manipulating the secondary pump power. The manipulated pump powers are used by the SAM gFHR model to determine the pump speeds, which drive the pump torque and head. 


The SAM gFHR model uses an adaptive time-step that is up to 25 seconds in steady-state phases and is approximately 5 seconds in transient changes. 
This causes variability in the run time, where simulations with numerous aggressive power changes will take longer to compute compared to a target power generation with fewer and smaller change rates. 
In addition, the SAM gFHR model may be prone to models errors where certain physical behaviors are not captured. 
Due to these limitations, it is infeasible for the digital twin to use the SAM gFHR model when it is online, operating simultaneously with the physical plant.
Thus, the surrogate models are used to represent current system dynamics and the SAM gFHR model is only used offline for data generation.
Since there is currently no experimental facility for the presented FHR plant, the SAM gFHR model is used to emulate the physical asset, providing measurements for the digital twin. 

\section{Digital Twin Framework}\label{sec:dt-frame}

A digital twin is created for the SAM gFHR reactor configuration with the goals of minimizing operations and maintenance costs as well as increasing safety awareness. 
The Gen-IV reactor is expected to meet desired power generations, using the digital twin to optimally manage identified Health Sensitive Components (HSCs) and abide by set operating constraints. 
For the FHR application, the HSCs are the primary pump and the intermediate pump. 
This section will first give an overview of the digital twin framework, defining its inputs and outputs, and its internal components.
Then each of the identified digital twin components are introduced, starting with the Physical and Virtual Assets and ending with the linking Virtual to Physical Module and the Physical to Virtual Module.

\subsection{Framework Architecture}
\begin{figure}
    \centering
    \includegraphics[width=\linewidth]{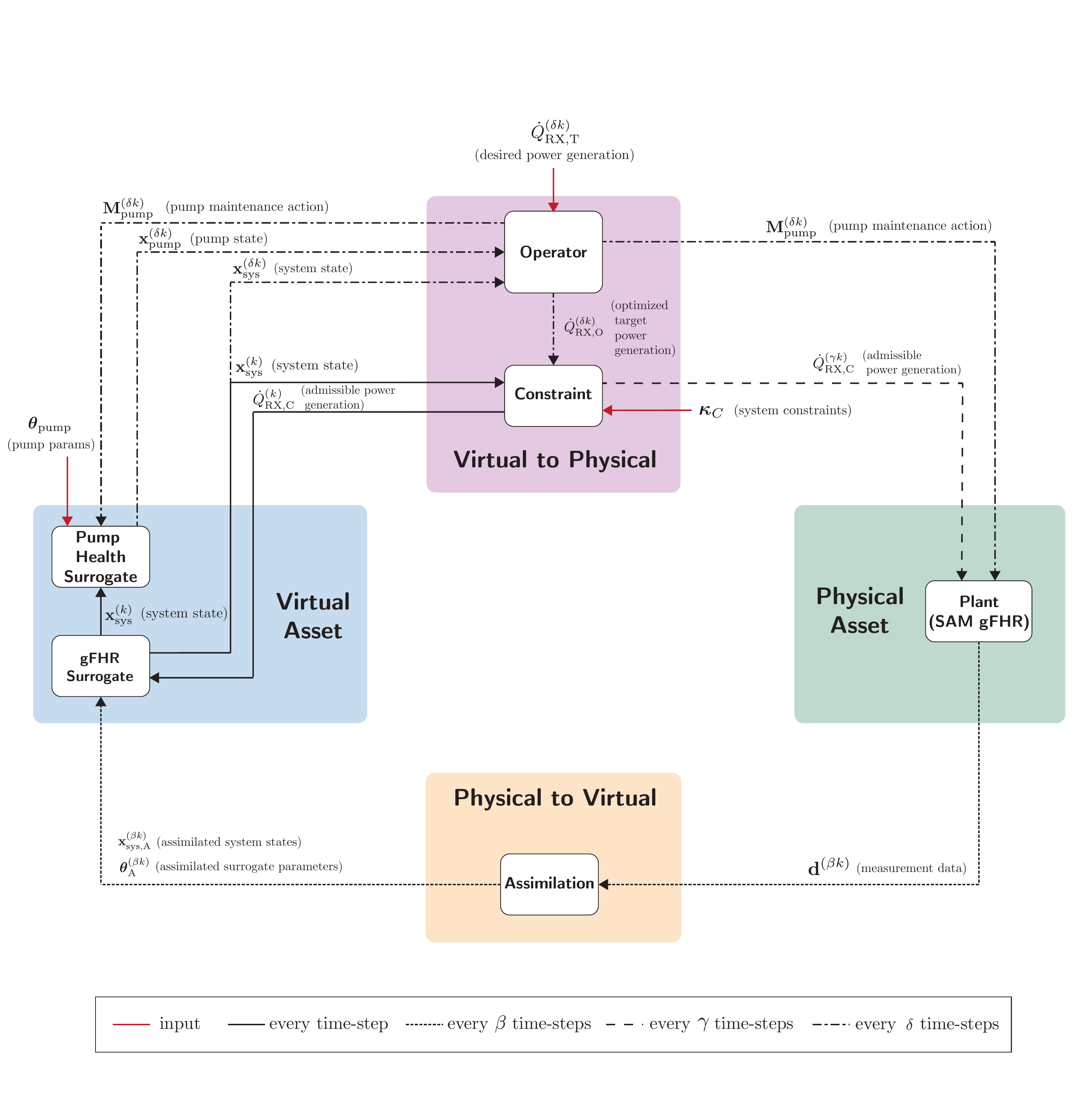}
    \caption{Digital Twin Framework. The digital twin framework is developed for the SAM gFHR reactor configuration, containing four modules: the Physical Asset, the Virtual Asset, the Virtual to Physical Module, and the Physical to Virtual Module.}
    \label{fig:DT-Framework}
\end{figure}

The digital twin framework as was modeled after the Mirrored Spaces Model (MSM), a PLM concept introduced by Michael Grieves~\cite{Grieves:05}.
The MSM has a real space and a virtual space, with linking mechanisms in both directions between the two. 
Data bridges the real space to the virtual space, and information/processes bridge the virtual space to the real space.
The digital twin framework developed in this work contains the four elements of this foundational structure: a Physical Asset, a Virtual Asset, a Physical to Virtual Model, and a Virtual to Physical Module.
All of which are united within the singular framework shown in Figure~\ref{fig:DT-Framework}.
The framework depicts the online operations of the digital twin, showing the lines of communications between the modules as it operates in real-time with the Physical Asset.

\subsubsection{Inputs}

There are three inputs that govern the operations of the FHR plant.
The first is the target power generation, which is the requested core power output defined by the discrete-time distribution $\dot{Q}_{\text{RX,T}}$.
This target power generation will simulate a \textit{load-following} mode, meaning the system is maneuverable, adapting its power generation to daily, seasonal and other power variations. 
The second input is the pump degradation parameters $\boldsymbol{\theta}_{\text{pump}}$. 
Since the system pumps are identified as the HSCs for this digital twin framework, the pump health analysis is monitored by the Virtual to Physical module, which uses information from the Pump Health Surrogate model within the Virtual Asset.
The parameters $\boldsymbol{\theta}_{\text{pump}}$ are supplied to the Virtual Asset and represent the expected degradation rates of the pumps.
Finally, the third input is the system constraints $\boldsymbol\kappa_{\text{C}}$, which is also given to the Virtual to Physical Module.
With this, the digital twin can enforce any number of operational restrictions and intervene before setpoints are sent to the Physical Asset. 

\subsubsection{Time-scales}
A digital twin is designed to evolve with a system, therefore, the framework must be robust in handling differing time-scales. 
Goals vary, ranging from short-term tasks that are managed over a small time horizon to long-term goals that must be kept in consideration over the entire system lifecycle. 
In addition to goals, just the monitoring of a system will have different timescale requirements depending on the task. 
For instance, high risk safety values may need to be reported every few seconds, but the effects of a component's degradation may only be noticeable over months, maybe even years.

As the framework shows in Figure~\ref{fig:DT-Framework}, there are multiple time-scales throughout the structure, varying across the different lines of communication between the modules.
This particular framework set-up features four time-scales: $k < \beta k < \gamma k < \delta k$ in ascending order, where the time-step $k\in\mathbb{Z}_{\geq}$ is scaled by $\beta, \gamma, \delta \in \mathbb{N}$.
The single time-step $k$ is the finest time margin and set by the Virtual Asset to ensure the surrogate simulations are refined enough for meaningful analysis.
The largest timescale $\delta$ means that task is only evaluated every $\delta$ time-steps, and is used for managing long-term goals that mainly pertain to the Virtual to Physical modules.

\subsection{Physical Asset}
The Physical Asset is the real-life entity that the digital twin is integrated into.
In this work, the Physical Asset is the reactor configuration in the physics-based SAM gFHR model as presented in Section~\ref{sec:FHR}.
The offline phase of this module involves preparing a system to be operative, beginning with research \& development; to planning \& construction; to testing \& qualification.
No experimental facility exists for this exact reactor configuration, therefore the SAM gFHR model emulates the physical plant.

The FHR plant is operated according to the target power and is notified of any maintenance actions by the Virtual to Physical modules.
As a result, the system will exhibit transient behaviors such controller activation, system state transitions, and inevitable component degradation. 
Observational measurements are then collected from the system to inform the Virtual Asset of the Physical Asset's current status and health.
The Physical Asset $F_{\text{P}}(\cdot)$ for the FHR plant is mathematically represented as follows:
\begin{align}
    \mathbf{d}^{(\beta k)} = F_{\text{P}} \left(\dot{Q}_{\text{RX,C}}^{(\gamma k)}, \mathbf{M}_{\text{pump}}^{(\delta k)}\right).
\end{align}
The target power profile $\dot{Q}_{\text{RX,C}}^{(\gamma k)}$ comes from the Virtual to Physical module, where the admissible power generation has been processed by two submodules in the Virtual to Physical component. 
As the Physical Asset operates, a new target power value is supplied every $\gamma$ time-steps, for which the reactor must operate at until the next target power generation value is specified. 
The maintenance notifications $\mathbf{M}_{\text{pump}}$ for each pump are determined by the Operator submodule in the Virtual to Physical module, which can only be replaced or refurbished every $\delta$ time-steps. 
Measurements $\mathbf{d}$ are available (or are taken) every $\beta$ time-steps and passed to the Physical to Virtual Assimilation submodule.

\subsection{Virtual Asset}
The Virtual Asset is the computerized model representation of the Physical Asset. 
This consists of a system-level model of digital system state values to describe the actual asset, and could also include additional supplemental sub-virtual models.
The Virtual Asset is a foundational component of the digital twin, providing continuous up-to-date status reports for monitoring and real-time predictions for informing future actions. 
Thus, in order for the digital twin to be effective and impactful, the Virtual Asset must be both computationally efficient and highly reliable.
The Virtual Asset consists of two submodules --- the gFHR Surrogate and the Pump Health Surrogate --- for the FHR application.
A detailed account of these surrogate developments are published in the report by Lim et al.~\cite{Lim:25}, and supplemental information to support this work is found in~\ref{app:gFHRsurrogate}.

This module interacts with virtually all aspects of the digital twin, but is heavily used by the Virtual to Physical modules to drive Physical Asset actions.
To keep the Virtual Asset up-to-date with the Physical Asset, the Physical to Virtual Module processes observational measurements $\mathbf{d}^{(\beta k)}$ for online adjustments. 
Adjustments involve correcting the Virtual Asset estimation of the digital states and recalibration of characteristic parameters.
Digital states are typically more detailed than the observable quantities, containing all the measurable states and more.
Parameters determine the Virtual Asset performance; for instance in physics-based models these could be characteristic quantities such as material properties or component sizes, and in data-driven models these could be trainable coefficients such as weights and biases values in a neural network.
The digital twin will leverage collected data to update both states and parameters for online correction, allowing the virtual models to evolve with the asset.

The gFHR Surrogate Model provides the system state-space estimation, and the Pump Health Surrogate is a component-level model that monitors the pump HSCs.
Within the digital twin framework, the Virtual Asset $F_{\text{V}}(\cdot)$ is mathematically represented by:
\begin{equation}
    \mathbf{x}_{\text{sys}}^{(k)}, \mathbf{x}_{\text{pump}}^{(k)} = F_{\text{V}} \left(\mathbf{x}_{\text{sys}}^{(k-1)}, \mathbf{x}_{\text{pump}}^{(k-1)}, \dot{Q}_{\text{RX,C}}^{(\gamma k)}; \boldsymbol{\theta}_{\text{sys}}^{(\beta k)},\boldsymbol{\theta}_{\text{pump}}, \mathbf{M}_{\text{pump}}^{(\delta k)}\right),
\end{equation}
where the output is the system state $\mathbf{x}_{\text{sys}}$ and pump health state $\mathbf{x}_{\text{pump}}$ at time-step $k$. 
The inputs to the Virtual Asset are the previous time-step states $\mathbf{x}_{\text{sys}}$ and $\mathbf{x}_{\text{pump}}$; the admissible target power distribution $\dot{Q}_{\text{RX,C}}$; the gFHR Surrogate Model parameters $\boldsymbol{\theta}_{\text{sys}}$; the pump degradation parameters $\boldsymbol{\theta}_{\text{pump}}$; and the pump maintenance actions $\mathbf{M}_{\text{pump}}$.
The pump degradation parameters $\boldsymbol{\theta}_{\text{pump}}$ are the deterministic values set by the user, and the maintenance actions $\mathbf{M}_{\text{pump}}$ are set by the Virtual to Physical Operator submodule that can be specified every $\delta$ time-steps.
The gFHR Surrogate Model parameters $\boldsymbol{\theta}_{\text{sys}}$ are the trainable coefficients in the data-driven model; and will be explicitly identified in~\ref{app:gFHRsurrogate} along with $\mathbf{x}_{\text{sys}},\mathbf{x}_{\text{pump}}$ and $\boldsymbol{\theta}_{\text{pump}}$.

\subsection{Virtual to Physical Module}
The role of the Virtual to Physical Module is to drive inputs for the Physical Asset such that system goals are met.
In this work, there are three goals for the FHR plant: i) aim to meet the desired power generation requested for the supervised plant as required by its associated energy grid, ii) reduce operations and maintenance (O\&M) costs, and iii) satisfy safety envelops.
Manipulatable actions to reduce O\&M costs are optimally setting the target power generation and best managing wear-and-tear, health, and predictive maintenance.
Safety assurance is provided by continuously monitoring of the system through the Virtual Asset and ensuring the status of the HSCs remain in an acceptable region.
Two submodules make up this module, the Operator submodule and the Constraint submodule, whose methods will now be introduced.

\subsubsection{Operator}

Integral system wear-and-tear accumulates overtime, typically irreversibly and stochastically, which is driven by system operation and dynamics.
Maintenance is scheduled to restore (either partially or completely) component health, system efficiency, and satisfy regulatory constraints. 
Coordinating maintenance for various heterogeneous components in a stochastic environment is a complex problem, where multiple expected rates of degradation and uncertainty must be considered.

The Operator module functions as the operations and maintenance supervisor.
It uses a Reinforcement Learning (RL) based supervision policy developed by Pylorof and Garcia~\cite{PylorofGarcia:22} with the Soft Actor Critic~\cite{haarnoja:18} algorithm. 
This RL-based supervision uses an approach inspired by the \textit{receding horizon principle}, where a policy is constructed with a certain reasoning horizon (months/years) but is applied for a shorter horizon (weekly/monthly). 
As the digital twin is reconfigured with new information, the policy is reconstructed as events in the reasoning horizon move closer to the actual applied horizon. 
The RL solution is capable of handling both continuous and discrete setpoints with embedded predictive knowledge of the dynamics-driven degradation and running test scenarios through the digital twin.

Given the desired power generation $\dot{Q}_{\text{RX,T}}$ every $\delta$ time-steps (as tentatively-requested for generation by the energy grid where the supervised generation units is located), the Operator submodule determines the optimized target power generation $\dot{Q}_{\text{RX,O}}$ that best balances competitive objectives such as trying to closely meet the power generation requested by the grid to the supervised unit under consideration while accommodating health and maintenance objectives. 
The module also uses wear-and-tear dynamics provided about the defined HSCs to schedule plant shutdown for maintenance actions $\mathbf{M}_{\text{pump}}$. 
Given the current pump health status $\mathbf{x}_{\text{pump}}$, the Operator submodule is mathematically represented as:
\begin{equation}
    \dot{Q}_{\text{RX,O}}^{(\delta k)}, \mathbf{M}_{\text{pump}}^{(\delta k)} = F_{O}\left(\dot{Q}_{\text{RX,T}}^{(\delta k)}, \mathbf{x}_{\text{pump}}^{(k)}\right).
\end{equation}

\begin{figure}
    \centering
    \includegraphics[width=\linewidth]{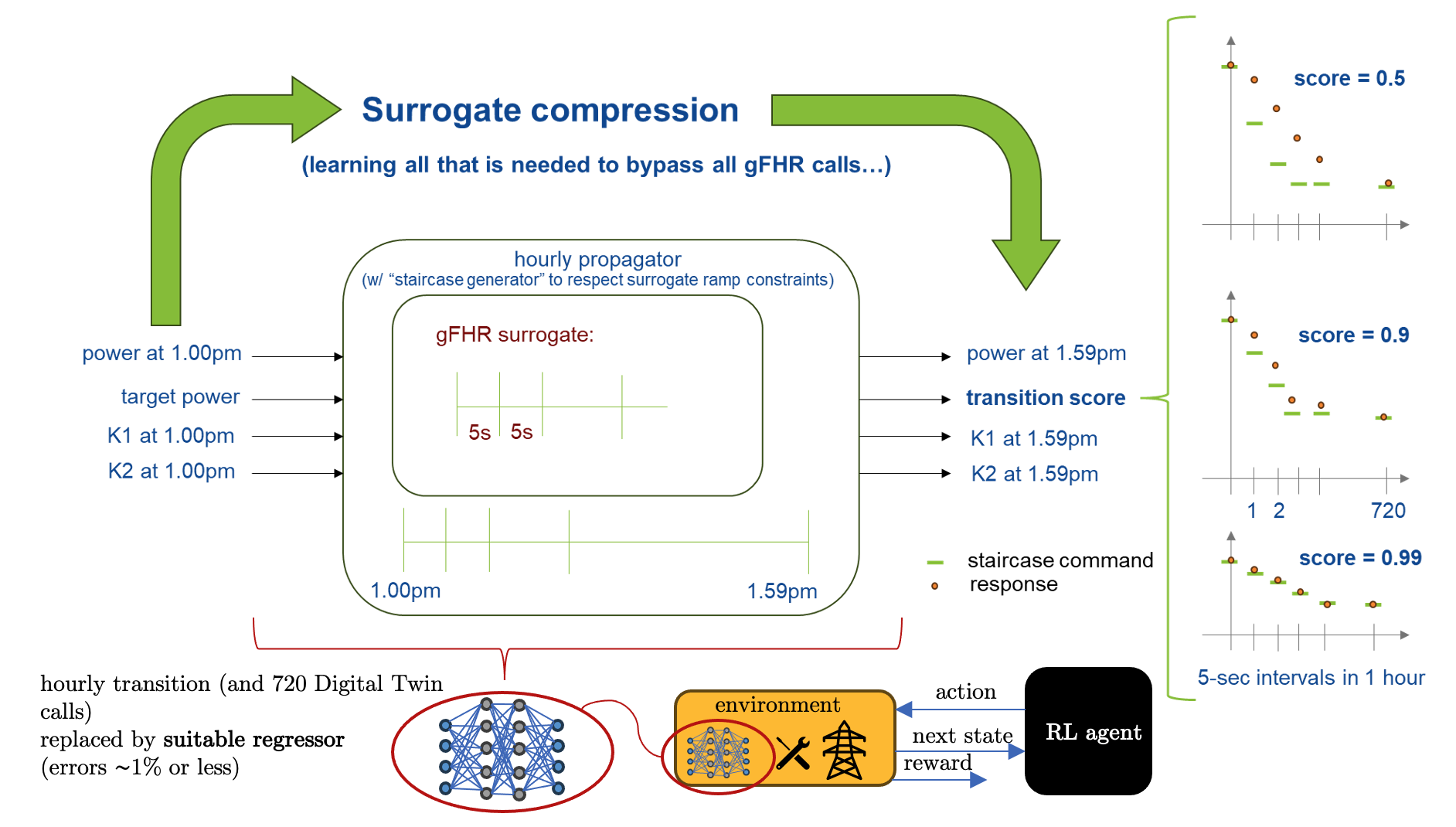}
    \caption{Surrogate Compression of the Virtual Asset to enable Reinforcement Learning training. The compressed model only outputs the power and pump degradation rates $K_i$ on an hourly timescale. }
    \label{fig:surr-compress}
\end{figure}

It is known that machine learning models require generous amounts of data for training, and using the higher-fidelity SAM gFHR model is not feasible due to its long run-times. 
Therefore, the Operator submodules will use data generated from the Virtual Asset gFHR Surrogate and the Pump Health Surrogate.
The fine-grained Virtual Asset timescale $k$ is re-scoped to the Operator timescale $\delta k$ using a surrogate compression, where data generated from the Virtual Asset is used to train a neural network.
In this compressed surrogate, shown in Figure~\ref{fig:surr-compress}, the inputs are the: target power, power at the beginning of a given time period, and the pump degradation rates at the top of the hour (e.g., xx:00); and the outputs are: the power, pump degradation rates at the end of the hour (e.g., xx:59), and the transition score.
The transition score is low when a power transition results in larger stressing dynamics on a system, typically this is a big power step where the system takes a while to reach the target.
A high transition scores are placed on small power steps, where the transitions are fast.
Using the power predicted by the gFHR Surrogate, the degradation rates predicted by the Pump Health Surrogate, the resulting surrogate compression generates 60 months of load-following evaluations in approximately one CPU second, enabling Deep RL iterations. 
The Soft-Actor Critic RL agent is built with the python \texttt{Stable-Baselines3}~\cite{stable-baselines3} library.

\subsubsection{Constraint}

Assuring the safety of a system involves limiting operations to stay within an acceptable region.
The Constraint submodule ensures that the plant is operating under conditions that do not violate set system constraint(s), intervening only when necessary.
It uses the Reference Governor (RG)~\cite{BempoardMosca:94}, a constraint enforcement algorithm that enforces \textit{pointwise-in-time} state and control constraints~\cite{kolmanovsky:14}.
This Constraint submodule was developed and constructed specifically for the SAM gFHR model by Dave et al.~\cite{Dave:2023}. In the SAM gFHR model, the following three constraints are set:
\begin{enumerate}
    \item Primary Pump Mass Flow Rate $\dot{m}_{P,p} \geq \dot{m}_{P,p}^{\text{min}} = 720$ kg/s;
    \item Secondary Pump Mass Flow Rate $\dot{m}_{P,s} \geq \dot{m}_{P,s}^{\text{min}}=1000$ kg/s; and
    \item IHX Secondary Outlet Temperature $T_{\text{ihx,s,out}} \leq T_{\text{ihx,s,out}}^{\text{max}} = 890$ K.
\end{enumerate}
The bounds are chosen to ensure that the FHR reactor is operating well within the 50\% to 100\% power range. 
The design choice process considered the accuracy of the gFHR surrogate predictions and selecting states that could be realistically monitored with sensors in the reactor setting. 
It is noted that in a real-world system, the types, and number of system states could be different from the ones that are selected for this work and the three chosen constraints are to showcase the RG implementation.

\begin{figure}
    \centering
    \includegraphics[width=0.5\linewidth]{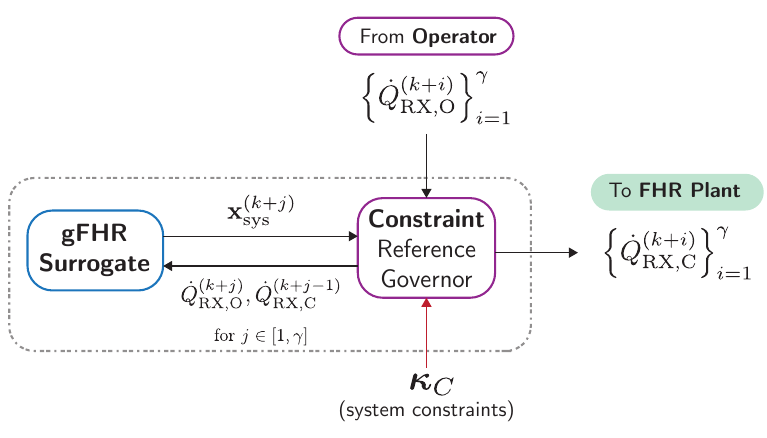}
    \caption{Constraint Sub-Module Functionality. Using the Reference Governor (RG) algorithm, the Constraint submodule interacts with the Virtual Asset gFHR Surrogate to enforce system constraints.}
    \label{fig:FHR-RG-framework}
\end{figure}

Within the digital twin framework, the RG is represented functionally by $F_{\text{C}}(\cdot)$, which outputs the admissible power generation $\dot{Q}_{\text{RX,C}}^{(k)}$ given the Operator submodule set optimized target power generation $\dot{Q}_{\text{RX,O}}$, the previous admissible power $\dot{Q}_{\text{RX,C}}^{(k-1)}$, and the constraints $\boldsymbol{\kappa}_{\text{C}}$. 
This is mathematically written as:
\begin{equation}
    \dot{Q}_{\text{RX,C}}^{(k:k+\gamma)} = F_{\text{C}} \left(\dot{Q}_{\text{RX,O}}^{(k:k+\delta)}, \boldsymbol{\kappa}_{\text{C}}\right).
\end{equation}
At each time-step, the RG uses the gFHR Surrogate to evaluate the system response behavior $\mathbf{x}_{\text{sys}}^{(k)}$ under the conditions set by the requested input $\dot{Q}_{\text{RX,O}}$.
If the system surrogate model predicts a violation to the constraints, then the RG will intervene and modify the input to $\dot{Q}_{\text{RX,O}}$.
This process is depicted in Figure~\ref{fig:FHR-RG-framework}.

\begin{figure}
    \centering
    \includegraphics[width=\linewidth]{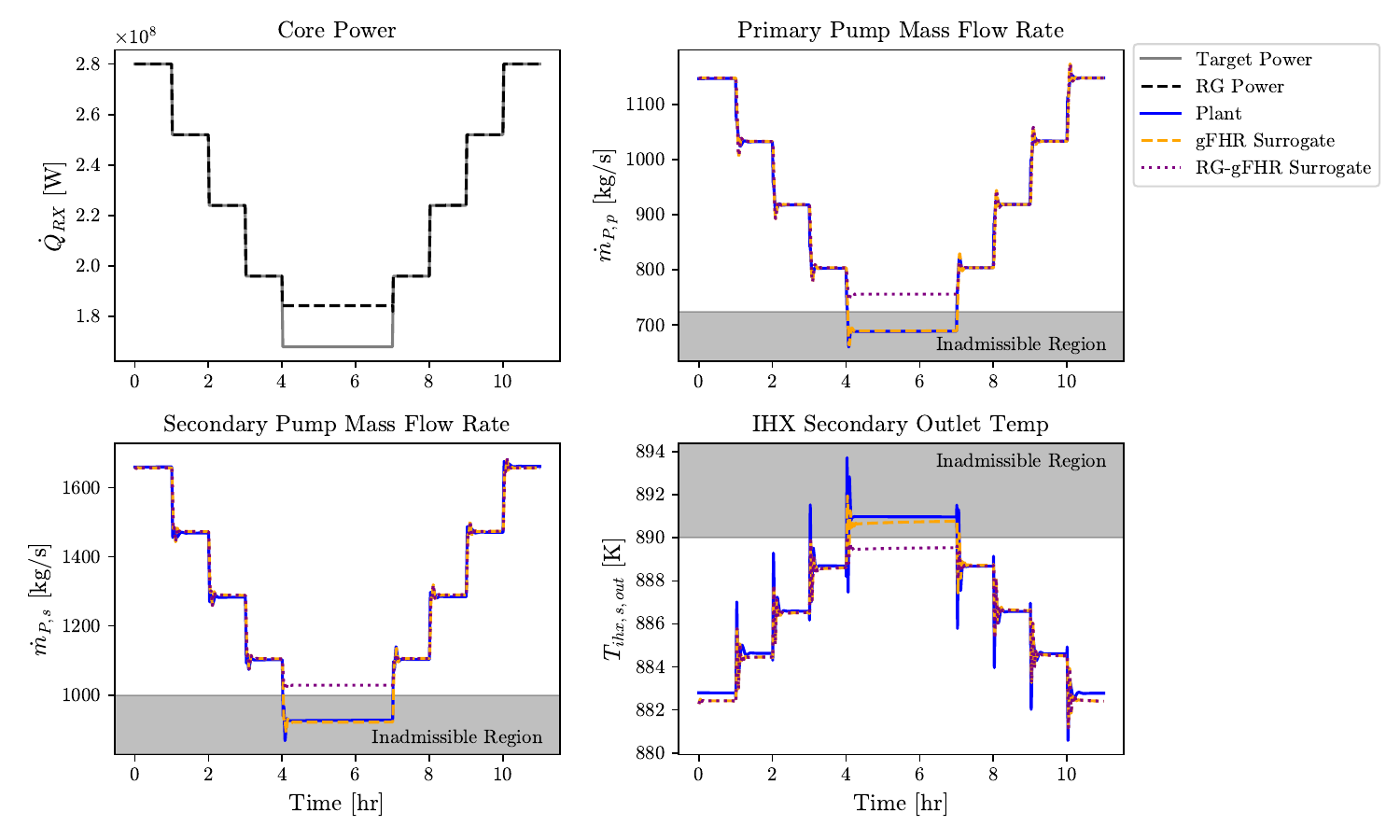}
    \caption{Demonstration of the Reference Governor (RG) in the Constraint submodule intervening in a load-following case.}
    \label{fig:RG_example}
\end{figure}

To demonstrate the RG, the load-following case in Figure~\ref{fig:RG_example} is considered. 
As shown in the results, the user's requested power distribution request would result in $T_{ihs,s,out}$ going beyond the set maximum, thus the RG intervenes and adjusts the target power. 
In this example, the RG successfully intervenes based off the surrogate model prediction.
However, the surrogate model is slightly underpredicting the micro-dynamic response of the temperature, resulting in the input actually violating the constraint when simulated in SAM.
The effectiveness of the RG is directly dependent on the accuracy of the surrogate model, and in this case the errors in the surrogate model prediction in the power transition regions results in a fault in the constraint. 

\subsection{Physical to Virtual Module}
The transient response of the Physical Asset is communicated to the Virtual Asset module through collected data. 
Data measurements are processed by the Physical to Virtual Module through data assimilation, activating online adaptation of the Virtual Asset. 
This online adaptation is necessary for two reasons, one for fine-tuning the virtual models and two, for updating them when the system undergoes a change.
In the offline preparation, the Virtual Module may only be exposed to data that represents population-averaged dynamics, thus further online corrections can tailor the digital twin to an individual system's unique characteristics.
Additionally, it is expected that the Physical Asset will change overtime, whether its expected usage effects or anomalous events. 
The wear-and-tear overtime will be visible through changes in the transient physical behaviors, with irreversible effects that can never be fully remedied. 
In such cases, the effected digital twin modules are notified through this module. 

For the FHR application, the Physical to Virtual Module contains the Assimilation submodule that uses Bayesian inference to correct the Virtual Module system states and parameters. 
The Physical to Virtual Module $F_{\text{PV}}(\cdot)$ is represented mathematically as:
\begin{equation}
    \mathbf{x}_{\text{sys},A}^{(\beta k)}, \boldsymbol{\theta}_{\text{sys,A}}^{(\beta k)} = F_{\text{PV}}\left(\mathbf{d}^{(\beta k)}\right),
\end{equation}
where measurements $\mathbf{d}$ from the Physical Asset are taken every $\beta$ time-steps. 
This module yields the assimilated states $\mathbf{x}_{\text{sys,A}}\subseteq \mathbf{x}_{\text{sys}}$ and the assimilated parameters $\boldsymbol{\theta}_{\text{sys,A}} \subseteq \boldsymbol{\theta}_{\text{sys}}$, both of which are a subset or equal to their full representations.
The assimilated values are then passed to the Virtual Asset to update the digital models every $\beta$ time-steps.

To set up the Assimilation submodule, consider a discrete non-linear system with the state realization $\mathbf{x} \in \mathbb{R}^{n_x}$ of the multivariate random variable $\mathbf{X}$ and output realization $\mathbf{y} \in \mathbb{R}^{n_y}$ of the multivariate random variable $\mathbf{Y}$:
\begin{align}
    \mathbf{x}^{(k+1)} &= \Phi\left(\mathbf{x}^{(k)}\right) + \boldsymbol\xi, \quad \boldsymbol\xi\sim \mathcal{N}(0,\boldsymbol{\Sigma}) \label{eq:kf_dynamics}\\
    \mathbf{y}^{(k)} &= h \left(\mathbf{x}^{(k)}\right) + \boldsymbol{\eta}, \quad \boldsymbol{\eta} \sim \mathcal{N}(0, \boldsymbol{\Gamma}),\label{eq:kf_obs}
\end{align}
where $\Phi(\cdot)$ is the dynamics model and $h(\cdot)$ is the output model. 
The covariance matrices $\boldsymbol{\Sigma}\in\mathbb{R}_{\geq}^{(n_x\times n_x)}$ and $\boldsymbol{\Gamma}\in\mathbb{R}_{\geq}^{(n_y\times n_y)}$ are for the dynamics model noise $\boldsymbol\xi$ and measurement noise $\boldsymbol\eta$ terms, respectively.
It is assumed that the initial state $\mathbf{x}^{(k)}$, the process noise $\boldsymbol{\xi}$, and the measurement noise $\boldsymbol{\eta}$ are independent and identically distributed (i.i.d).

In this dynamical system setting, we are interested in using Bayesian inference to solve the inverse problem of predicting hidden or unknown state values $\mathcal{X}^{(0:k)} = \left\{\mathbf{x}^{(0)}, \mathbf{x}^{(1)},\hdots, \mathbf{x}^{(k)}\right\}$ given observed measurements $\mathcal{Y}^{(1:k)} = \left\{\mathbf{y}^{(1)},\hdots, \mathbf{y}^{(k)}\right\}$. 
This is represented by the \textit{posterior} probability distribution $P\left(\mathcal{X}^{(0:k)}\mid \mathcal{Y}^{(k)}\right)$, where the states up to time-step $k$ are inferred using all data $\mathcal{Y}$ up to time-step $k$. 
This joint probability distribution is constructed using Bayes' rule:
\begin{equation}
    P\left(\mathcal{X}^{(0:T)}\mid \mathcal{Y}^{(1:T)}\right) = \frac{ P\left(\mathcal{Y}^{(1:T)}\mid\mathcal{X}^{(0:T)}\right) P\left(\mathcal{X}^{(0:T)}\right) } { P\left(\mathcal{Y}^{(1:T)}\right)},
\end{equation}
where $P\left(\mathcal{Y}^{(1:T)}\mid\mathcal{X}^{(0:T)}\right)$ is the \textit{likelihood} distribution of the measurements, $P\left(\mathcal{X}^{(0:T)}\right)$ is the \textit{prior} distribution defined by the dynamics model, and $P\left(\mathcal{Y}^{(1:T)}\right)$ is the \textit{evidence} normalization constant defined as:
\begin{equation}
    P\left(\mathcal{Y}^{(1:T)}\right) = \int P\left(\mathcal{Y}^{(1:T)}\mid \mathcal{X}^{(0:T)}\right)P\left(\mathcal{X}^{(0:T)}\right) d \mathbf{x}^{(0:T)}.
\end{equation}
Using this Bayesian approach in the digital twin framework allows the Virtual Asset to incorporate information from measurement data and embeds uncertainty quantification with the probabilistic representation.
The disadvantage of the full posterior distribution, however, is that it must be re-computed each time a new measurement is obtained. 
Additionally, as the time step increases, the dimensionality of the full posterior calculation increases and therefore increasing the computational complexity. 
For models that will run for long time horizons with many time steps, the computation of the integrals are at risk to eventually become intractable. 

To overcome this issue, Bayesian filtering uses a recursive approach to update the posterior distribution at each time-step.
Filtering assumes that the dynamics model is a sequential Markovian process, where the current state is only dependent on the previous state, thus probability distribution becomes:
\begin{equation}
    P\left(\mathbf{x}^{(k+1)} \mid \mathbf{x}^{(k)}, \hdots, \mathbf{x}^{(0)}\right) \triangleq P\left(\mathbf{x}^{(k+1)}\mid \mathbf{x}^{(k)}\right).
\end{equation}
The likelihood distribution of the observational is also assumed to only depend on the current state value:
\begin{equation}
    P\left(\mathbf{y}^{(k)}\mid \mathcal{X}^{(0:k)}\right) \triangleq P\left(\mathbf{y}^{(k)}\mid \mathbf{x}^{(k)}\right).
\end{equation}
Beginning with a prior distribution $P\left(\mathbf{x}^{(0)}\right)$, this sequential assimilation process is performed in two steps:
\begin{enumerate}
    \item Prediction: using the posterior distribution at time-step $k-1$ to predict the state at time-step $k$ with marginalization using the \textit{Kolmogorov} equation.
    \begin{align}
        P\left(\mathbf{x}^{(k)}\mid \mathcal{Y}^{(k-1)}\right) &= \int P\left(\mathbf{x}^{(k)},\mathbf{x}^{(k-1)}\mid\mathcal{Y}^{(k-1)}\right)d\mathbf{x}^{(k-1)} \\
        &=\int P\left(\mathbf{x}^{(k)}\mid\mathbf{x}^{(k-1)}\right) P\left(\mathbf{x}^{(k-1)}\mid\mathcal{Y}^{(k-1)}\right)d\mathbf{x}^{(k-1)}\notag
    \end{align}
    \item Update: given a measurement at time step $k$, apply Bayes' rule to update the posterior distribution of state at time $k$.
    \begin{equation}
        P\left(\mathbf{x}^{(k)}\mid \mathcal{Y}^{(1:k)}=\left\{\mathbf{y}^{(k)},\mathcal{Y}^{(1:k-1)}\right\}\right) = \frac{1}{Z^{(k)}} P\left(\mathbf{y}^{(k)}\mid \mathbf{x}^{(k)}\right)P\left(\mathbf{x}^{(k)}\mid \mathcal{Y}^{(1:k-1)}\right),
    \end{equation}
    where the normalization constant $Z^{(k)}$ is:
    \begin{equation}
        Z^{(k)} = \int P\left(\mathbf{y}^{(k)} \mid \mathbf{x}^{(k)}\right) P \left(\mathbf{x}^{(k)} \mid \mathcal{Y}^{(1:k-1)}\right) d \mathbf{x}^{(k)}.
    \end{equation}
\end{enumerate}

\subsubsection{Ensemble Kalman Filter}
The Kalman Filter (KF)~\cite{Kalman:60} is an algorithm used to solve the filtering prediction and update equations for a linear system. 
The Ensemble Kalman Filter (EnKR)~\cite{Evensen:94} applies a stochastic approach to the KF, using an ensemble of samples to produce a Monte Carlo estimation of the posterior mean and covariance. 
This ensemble approach makes the method suitable for nonlinear dynamics with high dimensional states space~\cite{Evensen:03} because instead of costly covariance matrix update step, the KF error statistics equation is replaced with an ensemble constructed estimate.
Additionally, the EnKF is non-intrusive and does not require any internal information about the dynamical system, such as Jacobian matrices.

The first step in Bayesian filtering is prediction. 
Assuming the initial prior distribution of $\mathbf{x}^{(0)}$ is Gaussian:
\begin{equation}
    p\left(\mathbf{x}^{(0)}\right) \triangleq \mathcal{N}\left(\boldsymbol\mu^{(0)},\mathbf{C}^{(0)}\right),
\end{equation}
and with the linear Gaussian assumption shown in Equation~\ref{eq:kf_dynamics}, the resulting prediction marginal equation for $\mathbf{x}^{(k)}$ will also be Gaussian.
At time-step $k$, the already computed filtering distribution of the previous state $\mathbf{x}^{(k-1)}$:
\begin{equation}
    p\left(\mathbf{x}^{(k-1)}\mid \mathcal{Y}^{(1:k-1)}\right) \triangleq \mathcal{N}\left(\boldsymbol\mu^{(k-1)}, \mathbf{C}^{(k-1)}\right),
\end{equation}
is available and utilized to determine the distribution of $\mathbf{x}^{(k)}$:
\begin{equation}
    p\left(\mathbf{x}^{(k)}\mid \mathcal{Y}^{(1:k-1)}\right)\triangleq\mathcal{N}\left(\boldsymbol\mu^{(k)}, \mathbf{C}^{(k)}\right),
\end{equation}
where $\boldsymbol\mu^{(k)}\in\mathbb{R}^{n_x}$ is the prediction mean vector and $\mathbf{C}^{(k)}\in\mathbb{R}_{\geq}^{(n_x\times n_x)}$ is the prediction covariance matrix.

With the EnKF, $n_m$ state realizations are stored in an ensemble matrix $\mathbf{X}^{(k)}\in\mathbb{R}^{(n_x\times n_m)}$:
\begin{equation}
    \mathbf{X}^{(k)} \triangleq \begin{bmatrix} \mathbf{x}^{(k)}_1& \ldots& \mathbf{x}^{(k)}_{n_m}\end{bmatrix}.
\end{equation}
To represent the error statistics at each time step, the mean vector $\boldsymbol{\mu}^{(k)}$ is approximated by the ensemble mean vector $\bar{\mathbf{x}}^{(k)}\in\mathbb{R}^{n_x}$ defined by:
\begin{equation}
    \boldsymbol{\mu} \approx \bar{\mathbf{x}}^{(k)} \triangleq \frac{1}{n_m} \sum_{j=1}^{n_m} \mathbf{x}^{(k)}_j.
\end{equation}
With the mean estimate, the state ensemble centered matrix $\bar{\mathbf{X}}^{(k)} \in \mathbb{R}^{(n_x \times n_m)}$ captures the sample deviations (or anomalies) from the mean:
\begin{equation}
    \bar{\mathbf{X}}^{(k)} \triangleq \begin{bmatrix} \mathbf{x}^{(k)}_1 - \bar{\mathbf{x}}^{(k)}& \ldots& \mathbf{x}^{(k)}_{n_m}-\bar{\mathbf{x}}^{(k)}\end{bmatrix}.
\end{equation} 
This state ensemble centered matrix is used to approximate the covariance matrix $\mathbf{C}^{(k)}$ with the ensemble error matrix $\bar{\mathbf{P}}^{(k)}\in\mathbb{R}^{(n_m\times n_m)}$ given by:
\begin{equation}
    \mathbf{C}^{(k)} \approx \bar{\mathbf{P}}^{(k)}\triangleq \frac{1}{n_m-1} \bar{\mathbf{X}}^{(k)}\bar{\mathbf{X}}^{(k)^T}.
\end{equation}

For the first step of filtering, a prediction ensemble is computed using the ensemble from the previous time-step.
With the ensemble of $n_m$ realizations $\boldsymbol{\xi}_j$ of the process noise random variable $\boldsymbol{\xi}$ are stored in matrix $\mathbf{V}^{(k)}\in\mathbb{R}^{(n_x\times n_m)}$:
\begin{equation}
    \mathbf{V}^{(k)} \triangleq \begin{bmatrix}
        \boldsymbol{\xi}_1 & \hdots & \boldsymbol{\xi}_{n_m}
    \end{bmatrix},
\end{equation}
the ensemble states are predicted following the non-linear dynamical model defined in Equation~\ref{eq:kf_dynamics}:
\begin{equation}
    \mathbf{X}^{(k+1)} = \Phi\left(\mathbf{X}^{(k)}\right) + \mathbf{V}^{(k)}.
\end{equation}

In the second step of filtering, the prediction ensemble is processed to obtain the filtering (or assimilated) ensemble $\mathbf{X}_a^{(k)}\in\mathbb{R}^{(n_x \times n_m)}$ when a new measurement $\mathbf{d}^{(k)}\in\mathbb{R}^{n_y}$ is given. 
Using the prediction ensemble of the state, the corresponding output ensemble matrix $\mathbf{Y}^{(k)}\in\mathbb{R}^{(n_y \times n_,)}$ is computed following the non-linear output model defined in Equation~\ref{eq:kf_obs}:
\begin{equation}
    \mathbf{Y}^{(k)} = h\left(\mathbf{X}^{(k)}\right) + \mathbf{E}^{(k)},
\end{equation}
where $\mathbf{E}^{(k)}\in\mathbb{R}^{(n_y \times n_m)}$ contains $n_m$ realizations $\boldsymbol{\eta}_j$ of the output noise random variable $\boldsymbol{\eta}$:
\begin{equation}
    \mathbf{E}^{(k)} \triangleq \begin{bmatrix}
        \boldsymbol{\eta}_1 & \hdots &\boldsymbol{\eta}_{n_m}
    \end{bmatrix}
\end{equation}
The output mean vector $\bar{\mathbf{y}}^{(k)}\in\mathbb{R}^{n_y}$ and the output ensemble centered matrix $\bar{\mathbf{Y}}^{(k)}\in\mathbb{R}^{(n_y \times n_m)}$ are defined as:
\begin{align}
    \mathbf{y}^{(k)} &\triangleq \frac{1}{n_m}\sum_{j=1}^{n_m} \mathbf{y}^{(k)}_j \\
    \bar{\mathbf{Y}}^{(k)} &\triangleq \begin{bmatrix} \mathbf{y}^{(k)}_1 - \bar{\mathbf{y}}^{(k)}& \ldots& \mathbf{y}^{(k)}_{n_m}-\bar{\mathbf{y}}^{(k)}\end{bmatrix}.
\end{align}
The samples in the ensemble state matrix $\mathbf{X}^{(k)}$ are updated to the ensemble state assimilated matrix $\mathbf{X}_a^{(k)}\in\mathbb{R}^{(n_x\times x_m)}$ by:
\begin{align}
    \bar{\mathbf{K}}^{(k)} &= \frac{1}{n_m-1}\bar{\mathbf{X}}^{(k)}\bar{\mathbf{Y}}^{(k)^T}\left(\frac{1}{n_m-1}\bar{\mathbf{Y}}^{(k)}\bar{\mathbf{Y}}^{(k)^T}\right)^{-1} \\
    \mathbf{X}_a^{(k)} &= \mathbf{X}^{(k)} + \bar{\mathbf{K}}^{(k)}\left(\mathbf{d}^{(k)}\mathbf{1}_{n_m}^T - \mathbf{H} \mathbf{X}^{(k)}\right),
\end{align}
where $\bar{\mathbf{K}}^{(k)}\in \mathbb{R}^{(n_x\times n_y)}$ is the ensemble approximated Kalman Gain matrix and $\mathbf{1}_{n_m}\in \mathbb{R}^{n_m}$ is a ones column vector. 
Finally, the updated mean vector is approximated by the updated ensemble mean $\bar{\mathbf{x}}_a^{(k)}\in\mathbb{R}^{n_x}$:
\begin{equation}
    \boldsymbol{\mu}^{(k)}_a \approx \bar{\mathbf{x}}_a^{(k)} \triangleq \frac{1}{n_m}\sum_{j=1}^{n_m} \mathbf{x}_{a_j},
\end{equation}
and the updated covariance matrix is approximated by the updated ensemble centered matrix $\bar{\mathbf{X}}_a^{(k)}$:
\begin{align}
    \mathbf{C}_a^{(k)} \approx \bar{\mathbf{X}}_a^{(k)} \triangleq \begin{bmatrix}\mathbf{x}_{a_1}^{(k)} - \bar{\mathbf{x}}_{a}^{(k)} & \hdots & \mathbf{x}_{a_{n_m}}^{(k)} - \bar{\mathbf{x}}_{a}^{(k)}\end{bmatrix}.
\end{align}

The EnKF acts as a reduced complexity model of the KF because the explicit computation of the covariance matrix is avoided through the ensemble low-rank approximation. 
The size of the ensemble $n_m<n_x$ (usually $n_m \ll n_x$) is a heuristic parameter that is large enough such that the error statistics are adequately modeled for data assimilation to be informative~\cite{burgers:98}.
As $n_m \rightarrow \infty$, the EnKF mean and covariance estimate converge to the KF solution.
However, with the finite $n_m$ estimation, the EnKF is susceptible to some unfavorable consequences such as underestimation of the uncertainty due to too few samples.

\subsubsection{State Estimation}

The EnKF is implemented for state estimation of the gFHR Surrogate model system state vector $\mathbf{x}_{\text{sys}}\in\mathbb{R}^{42}$. 
Since the gFHR surrogate model requires two time-delays (defined in~\ref{app:gFHRsurrogate}), the estimation state vector $\mathbf{x}^{(k)}\in\mathbb{R}^{84}$ stacks the two system state vectors at time-steps $k$ and $k-1$. 
The EnKF state vector $\mathbf{x}^{(k)}$ is therefore defined as:
\begin{equation}
    \mathbf{x}^{(k)} \triangleq \begin{bmatrix} \mathbf{x}_{\text{sys}}^{(k-1)} & \mathbf{x}_{\text{sys}}^{(k)} \end{bmatrix}^T.
\end{equation}
For state estimation, the EnKF dynamics function is the gFHR Surrogate model $F_{\text{sys}}(\cdot)$ except the two time-delays are replaced with the defined EnKF state vector:
\begin{align}
    \Phi\left(\mathbf{x}^{(k)}\right) &\triangleq F_{\text{sys}}\left(\mathbf{x}^{(k)}, \mathbf{u}_{\text{sys}}^{(k+1)}; \boldsymbol{\theta}_{\text{sys}}^{(\beta k)}\right). \label{eq:stateDynamics}
\end{align}
Note that the EnKF filter only estimates the system state variable and all other inputs to the dynamics model in the above equation are deterministic. 

For this study, the core power $\dot{Q}_{RX}$ is the only observed quantity, thus the output model is the following linear operator:
\begin{align}
    h\left(\mathbf{x}^{(k)}\right) &\triangleq \mathbf{H}\mathbf{x}^{(k)}\\
    \mathbf{H} &\triangleq \text{diag}(\mathbf{h}) \\
    h_i &\triangleq \left\{ \begin{array}{ll}
      1 & \text{if }x_i = \dot{Q}_{RX} \\
      0 & \text{otherwise} \\
      \end{array} \right. .
\end{align}

The EnKF state estimation is evaluated in a toy case where the core power is observed for the steady-state reactor power at 100\% for one hour. 
Using $n_m = 25$ samples, the ensemble is initialized with the initial covariance $\mathbf{C}^{(k)} = 1\times 10^{-6} \mathbb{I}_{n_x}$, where $\mathbb{I}_{n_x}\in \mathbb{R}^{(n_x \times n_x)}$ is the identity matrix. 
The dynamics model noise covariance matrix is $\mathbf{\Sigma} = 1\times 10^{-9} \mathbb{I}_{n_x}$ and the observation noise covariance matrix is $\mathbf{\Gamma} = 1\times10^{-10} \mathbb{I}_{n_x}$. 
Data is observed every $\beta=20$ steps.

The core energy state prediction results are shown in Figure \ref{fig:enkf-state} where the data, target core power, the gFHR Surrogate prediction, and the EnKF state estimation are plotted. 
Here, the originally trained gFHR Surrogate model is over-predicting core power, so using the EnKF state estimation is valuable in correcting state values. 
However, with this state-only assimilation, the EnKF does not improve the subsequent prediction performance and is actually introducing an oscillatory-like behavior. 
The ineffectiveness of the state-only approach is because while the state values are corrected, the gFHR Surrogate is not and is continuing to overpredict the reactor power in between observations.
In order to make the virtual model accurate, the surrogate's parameters need to also be assimilated. 

\subsubsection{State Parameter Estimation}

With state only estimation, the EnKF is limited and can only apply the corrections to the state themselves. 
To overcome this, a state-parameter estimation is achieved using the \textit{state augmentation approach}~\cite{sarkka:13} where the assimilation parameters, contained in the vector $\boldsymbol{\theta}_{\text{A}}\in\mathbb{R}^{n_p}$, are augmented to the EnKF state vector.
This allows the trainable parameters of in the gFHR Surrogate to be corrected while the digital twin is running simultaneously with the Physical Asset.
The state vector is redefined as $\hat{\mathbf{x}}\in \mathbb{R}^{n_{\bar{x}}}$:
\begin{align}
    \hat{\mathbf{x}}^{(k+1)} &\triangleq \left[\begin{array}{c|c} \mathbf{x}^{(k)}& \boldsymbol\theta_{\text{A}}^{(k)} \end{array}\right]^T \notag\\
    &= \left[\begin{array}{cc|c} \mathbf{x}_{\text{sys}}^{(k-1)} & \mathbf{x}_{\text{sys}}^{(k)} & \boldsymbol{\theta}_{\text{A}}^{(k)} \end{array}\right]^T,
\end{align}
where $n_{\hat{x}} = n_x + n_p$ and the EnKF state vector now including both the state $\mathbf{x}$ and the augmented parameters $\boldsymbol{\theta}_{\text{A}}$.
With the gFHR surrogate model system states, the EnKF augmented state vector has $n_{\hat{x}}=84+n_p$ variables.

The dynamics model for the state-space is the gFHR Surrogate, shown in Equation~\ref{eq:stateDynamics} and the parameters are modeled with assumed random-walk dynamics:
\begin{align}
    \boldsymbol{\theta}^{(k)} = \boldsymbol{\theta}^{(k-1)} + \tau, 
\end{align}
where $\tau\sim\mathcal{N}(0,\boldsymbol{\Sigma}_{\tau})$ is the white-noise random variable distributed by a multivariate Gaussian distribution with a zero mean and covariance matrix $\boldsymbol{\Sigma}_{\tau} \ll 1$. 
It is noted that this ``hack" can lead to numerical instabilities if $\tau$ is too small because it causes matrix singularly issues, which will inhibit the matrix inversion computation in the Kalman Gain matrix~\cite{sarkka:13}. 

With this approach, the gFHR surrogate model parameters can be added to the estimation with the goal of improving the virtual asset capabilities. However, the gFHR surrogate model has over 500 trainable parameters, as discussed in~\ref{app:gFHRsurrogate}, so it is avoided to add them all to the EnKF estimation vector. 
Since a select number of states will be observed such that $n_y < n_x$, only the parameters that will yield significant effects to the surrogate prediction need to be corrected. 
A Sobol indices analysis~\cite{sobol:01} is performed on the gFHR surrogate model in order to determine which parameter have the greatest effects on the measurement quantity, the core energy. 
The Sobol analysis procedure and results are presented in~\ref{app:sobol}, where the parameter vector is determined to be:
\begin{equation}
    \boldsymbol{\theta}_a^{(k)} \triangleq \begin{bmatrix} \mathbf{A}_{\text{VI},0,0}^{(k)} & \mathbf{A}_{\text{VI},0,1}^{(k)} & \mathbf{A}_{\text{VI},0,2}^{(k)} & \mathbf{A}_{\text{VI},0,3}^{(k)} & \mathbf{B}_{\text{VI},0,0}^{(k)} & \mathbf{B}_{\text{VII},2,0}^{(k)} & \mathbf{B}_{\text{VII},2,1}^{(k)} & \mathbf{B}_{\text{VII},2,2}^{(k)}\end{bmatrix}^T.
\end{equation}

\begin{figure}
    \centering
    \begin{subfigure}{0.45\textwidth}
        \centering
        \includegraphics[width=\textwidth]{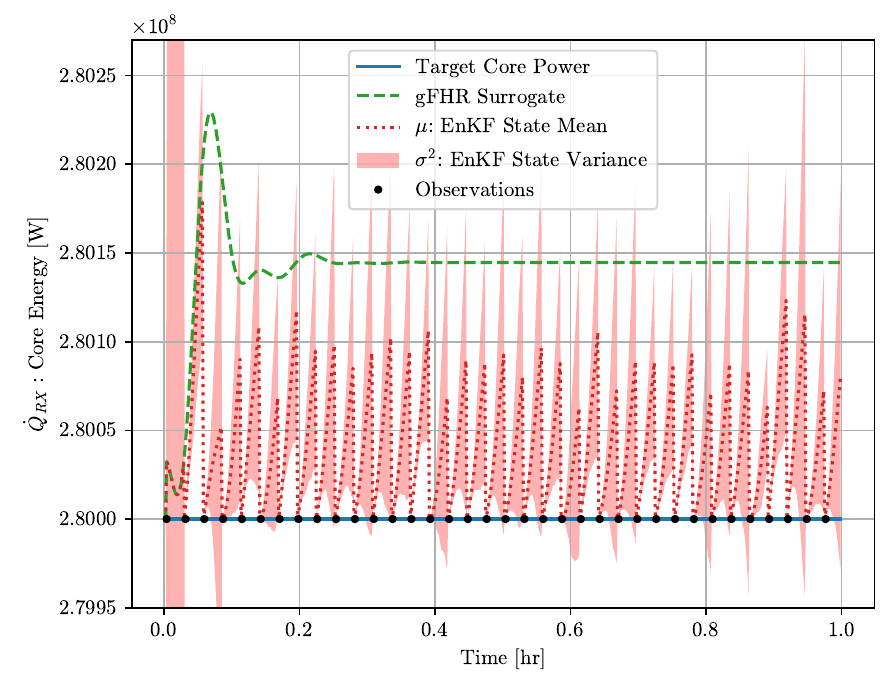}
        \caption{EnKF state estimation}
        \label{fig:enkf-state}
    \end{subfigure}
    ~
    \begin{subfigure}{0.45\textwidth}
        \centering
        \includegraphics[width=\textwidth]{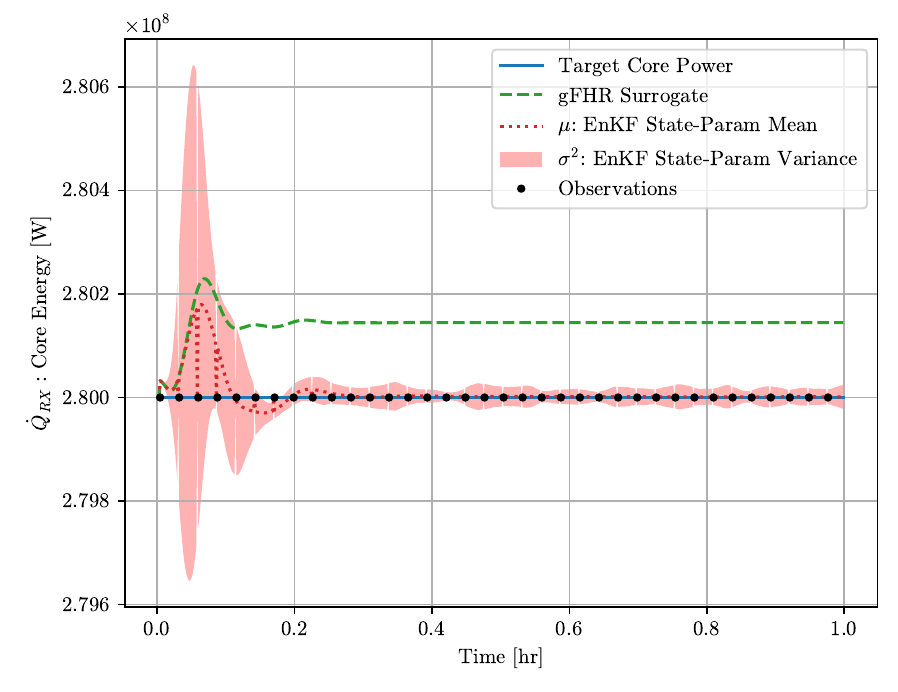}
        \caption{EnKF state-parameter estimation}
        \label{fig:enkf-stateparam}
    \end{subfigure}
    \caption{Core power EnKF estimation for steady-state simulation at 100\% power (280 MWs) over one hour}
    \label{fig:enkf-core}
\end{figure}

\begin{figure}
    \centering
    \includegraphics[width=0.8\textwidth]{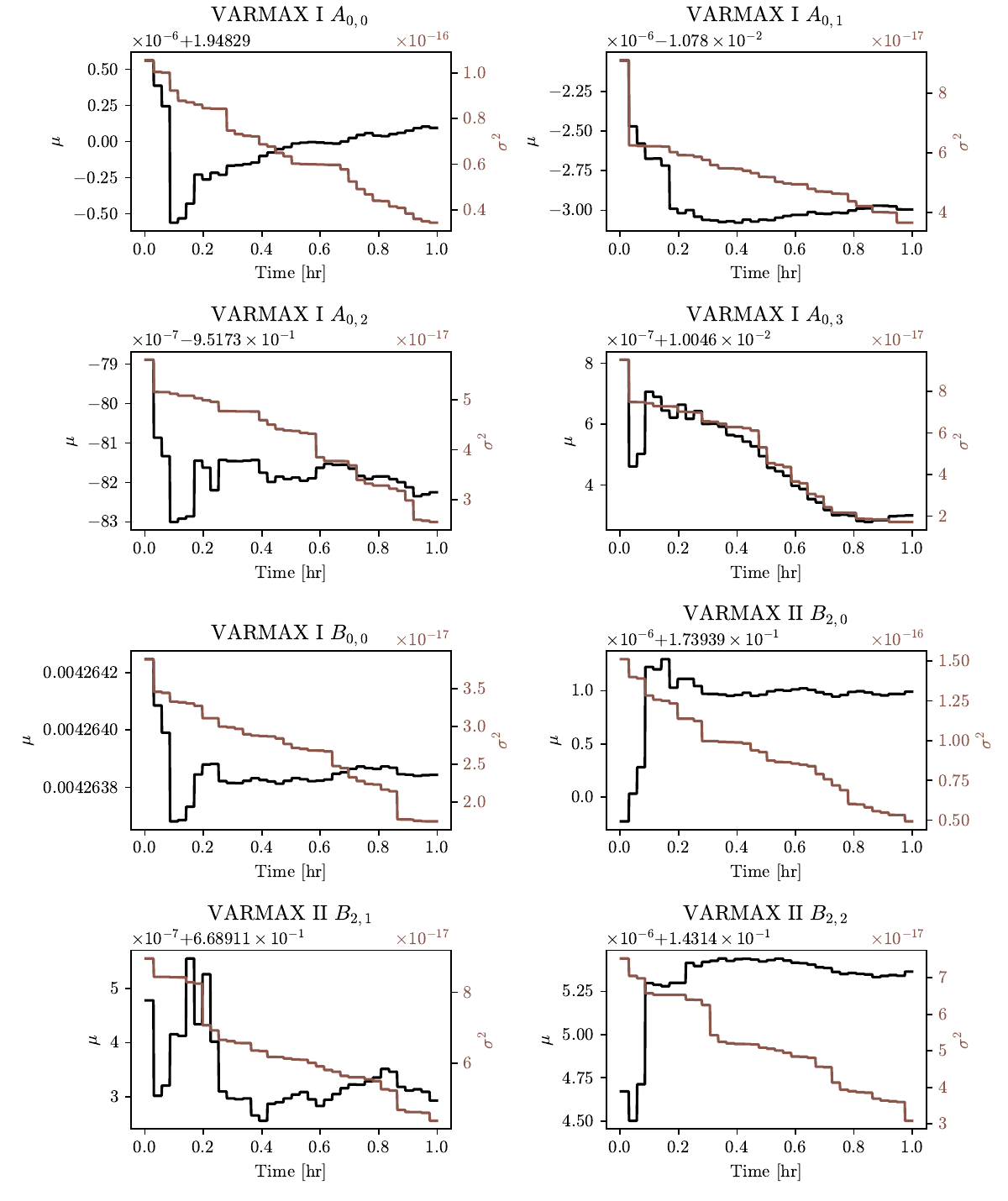}
    \caption{EnKF state-parameter estimation, VARMAX parameter estimation}
    \label{fig:enkf-stateparam-param}
\end{figure}

Using the same simulation set-up as the state-only estimation, the core power observations occur every $\beta=20$ steps at 100\% power steady-state for one hour. 
The ensemble contains $n_m = 20$ samples, with an initial covariance $\mathbf{C}^{(k)} = 1\times 10^{-6} \mathbb{I}_{n_x}$, where $\mathbb{I}_{n_x}\in \mathbb{R}^{(n_x \times n_x)}$ is the identity matrix. 
The dynamics model noise covariance matrix is $\mathbf{\Sigma} = 1\times 10^{-9} \mathbb{I}_{n_x}$ and the observation noise covariance matrix is $\mathbf{\Gamma} = 0 \mathbb{I}_{n_x}$. 
The core energy state-parameter prediction results are shown in Figure \ref{fig:enkf-stateparam} in comparison to the state-only estimation; where the data, target core power, the gFHR Surrogate prediction, and the EnKF state-parameter estimation are plotted. 
In this plot, it is seen that assimilating the parameters of the surrogate model does improve the EnKF performance with better accuracy and reducing the uncertainty. 
Figure \ref{fig:enkf-stateparam-param} shows the parameter estimation of the simulation, including both the mean and variance time-history. 
For all the parameters, the ensemble mean estimate changes with each new observation and the ensemble spread is decreasing as more information about the system becomes available. 
The Assimilation module uses the EnKF state-parameter estimation to incorporate data from the Physical Asset into the Virtual Asset via the gFHR Surrogate submodule.

\section{Digital Twin Demonstrations}\label{sec:demo}

The developed digital twin framework is showcased in three demonstrations.
Since no experimental facility was available for this study, the physics-based SAM gFHR emulates the FHR plant Physical Asset. 
The first is a long-term simulation showcasing the framework's scalability and maintenance scheduling. 
The second showcases a short-term simulation with a higher data assimilation rate for refining the digital twin representation of small-scale dynamics. 
Lastly, the third case shows the re-calibration of the framework with a system shock caused by the steam generator. 
All cases are implemented in \texttt{Python} and run on workstation with 128 GB of RAM memory; with computing allocation maximized at four nodes.

In all demonstrations, the Operator submodule is called every $\delta = 1$ month to determine if pump maintenance is necessary and set the optimized target power generation for next month ahead. 
The power generation is structured in hourly steps, thus the target power will change at the top of each hour and remain there until the next hour. 
The primary pump degradation parameters are set to 10\% head loss after $4.6656\times 10^7$ seconds (1.47 years) degradation rate, $\sigma_D = 0.0$, $\sigma_I = 0.5\times10^{-3}$, $\alpha=1/\bar{V}_p$, and $\alpha_{\dot{m}} = 1/\bar{V}_p$; where $\bar{V}_p = 6.04\times 10^{-1}\text{ m}^3\text{/s}$ is the primary pump volumetric flow rate at 100\% reactor power. 
The secondary pump degradation parameters are set to 10\% head loss after $2.592\times10^7$ seconds (0.82 years) degradation rate, $\sigma_D = 0.0$, $\sigma_I = 1\times10^{-3}$, $\alpha=1/\bar{V}_s$, and $\alpha_{\dot{m}} = 1/\bar{V}_s$; where $\bar{V}_s = 9.37\times10^{-1} \text{ m}^3\text{/s}$ is the intermediate pump volumetric flow rate at 100\% reactor power. 

The Operator submodule defines a health monitoring function $\eta_O$ that tracks the status of the HSC through the pump degradation loss coefficients. 
At each time step, $\eta_O^{(k)}$ is computed by the function:
\begin{equation}
    \eta_{O}^{(k)} \triangleq 1 - \frac{K^{(k)}}{K_0},
\end{equation}
where $K^{(k)}$ is the current degradation loss coefficient and $K_0 \triangleq 3.125$ is set as the minimum degradation loss coefficient.
A maintenance policy is set such that the pumps do not fall below a minimum health monitoring score of $\eta_O > 0.2$. 

\subsection{Long-term horizon operations and maintenance planning}\label{subsec:demo1}

In this first case, a one-year simulation of the FHR plant is demonstrated for long-term operational planning and pump maintenance scheduling. 
For this simulation, the power grid, where the supervised generation unit is connected to, provides the Operation submodule desired power generation on a monthly basis ($\delta = 1$ month) and requests that admissible power generations are sent to the physical plant every hour ($\gamma = 1$ hour). 
At the end of each power generation, a data measurement of the reactor's current power $\dot{Q}_{RX}^{(\beta k)}$ is sent back to the Virtual Asset, where the gFHR Surrogate states and parameters are updated every $\beta = 1$ hour.
Plant operations begin on January 1, 2018, at 100\% reactor power and with partially degraded pumps.
Both the primary and secondary pumps begin with pump loss degradation coefficients of $K_{p}^{(0)} = K_{s}^{(0)} = 1.15$.
For the EnKF, the states are assumed to have an initial covariance of $\mathbf{C}^{(0)} = 10^{-8}$ and the initial parameters have an initial covariance of $\mathbf{C}^{(0)}=10^{-16}$. 
Using $n_m = 20$ samples, the dynamics model is set to have an uncertainty of $\boldsymbol\Sigma = 10^{-15}$, a parameter white noise of $\boldsymbol\Sigma_{\tau} = 10^{-30}$ and a measurement noise of $\boldsymbol{\Gamma} = 10^{-30}$.

Since the power transitions at the top of each hour, measurements of the system are taken at the half hour mark in the steady-state region, making the assimilation frequency $\beta = 1$ hour beginning at $t = 30$ minutes.
The Virtual Asset system states and pump states are reported every five seconds, matching the gFHR Surrogate model time step, with assimilation updates every $\beta = 1$ hour. 
The Assimilation submodule, the Constraint submodule, and the Physical Asset data measurements are all synced to operate in conjunction at every hour.
Meaning that the observation is taken at the end of the admissible power generation so that the Constraint submodule is using the most updated gFHR surrogate to determine the next power generation request.

\begin{figure}
    \centering
    \includegraphics[width=\linewidth]{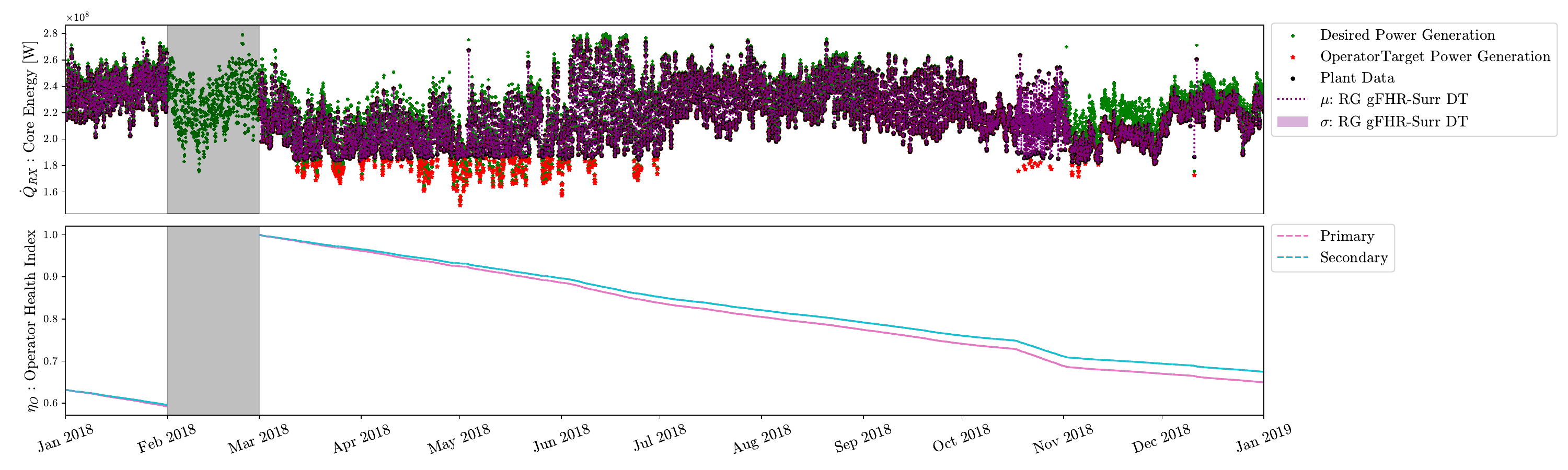}
    \caption{One-year operation of the digital twin framework. The top plot shows the observational state, the core power; plotting the desired power generation, the optimized target power generation, the virtual state representation, and the measurement data. The bottom plot tracks the Operator submodule health index score of each pump. }
    \label{fig:dt-full-month}
\end{figure}

The twelve-month digital twin demonstration is plotted in Figure~\ref{fig:dt-full-month}, showing the digital twin estimation of the core power $\dot{Q}_{RX}$ and the Operation submodule HSC health scores $\eta_O$. 
Since the demonstration was initialized with used pumps, after one-month the plant shuts down for the maintenance of both pumps and begins operations again on March 1, 2018. 
Once operations resume, the plant continues to run until the next pump replacement. 
The functionality of the RG in the Constraint submodule is also shown in March 2018 and April 2018, where the plant is restricted from reaching the Operator set optimized target power generation. 

\begin{figure}
    \centering
    \includegraphics[width=\linewidth]{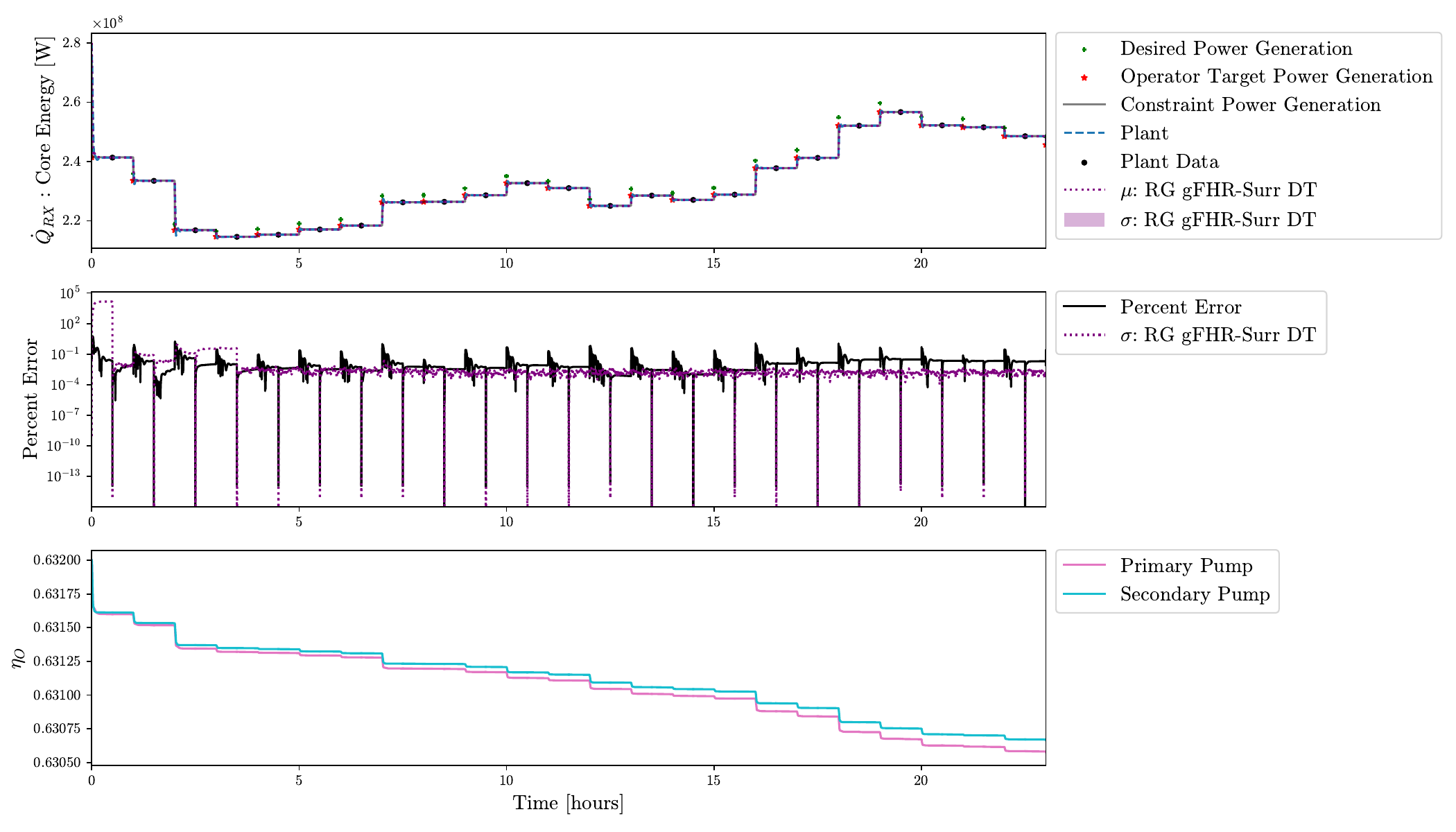}
    \caption{Zoomed in view of the first 24 hours in the one-year demonstration. The top plot shows the core energy generation, the middle plot shows the core energy percent error and ensemble variance, and the bottom plot shows the Operator submodule pump health index score.}
    \label{fig:first-24-states}
\end{figure}

\begin{figure}
    \centering
    \includegraphics[width=0.9\linewidth]{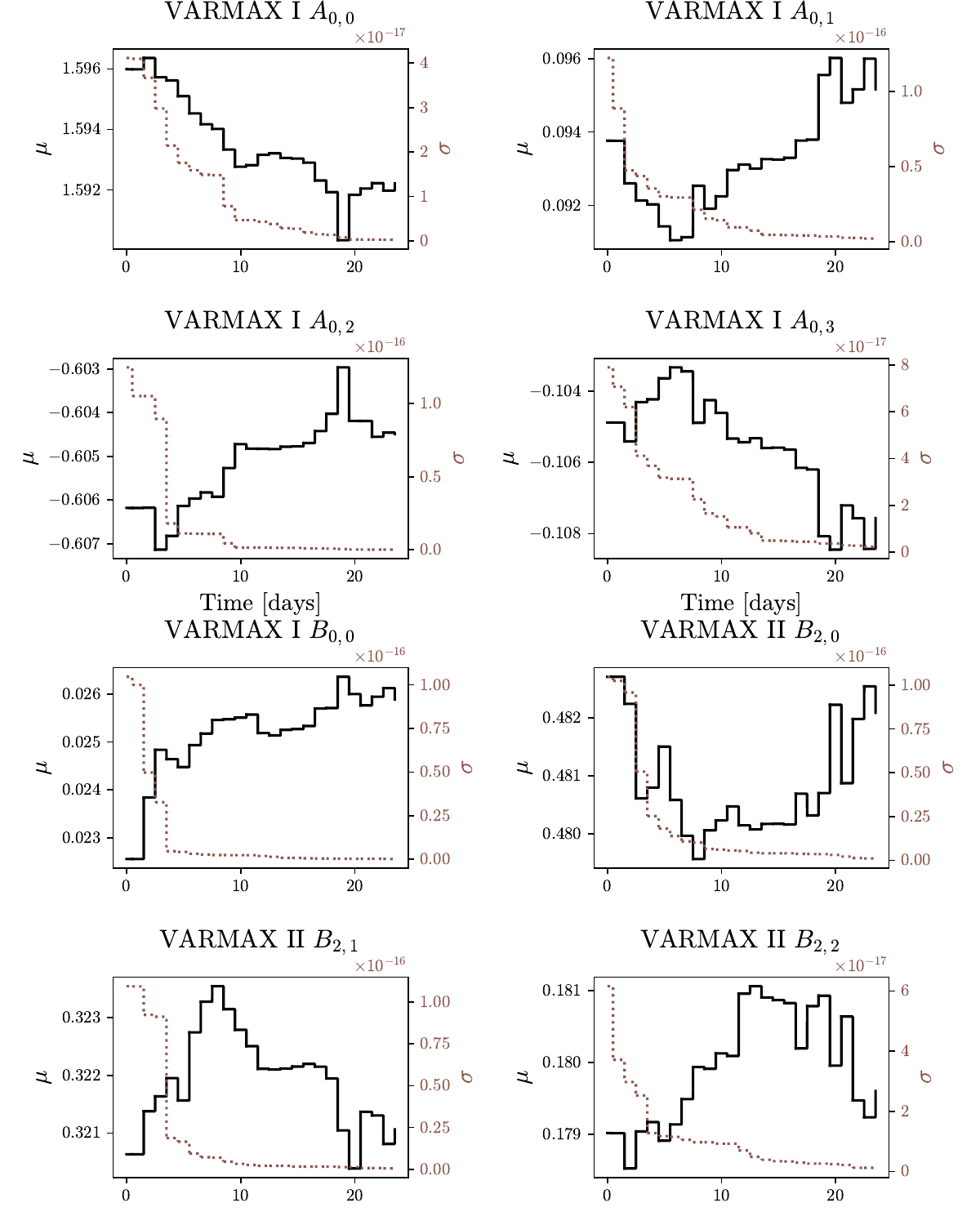}
    \caption{The gFHR Surrogate model parameters augmented to the EnKF state-parameter estimation. Zoomed in view for the first 24 hours in the one-year digital twin demonstration. For each of the eight parameters, the mean is plotted on the left axis and the variance is plotted on the right axis.}
    \label{fig:first-24-params}
\end{figure}

A look at the digital twin at an hourly scale is shown in Figure~\ref{fig:first-24-states}, where the first 24 hours of digital twin's virtual representation is compared to the physics-based SAM gFHR model. 
In the top subplot, the timescale interaction between modules is apparent, where the location of the data measurement and the Constraint submodule set power generation are synced on the same hourly frequency. 
The core energy percent error is show in the middle subplot in reference to the physics-based SAM gFHR model. 
The percent error stays below $1\%$ for the entire simulation, which is expected due to the accuracy that the gFHR Surrogate model provided a priori with the initial offline training. 
The percent error is the lowest when a measurement is obtained and the state is corrected, and the highest percent error is in the power transition regions.
The variance of EnKF starts high with an $10^3$ order of magnitude, but is quickly reduced after just under a dozen observations to stay around a $10^{-2}$ magnitude for the reminder of the simulation.
Lastly, the bottom subplot in Figure~\ref{fig:first-24-states} shows the close-up view of the pump health which depicts the expected behavior in the health degrading with time, pump flow rate,
and change in pump flow rate. 
It is visibly evident here that the plant's power transitions largely impact the pump's degradation rates, with higher slops of degradation occurring when the plant is changing power.

Having high prediction accuracy in the surrogate model before the digital twin comes online is beneficial because the digital twin will only need to make minimal changes to keep the virtual representation accurate. 
This is further highlighted in Figure~\ref{fig:first-24-params}, where the ensemble mean and covariance of the eight augmented parameters are plotted.
Here, the parameter mean distribution takes a step-like shape with little changes between observations.
Additionally, since the parameters were initialized with a very small ensemble covariance, the covariance remains small and continues to shrink as the more measurements increase confidence. 

The gFHR surrogate and the digital twin framework were designed in order to perform plant analysis over long periods of time, focusing on timescales of months and years.
This digital twin estimation is shown to provide accurate system representations for long-term plant analysis, including strategically scheduling pump maintenance on a month-by-month basis. 
However, as indicated in the percent error plot in Figure~\ref{fig:first-24-states}, the estimation of the power transitions regions is where the higher errors are concentrated.

\subsection{Short-term horizon with high-frequency measurements}\label{subsec:demo2}

As observed in the first case, the digital twin had larger percent errors in regions where the plant is transitioning power. 
The original gFHR Surrogate model was designed to be used over long time horizons in order to see the effects of plant operation on component degradation, and did not emphasize accuracy on the smaller time scales. 
Even with errors in these power transition regions, the mean percent error over a year-time scale was under $1\%$.
These higher errors are attributed to the surrogate model prediction frequency mismatching of the micro-dynamics responses as the control system is actuating to the target power.

In this case, observations are collected at a smaller time-step frequency to apply corrections to the surrogate in these transition regions.
Measurements are collected at every $\beta = 50$ time-steps, which converts to every 250 seconds (4.16 minutes) with the five-second constant time-step. 
The Constraint submodule timescale remains at $\gamma = 1$ hour, thus it will use the current state of the digital model every one hour.
Plant operation begins with brand-new pumps ($\eta_O$ = 1.0) and goes through four power changes over a 3.5 hour time frame.
The EnKF ensemble is initialized with the system state vector at 100\% reactor power operating conditions using an initial covariance of $\mathbf{C}^{(0)}= 10^{-8}$ and the parameters set at their initial offline trained values with a covariance of $\mathbf{C}^{(0)} = \times 10^{-16}$.
For the processes noises, the dynamics noise is $\boldsymbol\Sigma = 10^{-15}$, the parameter white noise is $\boldsymbol\Sigma_{\tau} = 10^{-30}$, and the measurement noise is $\boldsymbol{\Gamma} = 10^{-30}$.

\begin{figure}
    \centering
    \begin{subfigure}[t]{0.95\textwidth}
        \centering
        \includegraphics[width=\linewidth]{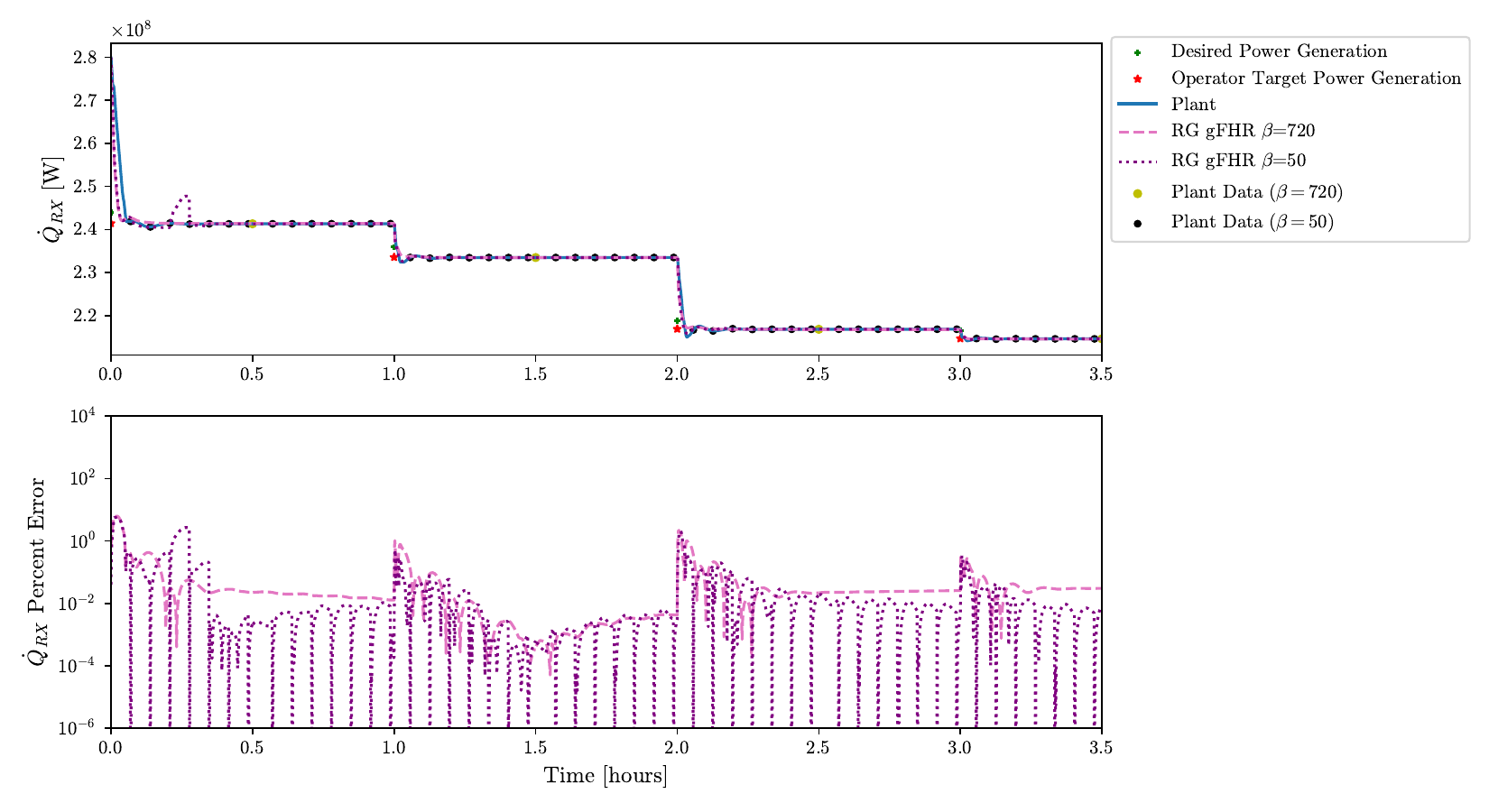}
    \end{subfigure}
    \begin{subfigure}[t]{0.95\textwidth}
        \centering
        \includegraphics[width=\linewidth]{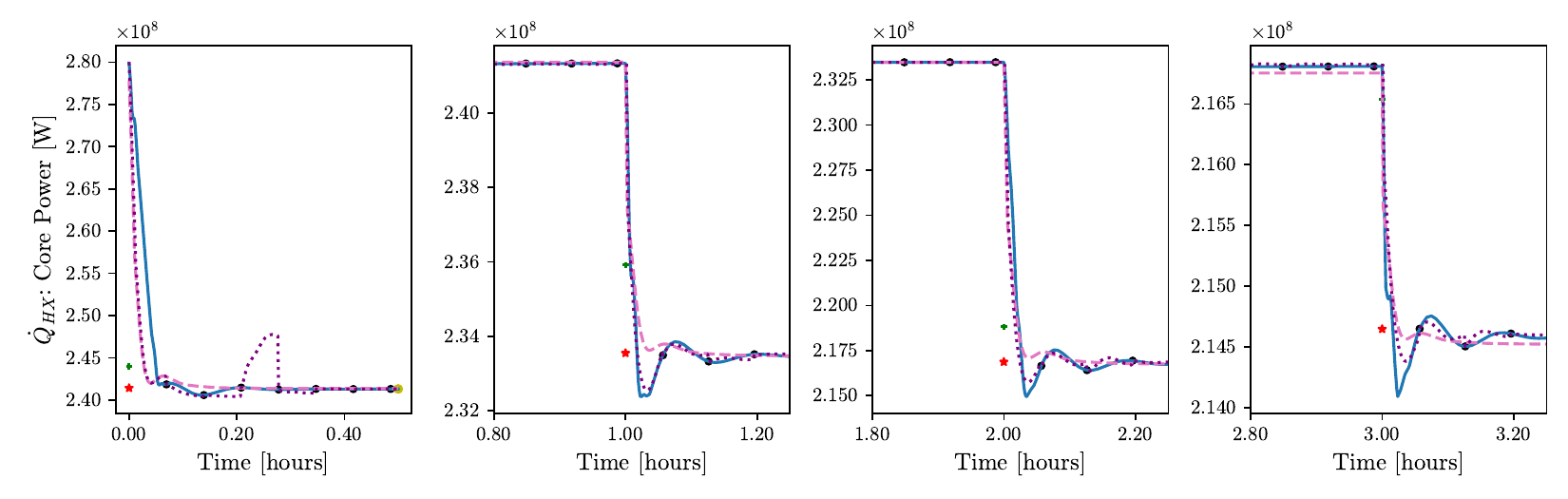}
    \end{subfigure}
    \caption{Short-term horizon with high-frequency measurement case. The top plot shows the core power estimation, displaying the predictions using $\beta=720$ and $\beta=50$ assimilation time-steps. The middle plot shows the core power percent error of each assimilation time-step estimation. The bottom plots show zoomed in views of the two assimilation time-step estimations in the power change regions.}
    \label{fig:st-power}
\end{figure}

Figure~\ref{fig:st-power} shows the comparison between the observations steps of $\beta=720$ (1 hour) and $\beta=50$ (4.2 minutes), plotting the core power estimation, the percent error, and close-up views of each power transition.
Looking at the entire core power simulation, the differences in the two measurement frequencies looks virtually identical, with each simulation load-following appearing to be in unison.
In the first hour, the $\beta=720$ is out-performing the $\beta=50$ prediction, however, as the $\beta=50$ gains more data, it's prediction accuracy surpasses the $\beta=720$ prediction. 
From the plots showing the zoomed in view of each energy transition ramp, the $\beta=50$ estimation is doing better at matching its oscillations of the SAM gFHR data.
In the second power ramp, the corrections added some oscillation amplitude while attempting to adjust, which propagate further into the steady-state region, causing both case to have similar errors in this transition. 
However, future adjustments to the gFHR surrogate correct the estimation such that the same behavior was not repeated in the following two power ramps, where the $\beta=50$ time-step resulted in smaller errors.

In terms of computational costs, both simulations complete in similar run times. 
Excluding the time to run the physics-based SAM gFHR module emulation of the Plant, for 3.5 hours of simulation the $\beta=720$ simulation ran in 45 CPU seconds and the $\beta=50$ simulation ran in 47 CPU seconds. 
This highlights the advantages of the EnKF for this high-dimensional problem, where an ensemble that is roughly a quarter of the size of the estimated state vector can still output predictions efficiently.
Even with the additional analysis steps, this does not affect the performance of the framework in a significant scale, making it an accessible parameter for the user to adjust if deemed necessary.

\subsection{Tailoring for System Shock Capturing}\label{subsec:demo3}

A large part of the Virtual Asset's role is to provide a digital representation that best serves the other digital twin tasks, which in this framework are the Operator and Constraint submodules.
In the current surrogate model, it is assumed that the system behaves in the same condition in which it was trained with. 
However, real-world systems are susceptible to events that would change the Physical Asset dynamics or operating conditions.
Therefore, the digital twin framework is tested and modified to model a system shock, showcasing the re-calibration capability of the surrogate model through the Assimilation submodule.

As stated in Section~\ref{subsec:samfhr}, the heat transfer in the SG is not modeled, thus it is assumed that the exiting temperature is 773.15 K and a boundary condition is set on the IHX secondary inlet temperature $T_{ihx,s,in}$. 
For the shock in this case, a power outage in the steam generator is modeled by a change in this exiting temperature. 
The gFHR surrogate model that has been used up to this point will not be adequate in capturing this system shock because of the additional knowledge that is now required to model the SG temperature change.
The first issue is the current Virtual Asset has no way of detecting the shock since the only observable state is the core power, which is not impacted by the shock since the control system is set up to still meet the target power distribution. 
The second issue is the Virtual Asset was not trained with data containing the shock transient, thus it will not show the correct response dynamics. 

Two updates are made to enable the Virtual Asset to capture this shock. 
The first update is within the Physical to Virtual Module, where another state is added to the observation vector so that the gFHR Surrogate can detect the shock occurrence.
Two observations ($n_y = 2$) are defined, the core energy and the secondary pump mass flow rate, so measurement vector $\mathbf{y}\in\mathbb{R}^{(n_y)}$, measurement noise covariance matrix $\boldsymbol{\Gamma}$, and measurement model $\mathbf{H}\in\mathbb{R}^{(n_y\times n_x + n_p)}$ for this case are:
\begin{align}
    \mathbf{y} &\triangleq \begin{bmatrix} \dot{Q}_{RX} & \dot{m}_{P,s}\end{bmatrix}^T \\
    \boldsymbol{\Gamma} &\triangleq \begin{bmatrix} 10^{-15} & 0 \\ 0 & 10^{-10}\end{bmatrix} \\
    H_{i,j} &\triangleq \left\{ \begin{array}{ll}
      1 & \text{if }\hat{x}_j = \dot{Q}_{RX} \text{ and }i=1\\
      1 & \text{if }\hat{x}_j= \dot{m}_{P,s} \text{ and }i=2 \\
      0 & \text{otherwise} \\
      \end{array} \right.  \text{ for } j\in[1,n_x+n_p].
\end{align}

Next, the gFHR surrogate model must show the expected changes in the system when the SG boundary condition is increased.
The updated surrogate is trained with data containing several SG temperature changes and reconstructed according to the procedure described in~\ref{app:gFHRSurrShock}. 
The EnKF state-parameter estimation set-up is slightly modified to accommodate this new surrogate model, including selecting new parameters to augment the EnKF state vector. 
With the two output states, a Sobol analysis is performed in order to determine the updated assimilation parameter vector $\boldsymbol{\theta}_A$ with $n_p = 6$:
\begin{equation}
    \boldsymbol{\theta}^{(k)}_A \triangleq \begin{bmatrix} \mathbf{A}_{\text{VI},0,1}^{(k)} & \mathbf{A}_{\text{VI},0,3}^{(k)} & \mathbf{A}_{\text{VI},1,1}^{(k)} & \mathbf{A}_{\text{VI},1,3}^{(k)} & \mathbf{B}_{\text{VII},0,0}^{(k)} & \mathbf{B}_{\text{VII},1,0}^{(k)} \end{bmatrix}^T.
\end{equation}
The results of the Sobol analysis can be found in~\ref{app:sobol2}

\begin{figure}
    \centering
    \includegraphics[width=0.9\linewidth]{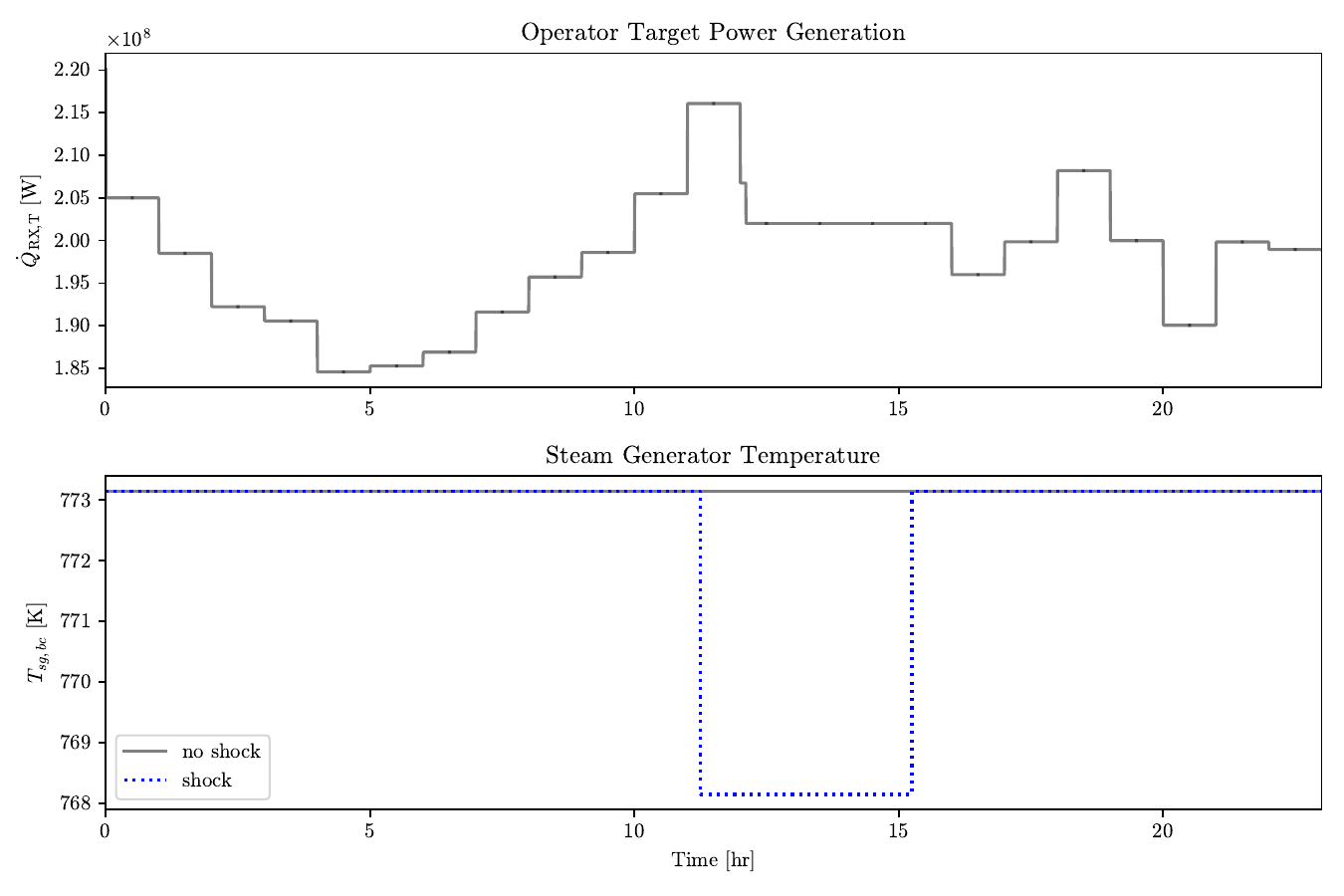}
    \caption{Steam generator shock case. The top plot shows the 24-hour Operator target power generation. The bottom plot shows the steam generator temperature with and without the temperature shock change.}
    \label{fig:sg-inputs}
\end{figure}

Now that the gFHR Surrogate has been updated, the framework is demonstrated on a 24-hour case with admissible power generations sent to the plant every $\gamma = 1$ hour, and observations sent for data assimilation every $\beta = 1$ hour. 
The EnKF initial state begins with the system states at 79\% full power and the parameters at their initial offline trained values. 
The states have an initial variance of $\mathbf{C}^{(0)} = 10^{-9}$ and the parameters have an initial variance of $\mathbf{C}^{(0)}=10^{-16}$.
Using $n_m = 25$ samples, the ensemble is propagated via the dynamics model with a model noise $\boldsymbol{\Sigma} = 10^{-8}$ and a parameter noise $\boldsymbol{\Sigma}_{\tau} = 10^{-12}$. 
Figure~\ref{fig:sg-inputs} displays the Operator target power generation and the steam generator temperature shock.

\begin{figure}
    \centering
    \includegraphics[width=\linewidth]{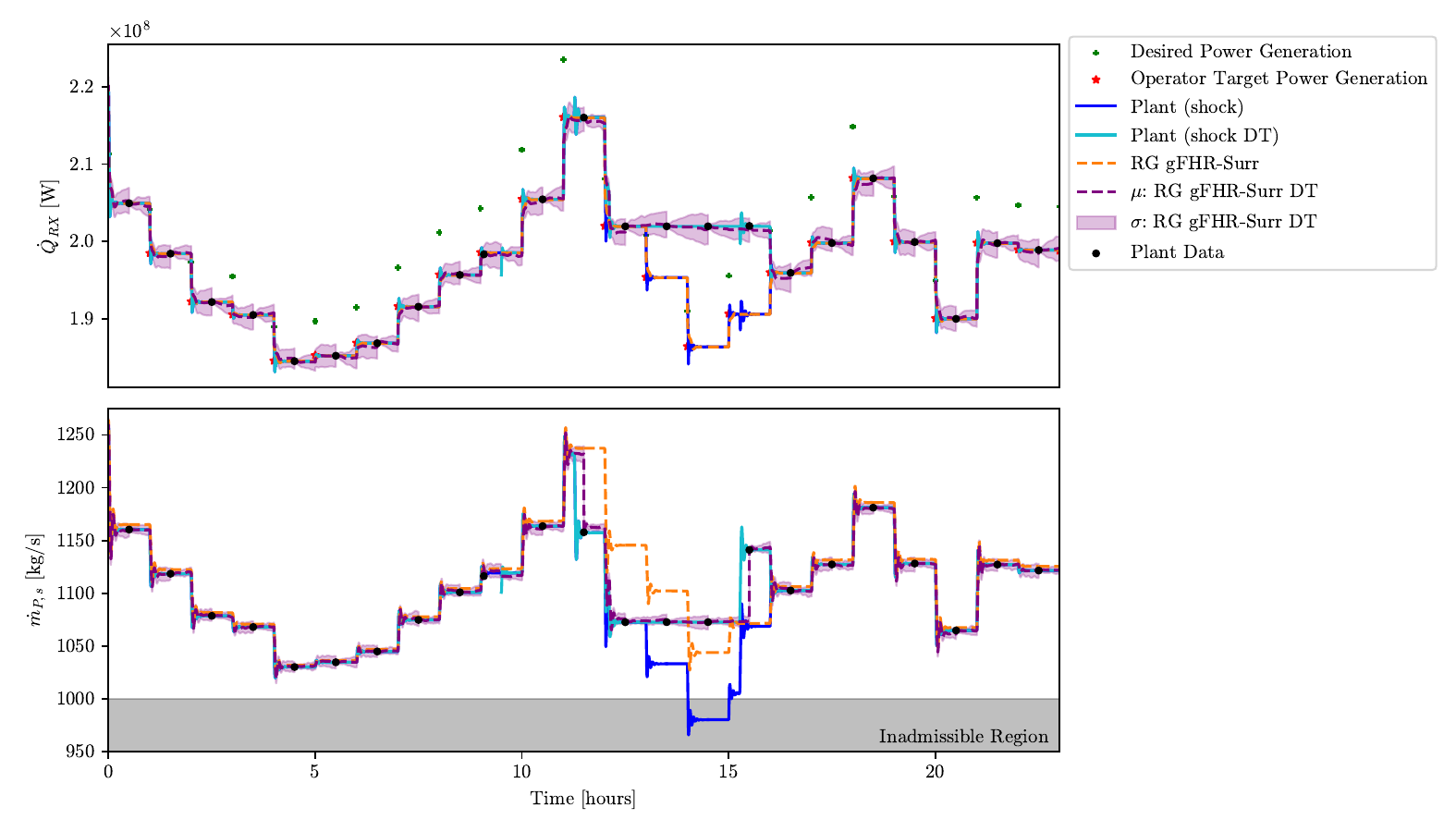}
    \caption{Digital twin observed state estimation with the gFHR Surrogate, modified for the steam generator shock case. Shown is the FHR plant simulations without the digital twin (Plant shock) and with the digital twin (Plant shock DT). Additionally, the non-adaptive gFHR Surrogate (RG gFHR-Surr) and the adaptive gFHR Surrogate via the digital twin (RG gFHR-Surr DT) estimations are both plotted.}
    \label{fig:sg-measurements}
\end{figure}

\begin{figure}
    \centering
    \includegraphics[width=\linewidth]{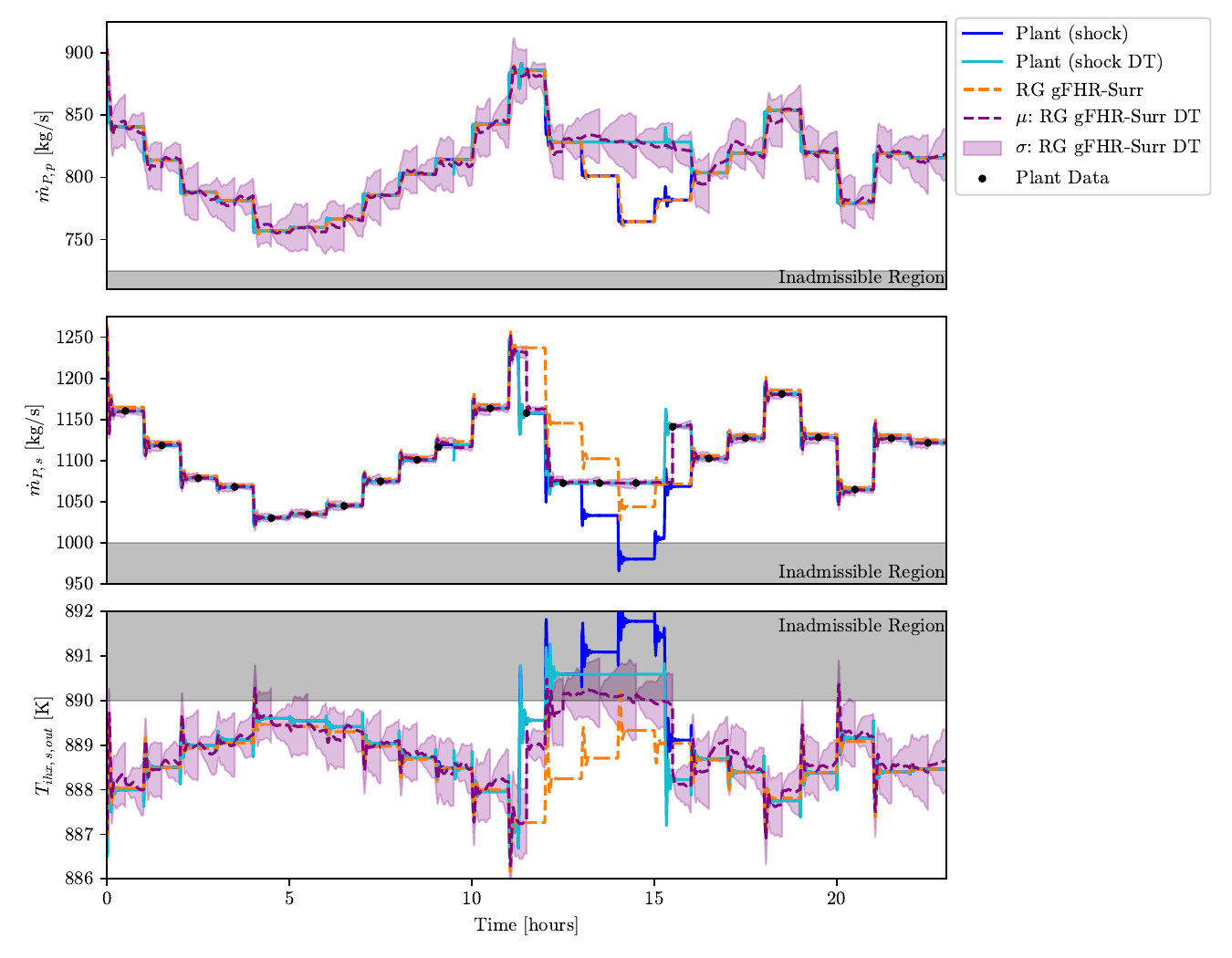}
    \caption{Digital twin constraint state estimation with the gFHR Surrogate modified for the steam generator shock case. Shown is the FHR plant simulations without the digital twin (Plant shock) and with the digital twin (Plant shock DT). Additionally, the non-adaptive gFHR Surrogate (RG gFHR-Surr) and the adaptive gFHR Surrogate via the digital twin (RG gFHR-Surr DT) estimations are both plotted.}
    \label{fig:sg-constraints}
\end{figure}

\begin{figure}
    \centering
    \includegraphics[width=\linewidth]{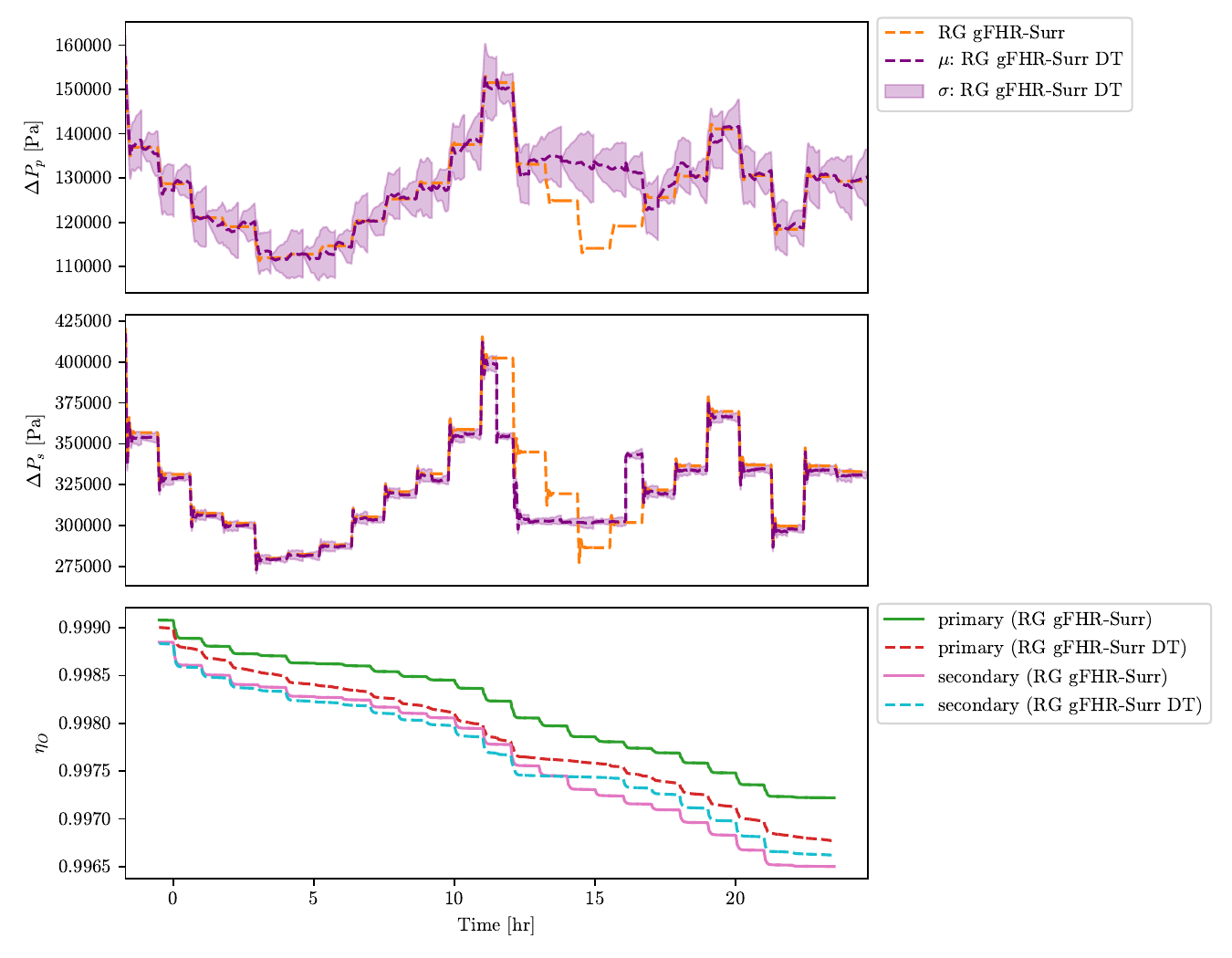}
    \caption{Digital twin estimation of pump state values using the gFHR Surrogate, modified for the steam generator shock case. Shown is the non-adaptive gFHR Surrogate (RG gFHR-Surr) and the adaptive gFHR Surrogate via the digital twin (RG gFHR-Surr DT) estimations for the primary pump head $\Delta P_p$ and the secondary pump head $\Delta P_s$. The bottom plot shows the operator module pump health score for each pump with and without the system shock.}
    \label{fig:sg-pump}
\end{figure}

The core energy and the secondary pump mass flow rate, the digital twin measurement estimations, are shown in Figure~\ref{fig:sg-measurements}.
The desired power generation and Operator target power generation are plotted along with the SAM gFHR simulated plant behavior under the system shock with and without digital twin intervention. 
The digital twin is shown to successfully model the shock indicated by the secondary pump mass flow rate $\dot{m}_{P,s}$ with the EnKF enabling the re-calibration of the model. 
This additionally seen with the constraint values in Figure~\ref{fig:sg-constraints} where the digital twin shock, that is communicated through the measurements, attempts to stay within the admissible region and while simultaneously updating the non-observable states during the shock.

The adaptability of the gFHR Surrogate through the digital twin is highlighted in the estimation of the constraint variables in Figure~\ref{fig:sg-measurements}. 
Due to the system shock, the updated gFHR Surrogate supplied this information to the Constraint submodule and intervened with a different admissible power generation at hour 13. 
There are still instances where the limit on $T_{ihs,s,out}$ is violated. 
The first is around hour 12.5 when the shock initially happens, and this is an input driven by the change in the SG temperatures that the RG has no prior knowledge of.
All the following violations occur because of the inaccuracies in the gFHR Surrogate and RG admissible region $O_{\infty}$.
In this case, micro-dynamics response of temperatures in the IHX are captured correctly when the system is in its normal conditions, but under the shock condition, the surrogate is under-predicting the temperature value; causing the RG to incorrectly scale the reference input. 
The RG's admissible region $O_{\infty}$ is also pre-determined in the offline training phase of the digital twin, however with this updated gFHR Surrogate, updating this region will improve accuracy.
A way to improve the RG is to add a buffer to the constraint values, which could account of expected errors or variances estimated by the EnKF.

The ability to capture the system response to the shock behavior is important for keeping the Virtual Asset updated, but also to inform the Operator submodule of the updated effects on the pumps. 
Figure~\ref{fig:sg-pump} plots the primary and secondary pump heads and the Operator health value to compare the difference in the pump degradation with and without the shock. 
The digital twin's ability to capture the shock's impacts on the degradation analysis is important to ensure the Operator has the most accurate and up-to-date information on the HSCs. 
With the shock, the pumps should not be degrading as much because the full target power generation is not realized, which is shown in the secondary pump.
However, the primary pump is actually predicted to degrade more with the shock, and this due to the change in mass flow rate effects on the degradation.
Due to the estimation noise that from the assimilation step in the EnKF algorithm, the digital twin is predicting more degradation in the primary pump.
To combat this, adding a measurement for the primary pump will reduce the noise and improve the health analysis results. 

In each of the three plots, the EnKF ensemble variance is plotted to visually show the digital twin's confidence in its state predictions. 
As expected, the variance bounds are smaller for the observable states because the filter is gaining direct information through measurements, while the unobserved state have higher uncertainty. 
Additionally, the uncertainty is smaller after observations due to the state assimilation correction but continues to grow with time until another measurement is made. 
This steady increase in uncertainty between measurements is due to the confidence in the dynamics models, which is applied through the model noise term $\boldsymbol{\xi}$.
In this case, the unobserved states are more susceptible to model noise, increasing the spread between the ensemble samples.

\section{Conclusion} \label{sec:conc}

In this work, a digital twin framework is designed for operational and maintenance planning of a Fluoride-salt-cooled High-temperature Reactor (FHR) that is health-, constraint-, and safety-aware.
The approach links a plant physical asset with its virtual counterpart to strategically drive plant policies (virtual to physical) while simultaneously adapting its computational representation (physical to virtual) throughout its operational lifetime.
Given a desired power generation, a reinforcement learning-based agent sets the optimized target power generation for the supervised plant power while also maintaining the health sensitive components (HSC), i.e., the system pumps.
Any set system constraints are checked and enforced with the Reference Governor (RG) algorithm, an augmentative controller designed to intervene if necessary.
The set points are then sent to the plant, where data is collected for assimilation with the Ensemble Kalman Filter (EnKF) to adapt the virtual asset online.
Data integration enables online accuracy enhancement and recalibration capability of the computational model, providing simultaneous simulation for future operative exploration.

The Virtual to Physical submodules, the operator, and the constraint, make smart plant input conditions based on a virtual plant surrogate model that evolves with the physical FHR plant.
The plant surrogate model is a hybrid physics-based and data-driven model that evolves using Bayesian filtering, where the surrogate tailored as more plant information, in the form of data, is exposed. 
For this non-linear and high-dimensional computational model, the Ensemble Kalman Filter is an appropriate approach that provides probabilistic estimation with uncertainty quantification. 
In addition to using Bayesian filtering to estimate the system states, selected trainable parameters of the data-driven models within the surrogate are also estimated to change the surrogate model itself. 
The augmented parameters are selected through a Sobol indices global sensitivity analysis that considers the physical plant set-up.
This assimilation completes the digital twin loop, enabling the end-to-end flow of information.

The capabilities of the digital twin are displayed through a series of demonstrations. 
First, the robustness is showcased in a long-term four-month operational period that is carried out for a study of maintenance planning. 
Second, the assimilation frequency is used to improve the virtual model accuracy on small time scales for dynamical response behavior.
Lastly, to capture a system shock caused by the steam generator, a new surrogate model that is trained with updated information on the response dynamics is easily replaced into the framework.
All three cases provide evidence of the user being able to adjust the framework parameters and set-up in order to capture certain dynamics, make requested decisions and control the experimentation.
This includes characteristics such as picking what the measurable quantities are, setting degradation rates, and controlling the rate at which information is passed between the physical and the virtual.
All of which relied on some incorporation of application specific knowledge, further emphasizing that digital twins must be designed in a model-specific approach and cannot solely rely on black-box machine learning and AI models.
This close-knit integration of physics-based knowledge improves accuracy and ultimately adds to the reliability and resiliency of the framework to inform a real design process or experiments.

The developed framework, including all of its components, were designed for the health-aware and constraint-informed operation of a nuclear plant reactor configuration, however the structure and methods are general and can be extended to any nuclear energy system.
Further development of this digital twin framework includes continued improvement for the modeling of the FHR plant. 
As more information and features because available about the plant, the need to explore other surrogate model methods may be necessary for improving predictions for small timescale decisions. 
This was an area of exposure seen in the framework, where the transition dynamics of the power have a higher percent error compared to the steady-state regions. 
The surrogate model was originally designed to capture dynamics for long-term accuracy, where the performance is evaluated over months and years for component maintenance analysis.
However, since the accuracy was only evaluated on an average performance rather than a detailed performance, it is not reliable for specific instance --- which is seen in the surrogate for the power transition regions.
This excludes it from functionalities that rely on individual accuracy, such as designing smart controllers and operations in regions especially close to system constraints. 
Another aspect pertains to the actual digital twin itself, including improving the design with a human-operator and physical experimental plant involvement. 
There are some operational functions of the plant that degrade with irreversible effects, at which a human or computerized monitor may decide that the current reinforcement learning policy can no longer adequately drive the system.
Thus, the plant would need to come offline in order to update the virtual model and other virtual to physical assets. 

\section*{CRediT}
\textbf{Jasmin Lim}: Conceptualization, Data curation, Formal analysis, Investigation, Methodology, Software, Visualization, Writing - original draft, Writing - review and editing.
\textbf{Dimitrois Pylorof}: Conceptualization, Methodology, Software, Validation, Writing - review and editing.
\textbf{Humberto Garcia}: Conceptualization, Funding acquisition, Methodology, Project administration, Writing - review and editing.
\textbf{Karthik Duraisamy}: Conceptualization, Funding acquisition, Project administration, Resources, Supervision, Writing - review and editing.

\section*{Acknowledgements}
This work was largely supported by the US Department of Energy (DOE) Advanced Research Projects Agency-Energy (ARPA-E) project titled \textit{SAFARI: Secure Automation for Advanced Reactor Innovation}; and supplemental support was provided by Los Alamos National Laboratory under the grant number AWD026741.

\bibliographystyle{elsarticle-num} 
\bibliography{references}

\begin{thebibliography}{10}
\expandafter\ifx\csname url\endcsname\relax
  \def\url#1{\texttt{#1}}\fi
\expandafter\ifx\csname urlprefix\endcsname\relax\def\urlprefix{URL }\fi
\expandafter\ifx\csname href\endcsname\relax
  \def\href#1#2{#2} \def\path#1{#1}\fi

\bibitem{GIF:02}
{U.S. DOE Nuclear Energy Research Advisory Committee and the Generation IV
  International Forum}, \href{https://www.osti.gov/biblio/859029}{A technology
  roadmap for generation iv nuclear energy systems} (2002).
\newblock \href {https://doi.org/10.2172/859029} {\path{doi:10.2172/859029}}.
\newline\urlprefix\url{https://www.osti.gov/biblio/859029}

\bibitem{mit:18}
{MIT Energy Initiative},
  \href{https://energy.mit.edu/research/future-nuclear-energy-carbon-constrained-world/}{The
  future of nuclear energy in a carbon-constrained world}, Tech. rep.,
  Massachusetts Institute of Technology (2018).
\newline\urlprefix\url{https://energy.mit.edu/research/future-nuclear-energy-carbon-constrained-world/}

\bibitem{Thellufusen:24}
J.~Z. Thellufsen, H.~Lund, B.~V. Mathiesen, P.~A. Østergaard, P.~Sorknæs,
  S.~Nielsen, P.~T. Madsen, G.~B. Andresen,
  \href{https://www.sciencedirect.com/science/article/pii/S0306261924010882}{Cost
  and system effects of nuclear power in carbon-neutral energy systems},
  Applied Energy 371 (2024) 123705.
\newblock \href
  {https://doi.org/https://doi.org/10.1016/j.apenergy.2024.123705}
  {\path{doi:https://doi.org/10.1016/j.apenergy.2024.123705}}.
\newline\urlprefix\url{https://www.sciencedirect.com/science/article/pii/S0306261924010882}

\bibitem{IEA:25}
{International Energy Agency (IEA)},
  \href{https://www.iea.org/reports/the-path-to-a-new-era-for-nuclear-energy}{The
  path to a new era for nuclear energy} (2024).
\newline\urlprefix\url{https://www.iea.org/reports/the-path-to-a-new-era-for-nuclear-energy}

\bibitem{GIF:18}
{Generation IV International Forum}, {GIF} research and development outlook for
  generation iv nuclear energy systems - 2018, Tech. rep., {Generation IV
  International Forum} (2018).

\bibitem{Meyer:02}
T.~A. Meyer, G.~G. Elder, R.~Llovet,
  \href{https://doi.org/10.1115/ICONE10-22536}{Life cycle management: Managing
  the aging of critical nuclear plant components}, Vol. 10th International
  Conference on Nuclear Engineering, Volume 1 of International Conference on
  Nuclear Engineering, 2002, pp. 129--136.
\newblock \href
  {http://arxiv.org/abs/https://asmedigitalcollection.asme.org/ICONE/proceedings-pdf/ICONE10/35952/129/4537244/129\_1.pdf}
  {\path{arXiv:https://asmedigitalcollection.asme.org/ICONE/proceedings-pdf/ICONE10/35952/129/4537244/129\_1.pdf}},
  \href {https://doi.org/10.1115/ICONE10-22536}
  {\path{doi:10.1115/ICONE10-22536}}.
\newline\urlprefix\url{https://doi.org/10.1115/ICONE10-22536}

\bibitem{Grieves:05}
M.~W. Grieves, Product lifecycle management: the new paradigm for enterprises,
  International Journal of Product Development 2~(1-2) (2005) 71--84.
\newblock \href {https://doi.org/10.1504/IJPD.2005.006669}
  {\path{doi:10.1504/IJPD.2005.006669}}.

\bibitem{Grieves:17}
M.~Grieves, J.~Vickers, Digital Twin: Mitigating Unpredictable, Undesirable
  Emergent Behavior in Complex Systems, Springer International Publishing,
  Cham, 2017, pp. 85--113.
\newblock \href {https://doi.org/10.1007/978-3-319-38756-7\_4}
  {\path{doi:10.1007/978-3-319-38756-7\_4}}.

\bibitem{Jones:20}
D.~Jones, C.~Snider, A.~Nassehi, J.~Yon, B.~Hicks, Characterising the digital
  twin: A systematic literature review, CIRP Journal of Manufacturing Science
  and Technology 29 (2020) 36–52.
\newblock \href {https://doi.org/10.1016/j.cirpj.2020.02.002}
  {\path{doi:10.1016/j.cirpj.2020.02.002}}.

\bibitem{Rasheed:20}
A.~Rasheed, O.~San, T.~Kvamsdal, Digital twin: Values, challenges and enablers
  from a modeling perspective, IEEE Access 8 (2020) 21980--22012.
\newblock \href {https://doi.org/10.1109/ACCESS.2020.2970143}
  {\path{doi:10.1109/ACCESS.2020.2970143}}.

\bibitem{KRITZINGER:18}
W.~Kritzinger, M.~Karner, G.~Traar, J.~Henjes, W.~Sihn,
  \href{https://www.sciencedirect.com/science/article/pii/S2405896318316021}{Digital
  twin in manufacturing: A categorical literature review and classification},
  IFAC-PapersOnLine 51~(11) (2018) 1016--1022, 16th IFAC Symposium on
  Information Control Problems in Manufacturing INCOM 2018.
\newblock \href {https://doi.org/https://doi.org/10.1016/j.ifacol.2018.08.474}
  {\path{doi:https://doi.org/10.1016/j.ifacol.2018.08.474}}.
\newline\urlprefix\url{https://www.sciencedirect.com/science/article/pii/S2405896318316021}

\bibitem{Batty:24}
M.~Batty, Digital twins in city planning, Nature Computational Science 4 (2024)
  192--199.
\newblock \href {https://doi.org/10.1038/s43588-024-00606-7}
  {\path{doi:10.1038/s43588-024-00606-7}}.

\bibitem{Ge:25}
C.~Ge, S.~Qin, Urban flooding digital twin system framework, Systems Science \&
  Control Engineering 13~(1) (2025) 2460432.
\newblock \href {https://doi.org/10.1080/21642583.2025.2460432}
  {\path{doi:10.1080/21642583.2025.2460432}}.

\bibitem{Kraft:17}
J.~Kraft, S.~Kuntzagk, \href{https://doi.org/10.1115/GT2017-63336}{Engine
  fleet-management: The use of digital twins from a mro perspective}, Vol.
  Volume 1: Aircraft Engine; Fans and Blowers; Marine; Honors and Awards of
  Turbo Expo, 2017, p. V001T01A007.
\newblock \href
  {http://arxiv.org/abs/https://asmedigitalcollection.asme.org/GT/proceedings-pdf/GT2017/50770/V001T01A007/2431703/v001t01a007-gt2017-63336.pdf}
  {\path{arXiv:https://asmedigitalcollection.asme.org/GT/proceedings-pdf/GT2017/50770/V001T01A007/2431703/v001t01a007-gt2017-63336.pdf}},
  \href {https://doi.org/10.1115/GT2017-63336}
  {\path{doi:10.1115/GT2017-63336}}.
\newline\urlprefix\url{https://doi.org/10.1115/GT2017-63336}

\bibitem{Zohdi:22}
T.~I. Zohdi, A digital-twin and machine-learning framework for precise heat and
  energy management of data-centers, Computational Mechanics 69~(6) (2022)
  1501–1516.
\newblock \href {https://doi.org/10.1007/s00466-022-02152-3}
  {\path{doi:10.1007/s00466-022-02152-3}}.

\bibitem{Torzoni:24}
M.~Torzoni, M.~Tezzele, S.~Mariani, A.~Manzoni, K.~E. Willcox, A digital twin
  framework for civil engineering structures, Computer Methods in Applied
  Mechanics and Engineering 418 (2024) 116584.
\newblock \href {https://doi.org/10.1016/j.cma.2023.116584}
  {\path{doi:10.1016/j.cma.2023.116584}}.

\bibitem{Shrivastava:22}
C.~Shrivastava, T.~Berry, P.~Cronje, S.~Schudel, T.~Defraeye, Digital twins
  enable the quantification of the trade-offs in maintaining citrus quality and
  marketability in the refrigerated supply chain, Nature Food 3~(6) (2022)
  413–427.
\newblock \href {https://doi.org/10.1038/s43016-022-00497-9}
  {\path{doi:10.1038/s43016-022-00497-9}}.

\bibitem{Erol:20}
T.~Erol, A.~F. Mendi, D.~Doğan, The digital twin revolution in healthcare, in:
  2020 4th International Symposium on Multidisciplinary Studies and Innovative
  Technologies (ISMSIT), 2020, pp. 1--7.
\newblock \href {https://doi.org/10.1109/ISMSIT50672.2020.9255249}
  {\path{doi:10.1109/ISMSIT50672.2020.9255249}}.

\bibitem{Lal:20}
A.~Lal, G.~Li, E.~Cubro, S.~Chalmers, H.~Li, V.~Herasevich, Y.~Dong, B.~W.
  Pickering, O.~Kilickaya, O.~Gajic, Development and verification of a digital
  twin patient model to predict specific treatment response during the first 24
  hours of sepsis, Critical Care Explorations 2~(11) (2020) e0249.
\newblock \href {https://doi.org/10.1097/CCE.0000000000000249}
  {\path{doi:10.1097/CCE.0000000000000249}}.

\bibitem{Yadav:21}
V.~Yadav, A.~V.~G. Vivek~Agarwal, R.~D. Hays, C.~S.~R. Adam J.~Pluth, H.~Zhang,
  P.~Jain, P.~Ramuhalli, D.~Eskins, J.~Carlson, R.~G. Lozada, C.~Ulmer,
  R.~Iyengar, Technical challenges and gaps in digital-twin-enabling
  technologies for nuclear reactor applications, Tech. Rep. INL/EXT-21-65316,
  Idaho National Laboratory (November 2021).

\bibitem{Patterson:16}
E.~A. Patterson, R.~J. Taylor, M.~Bankhead, A framework for an integrated
  nuclear digital environment, Progress in Nuclear Energy 87 (2016) 97–103.
\newblock \href {https://doi.org/10.1016/j.pnucene.2015.11.009}
  {\path{doi:10.1016/j.pnucene.2015.11.009}}.

\bibitem{Bowman:22}
D.~Bowman, L.~Dwyer, A.~Levers, E.~A. Patterson, S.~Purdie, K.~Vikhorev, A
  unified approach to digital twin architecture—proof-of-concept activity in
  the nuclear sector, IEEE Access 10 (2022) 44691--44709.
\newblock \href {https://doi.org/10.1109/ACCESS.2022.3161626}
  {\path{doi:10.1109/ACCESS.2022.3161626}}.

\bibitem{Kochunas:21}
B.~Kochunas, X.~Huan, Digital twin concepts with uncertainty for nuclear power
  applications, Energies 14~(1414) (2021) 4235.
\newblock \href {https://doi.org/10.3390/en14144235}
  {\path{doi:10.3390/en14144235}}.

\bibitem{Lin:21}
L.~Lin, H.~Bao, N.~Dinh,
  \href{https://www.sciencedirect.com/science/article/pii/S0306454921002383}{Uncertainty
  quantification and software risk analysis for digital twins in the nearly
  autonomous management and control systems: A review}, Annals of Nuclear
  Energy 160 (2021) 108362.
\newblock \href {https://doi.org/https://doi.org/10.1016/j.anucene.2021.108362}
  {\path{doi:https://doi.org/10.1016/j.anucene.2021.108362}}.
\newline\urlprefix\url{https://www.sciencedirect.com/science/article/pii/S0306454921002383}

\bibitem{stewart:25}
R.~Stewart, E.~Treviño, A.~Shields, K.~Heaps, J.~Darrington, Q.~Williams,
  C.~Pope, J.~Scott, B.~Baker, J.~Palmer, B.~Vainqueur, T.~S. Palmer,
  C.~Palmer, S.~Bays, M.~Schanfein, G.~Reyes, C.~Ritter,
  \href{https://www.sciencedirect.com/science/article/pii/S0306454924007047}{The
  agn-201 digital twin: A test bed for remotely monitoring nuclear reactors},
  Annals of Nuclear Energy 213 (2025) 111041.
\newblock \href {https://doi.org/https://doi.org/10.1016/j.anucene.2024.111041}
  {\path{doi:https://doi.org/10.1016/j.anucene.2024.111041}}.
\newline\urlprefix\url{https://www.sciencedirect.com/science/article/pii/S0306454924007047}

\bibitem{Ndum:24}
Z.~N. Ndum, J.~Tao, Y.~Liu, J.~Ford, V.~Vlassov, N.~Morton, J.~Grissom,
  P.~Tsvetkov, S.~Adu, A digital twin-based simulator for small modular and
  microreactors, in: 2024 Winter Simulation Conference (WSC), 2024, pp.
  2963--2974.
\newblock \href {https://doi.org/10.1109/WSC63780.2024.10838736}
  {\path{doi:10.1109/WSC63780.2024.10838736}}.

\bibitem{Rivas:25}
G.~K.~D. Andy~Rivas, J.~Hou,
  \href{https://doi.org/10.1080/00295639.2024.2372515}{A system predictive
  maintenance framework for advanced reactors using a data-driven digital
  twin}, Nuclear Science and Engineering 199~(3) (2025) 358--387.
\newblock \href
  {http://arxiv.org/abs/https://doi.org/10.1080/00295639.2024.2372515}
  {\path{arXiv:https://doi.org/10.1080/00295639.2024.2372515}}, \href
  {https://doi.org/10.1080/00295639.2024.2372515}
  {\path{doi:10.1080/00295639.2024.2372515}}.
\newline\urlprefix\url{https://doi.org/10.1080/00295639.2024.2372515}

\bibitem{thelen:22}
A.~Thelen, X.~Zhang, O.~Fink, Y.~Lu, S.~Ghosh, B.~D. Youn, M.~D. Todd,
  S.~Mahadevan, C.~Hu, Z.~Hu, A comprehensive review of digital twin—part 1:
  modeling and twinning enabling technologies, Structural and Multidisciplinary
  Optimization 65~(12) (2022) 354.

\bibitem{Liu:24}
Y.~Liu, B.~Wang, S.~Tan, T.~Li, W.~Lv, Z.~Niu, J.~Li, P.~Gao, R.~Tian,
  \href{https://www.sciencedirect.com/science/article/pii/S0029549324007556}{Applications
  of deep reinforcement learning in nuclear energy: A review}, Nuclear
  Engineering and Design 429 (2024) 113655.
\newblock \href
  {https://doi.org/https://doi.org/10.1016/j.nucengdes.2024.113655}
  {\path{doi:https://doi.org/10.1016/j.nucengdes.2024.113655}}.
\newline\urlprefix\url{https://www.sciencedirect.com/science/article/pii/S0029549324007556}

\bibitem{Nguyen:24}
K.~H.~N. Nguyen, A.~Rivas, G.~K. Delipei, J.~Hou,
  \href{https://www.mdpi.com/2673-4362/5/3/15}{Reinforcement learning-based
  control sequence optimization for advanced reactors}, Journal of Nuclear
  Engineering 5~(3) (2024) 209--225.
\newblock \href {https://doi.org/10.3390/jne5030015}
  {\path{doi:10.3390/jne5030015}}.
\newline\urlprefix\url{https://www.mdpi.com/2673-4362/5/3/15}

\bibitem{Tunkle:25}
L.~Tunkle, K.~Abdulraheem, L.~Lin, M.~I. Radaideh,
  \href{https://arxiv.org/abs/2504.00156}{Nuclear microreactor control with
  deep reinforcement learning} (2025).
\newblock \href {http://arxiv.org/abs/2504.00156} {\path{arXiv:2504.00156}}.
\newline\urlprefix\url{https://arxiv.org/abs/2504.00156}

\bibitem{Kim:24}
J.~Kim, J.~Seo, \href{https://arxiv.org/abs/2409.08231}{Design optimization of
  nuclear fusion reactor through deep reinforcement learning} (2024).
\newblock \href {http://arxiv.org/abs/2409.08231} {\path{arXiv:2409.08231}}.
\newline\urlprefix\url{https://arxiv.org/abs/2409.08231}

\bibitem{Zhong:23}
X.~Zhong, L.~Zhang, H.~Ban, Deep reinforcement learning for class imbalance
  fault diagnosis of equipment in nuclear power plants, Annals of Nuclear
  Energy 184 (2023) 109685.

\bibitem{Park:20}
J.~Park, T.~Kim, S.~Seong,
  \href{https://www.sciencedirect.com/science/article/pii/S014919701930232X}{Providing
  support to operators for monitoring safety functions using reinforcement
  learning}, Progress in Nuclear Energy 118 (2020) 103123.
\newblock \href {https://doi.org/https://doi.org/10.1016/j.pnucene.2019.103123}
  {\path{doi:https://doi.org/10.1016/j.pnucene.2019.103123}}.
\newline\urlprefix\url{https://www.sciencedirect.com/science/article/pii/S014919701930232X}

\bibitem{ZhaoSmidts:22}
Y.~Zhao, C.~Smidts,
  \href{https://www.sciencedirect.com/science/article/pii/S0951832022001922}{Reinforcement
  learning for adaptive maintenance policy optimization under imperfect
  knowledge of the system degradation model and partial observability of system
  states}, Reliability Engineering \& System Safety 224 (2022) 108541.
\newblock \href {https://doi.org/https://doi.org/10.1016/j.ress.2022.108541}
  {\path{doi:https://doi.org/10.1016/j.ress.2022.108541}}.
\newline\urlprefix\url{https://www.sciencedirect.com/science/article/pii/S0951832022001922}

\bibitem{HAO:24}
Z.~Hao, F.~{Di Maio}, E.~Zio,
  \href{https://www.sciencedirect.com/science/article/pii/S294992672300001X}{Monte
  carlo tree search-based deep reinforcement learning for flexible operation \&
  maintenance optimization of a nuclear power plant}, Journal of Safety and
  Sustainability 1~(1) (2024) 4--13.
\newblock \href {https://doi.org/https://doi.org/10.1016/j.jsasus.2023.08.001}
  {\path{doi:https://doi.org/10.1016/j.jsasus.2023.08.001}}.
\newline\urlprefix\url{https://www.sciencedirect.com/science/article/pii/S294992672300001X}

\bibitem{Bae:23}
J.~Bae, J.~M. Kim, S.~J. Lee,
  \href{https://www.sciencedirect.com/science/article/pii/S1738573323002723}{Deep
  reinforcement learning for a multi-objective operation in a nuclear power
  plant}, Nuclear Engineering and Technology 55~(9) (2023) 3277--3290.
\newblock \href {https://doi.org/https://doi.org/10.1016/j.net.2023.06.009}
  {\path{doi:https://doi.org/10.1016/j.net.2023.06.009}}.
\newline\urlprefix\url{https://www.sciencedirect.com/science/article/pii/S1738573323002723}

\bibitem{Jiang:22}
D.~Jiang, D.~Zhang, X.~Li, S.~Wang, C.~Wang, H.~Qin, Y.~Guo, W.~Tian, G.~H. Su,
  S.~Qiu, Fluoride-salt-cooled high-temperature reactors: Review of historical
  milestones, research status, challenges, and outlook, Renewable and
  Sustainable Energy Reviews 161 (2022) 112345.
\newblock \href {https://doi.org/10.1016/j.rser.2022.112345}
  {\path{doi:10.1016/j.rser.2022.112345}}.

\bibitem{Zhang:23}
S.~Zhang, H.-C. Lin, M.~Chen, S.~Shi, S.~Che, A.~Burak, X.~Sun, Q.~Lv, Design
  and construction of an integral-effect test facility flustfa for molten salt
  reactor applications, in: 20th International Topical Meeting on Nuclear
  Reactor Thermal Hydraulics (NURETH20), 2023, pp. 3156--3169.
\newblock \href {https://doi.org/10.13182/NURETH20-40278}
  {\path{doi:10.13182/NURETH20-40278}}.

\bibitem{Andreades:16}
C.~Andreades, J.~K.~C. Anselmo T.~Cisneros, A.~Y.~K. Chong, M.~Fratoni,
  S.~Hong, L.~R. Huddar, K.~D. Huff, J.~Kendrick, D.~L. Krumwiede, M.~R.
  Laufer, M.~Munk, R.~O. Scarlat, N.~Zweibau,
  \href{https://doi.org/10.13182/NT16-2}{Design summary of the mark-i
  pebble-bed, fluoride salt–cooled, high-temperature reactor commercial power
  plant}, Nuclear Technology 195~(3) (2016) 223--238.
\newblock \href {http://arxiv.org/abs/https://doi.org/10.13182/NT16-2}
  {\path{arXiv:https://doi.org/10.13182/NT16-2}}, \href
  {https://doi.org/10.13182/NT16-2} {\path{doi:10.13182/NT16-2}}.
\newline\urlprefix\url{https://doi.org/10.13182/NT16-2}

\bibitem{Zhao:23}
H.~Zhao, L.~Fick, A.~Heald, Q.~Zhou, S.~Richesson, N.~Sutton, B.~Haugh,
  Development, verification, and validation of an advanced systems code kp-sam
  for kairos power fluoride salt–cooled high-temperature reactor (kp-fhr),
  Nuclear Science and Engineering 197~(5) (2023) 813–839.
\newblock \href {https://doi.org/10.1080/00295639.2022.2106724}
  {\path{doi:10.1080/00295639.2022.2106724}}.

\bibitem{Lim:25}
J.~Y. Lim, J.~Li, D.~O’Grady, T.~Downar, K.~Duraisamy,
  \href{https://www.sciencedirect.com/science/article/pii/S0029549324007908}{A
  hybrid surrogate modeling framework for the digital twin of a
  fluoride-salt-cooled high-temperature reactor (fhr)}, Nuclear Engineering and
  Design 433 (2025) 113690.
\newblock \href
  {https://doi.org/https://doi.org/10.1016/j.nucengdes.2024.113690}
  {\path{doi:https://doi.org/10.1016/j.nucengdes.2024.113690}}.
\newline\urlprefix\url{https://www.sciencedirect.com/science/article/pii/S0029549324007908}

\bibitem{Dave:2023}
A.~J. Dave, T.~Lee, R.~Ponciroli, R.~B. Vilim, Design of a supervisory control
  system for autonomous operation of advanced reactors, Annals of Nuclear
  Energy 182 (2023) 109593.
\newblock \href {https://doi.org/10.1016/j.anucene.2022.109593}
  {\path{doi:10.1016/j.anucene.2022.109593}}.

\bibitem{Hu:21}
R.~Hu, L.~Zou, G.~Hu, D.~Nunez, T.~Mui, T.~Fei, Sam theory manual, Technical
  Report ANL/NSE-17/4 Rev. 1, Argonne National Laboratory (2021).

\bibitem{OGrady:21}
D.~O’Grady, T.~Mui, A.~Lee, L.~Zou, G.~Hu, R.~Hu,
  \href{https://www.osti.gov/servlets/purl/1781830/}{Sam code enhancement,
  validation, and reference model development for fluoride-salt-cooled
  high-temperature reactors}, Tech. Rep. ANL/NSE-21/15, 1781830, 167974,
  Argonne National Laboratory (Apr. 2021).
\newblock \href {https://doi.org/10.2172/1781830} {\path{doi:10.2172/1781830}}.
\newline\urlprefix\url{https://www.osti.gov/servlets/purl/1781830/}

\bibitem{Li:24}
J.~Li, T.~Downar, V.~Seker, D.~O’Grady, R.~Hu, N.~Satvat, S.~Kinast,
  \href{https://www.tandfonline.com/doi/full/10.1080/00295450.2024.2381282}{Neutronics
  and {Thermo}-{Fluids} {Simulation} of {Generic} {Pebble}-{Bed}
  {Fluoride}-{Salt}-{Cooled} {High}-{Temperature} {Reactor}}, Nuclear
  Technology (2024) 1--17Publisher: Taylor \& Francis.
\newblock \href {https://doi.org/10.1080/00295450.2024.2381282}
  {\path{doi:10.1080/00295450.2024.2381282}}.
\newline\urlprefix\url{https://www.tandfonline.com/doi/full/10.1080/00295450.2024.2381282}

\bibitem{Satvat:21}
N.~Satvat, F.~Sarikurt, K.~Johnson, I.~Kolaja, M.~Fratoni, B.~Haugh,
  E.~Blandford, Neutronics, thermal-hydraulics, and multi-physics benchmark
  models for a generic pebble-bed fluoride-salt-cooled high temperature reactor
  (fhr), Nuclear Engineering and Design 384 (2021) 111461.
\newblock \href {https://doi.org/10.1016/j.nucengdes.2021.111461}
  {\path{doi:10.1016/j.nucengdes.2021.111461}}.

\bibitem{Lee:23}
T.~Lee, R.~Ponciroli, A.~Dave, D.~O’Grady, R.~Vilim,
  \href{https://www.osti.gov/servlets/purl/1961558/}{Centrifugal pump model for
  system codes for advanced npp designs}, Tech. Rep. ANL/NSE-23/10, 1961558,
  181072, Argonne National Laboratory (Feb. 2023).
\newblock \href {https://doi.org/10.2172/1961558} {\path{doi:10.2172/1961558}}.
\newline\urlprefix\url{https://www.osti.gov/servlets/purl/1961558/}

\bibitem{Kennedy:1980}
W.~G. Kennedy, M.~C. Jacob, J.~C. Whitehouse, J.~D. Fishburn, G.~J. Kanupka,
  \href{https://www.osti.gov/biblio/6743330}{Pump two-phase performance
  program. volume 3. transient tests. final report. [pwr; bwr]}, Tech. Rep.
  EPRI-NP-1556(Vol.3), Combustion Engineering, Inc., Windsor, CT (United
  States) (Sep. 1980).
\newblock \href {https://doi.org/10.2172/6743330} {\path{doi:10.2172/6743330}}.
\newline\urlprefix\url{https://www.osti.gov/biblio/6743330}

\bibitem{Fanning:17}
T.~H. Fanning, A.~J. Brunett, T.~Sumner,
  \href{https://www.osti.gov/biblio/1352187}{The sas4a/sassys-1 safety analysis
  code system, version 5}, Tech. Rep. ANL/NE-16/19, Argonne National Laboratory
  (Jan. 2017).
\newblock \href {https://doi.org/10.2172/1352187} {\path{doi:10.2172/1352187}}.
\newline\urlprefix\url{https://www.osti.gov/biblio/1352187}

\bibitem{Hu:19}
G.~Hu, G.~Zhang, R.~Hu, \href{https://www.osti.gov/biblio/1499041}{Reactivity
  feedback modeling in sam}, Tech. Rep. ANL/NSE-19/1, Argonne National
  Laboratory (2 2019).
\newblock \href {https://doi.org/10.2172/1499041} {\path{doi:10.2172/1499041}}.
\newline\urlprefix\url{https://www.osti.gov/biblio/1499041}

\bibitem{Hu:20}
G.~Hu, D.~O'Grady, L.~Zou, R.~Hu,
  \href{https://www.osti.gov/biblio/1674975}{Development of a reference model
  for molten-salt-cooled pebble-bed reactor using sam}, Technical Report
  ANL/NSE-20/31, Argonne National Laboratory (9 2020).
\newblock \href {https://doi.org/10.2172/1674975} {\path{doi:10.2172/1674975}}.
\newline\urlprefix\url{https://www.osti.gov/biblio/1674975}

\bibitem{PylorofGarcia:22}
D.~Pylorof, H.~E. Garcia, A reinforcement learning approach to long-horizon
  operations, health, and maintenance supervisory control of advanced energy
  systems, Engineering Applications of Artificial Intelligence 116 (2022)
  105454.
\newblock \href {https://doi.org/10.1016/j.engappai.2022.105454}
  {\path{doi:10.1016/j.engappai.2022.105454}}.

\bibitem{haarnoja:18}
T.~Haarnoja, A.~Zhou, P.~Abbeel, S.~Levine,
  \href{https://proceedings.mlr.press/v80/haarnoja18b.html}{Soft actor-critic:
  Off-policy maximum entropy deep reinforcement learning with a stochastic
  actor}, in: J.~Dy, A.~Krause (Eds.), Proceedings of the 35th International
  Conference on Machine Learning, Vol.~80 of Proceedings of Machine Learning
  Research, PMLR, 2018, pp. 1861--1870.
\newline\urlprefix\url{https://proceedings.mlr.press/v80/haarnoja18b.html}

\bibitem{stable-baselines3}
A.~Raffin, A.~Hill, A.~Gleave, A.~Kanervisto, M.~Ernestus, N.~Dormann,
  \href{http://jmlr.org/papers/v22/20-1364.html}{Stable-baselines3: Reliable
  reinforcement learning implementations}, Journal of Machine Learning Research
  22~(268) (2021) 1--8.
\newline\urlprefix\url{http://jmlr.org/papers/v22/20-1364.html}

\bibitem{BempoardMosca:94}
A.~Bemporad, E.~Mosca, Constraint fulfilment in control systems via predictive
  reference management, in: Proceedings of 1994 33rd IEEE Conference on
  Decision and Control, Vol.~3, 1994, pp. 3017--3022 vol.3.
\newblock \href {https://doi.org/10.1109/CDC.1994.411327}
  {\path{doi:10.1109/CDC.1994.411327}}.

\bibitem{kolmanovsky:14}
I.~Kolmanovsky, E.~Garone, S.~Di~Cairano, Reference and command governors: A
  tutorial on their theory and automotive applications, in: 2014 American
  Control Conference, IEEE, 2014, pp. 226--241.

\bibitem{Kalman:60}
R.~E. Kalman, \href{https://doi.org/10.1115/1.3662552}{{A New Approach to
  Linear Filtering and Prediction Problems}}, Journal of Basic Engineering
  82~(1) (1960) 35--45.
\newblock \href {https://doi.org/10.1115/1.3662552}
  {\path{doi:10.1115/1.3662552}}.
\newline\urlprefix\url{https://doi.org/10.1115/1.3662552}

\bibitem{Evensen:94}
G.~Evensen,
  \href{https://agupubs.onlinelibrary.wiley.com/doi/abs/10.1029/94JC00572}{Sequential
  data assimilation with a nonlinear quasi-geostrophic model using monte carlo
  methods to forecast error statistics}, Journal of Geophysical Research:
  Oceans 99~(C5) (1994) 10143--10162.
\newblock \href
  {http://arxiv.org/abs/https://agupubs.onlinelibrary.wiley.com/doi/pdf/10.1029/94JC00572}
  {\path{arXiv:https://agupubs.onlinelibrary.wiley.com/doi/pdf/10.1029/94JC00572}},
  \href {https://doi.org/https://doi.org/10.1029/94JC00572}
  {\path{doi:https://doi.org/10.1029/94JC00572}}.
\newline\urlprefix\url{https://agupubs.onlinelibrary.wiley.com/doi/abs/10.1029/94JC00572}

\bibitem{Evensen:03}
G.~Evensen, The ensemble kalman filter: Theoretical formulation and practical
  implementation, Ocean dynamics 53 (2003) 343--367.

\bibitem{burgers:98}
G.~Burgers, P.~Jan~van Leeuwen, G.~Evensen, Analysis scheme in the ensemble
  kalman filter, Monthly weather review 126~(6) (1998) 1719--1724.

\bibitem{sarkka:13}
S.~Särkkä, Bayesian Filtering and Smoothing, Institute of Mathematical
  Statistics Textbooks, Cambridge University Press, 2013.
\newblock \href {https://doi.org/10.1017/CBO9781139344203}
  {\path{doi:10.1017/CBO9781139344203}}.

\bibitem{sobol:01}
I.~Sobol,
  \href{https://www.sciencedirect.com/science/article/pii/S0378475400002706}{Global
  sensitivity indices for nonlinear mathematical models and their monte carlo
  estimates}, Mathematics and Computers in Simulation 55~(1) (2001) 271--280,
  the Second IMACS Seminar on Monte Carlo Methods.
\newblock \href {https://doi.org/https://doi.org/10.1016/S0378-4754(00)00270-6}
  {\path{doi:https://doi.org/10.1016/S0378-4754(00)00270-6}}.
\newline\urlprefix\url{https://www.sciencedirect.com/science/article/pii/S0378475400002706}

\bibitem{Lutkepol:07}
H.~Lütkepohl, New Introduction to Multiple Time Series Analysis, Springer,
  Berlin, 2007.

\bibitem{Iwanaga2022}
T.~Iwanaga, W.~Usher, J.~Herman,
  \href{https://sesmo.org/article/view/18155}{Toward {SALib} 2.0: {Advancing}
  the accessibility and interpretability of global sensitivity analyses},
  Socio-Environmental Systems Modelling 4 (2022) 18155.
\newblock \href {https://doi.org/10.18174/sesmo.18155}
  {\path{doi:10.18174/sesmo.18155}}.
\newline\urlprefix\url{https://sesmo.org/article/view/18155}

\bibitem{Herman2017}
J.~Herman, W.~Usher, \href{https://doi.org/10.21105/joss.00097}{{SALib}: An
  open-source python library for sensitivity analysis}, The Journal of Open
  Source Software 2~(9) (jan 2017).
\newblock \href {https://doi.org/10.21105/joss.00097}
  {\path{doi:10.21105/joss.00097}}.
\newline\urlprefix\url{https://doi.org/10.21105/joss.00097}

\bibitem{saltelli:02}
A.~Saltelli,
  \href{https://www.sciencedirect.com/science/article/pii/S0010465502002801}{Making
  best use of model evaluations to compute sensitivity indices}, Computer
  Physics Communications 145~(2) (2002) 280--297.
\newblock \href {https://doi.org/https://doi.org/10.1016/S0010-4655(02)00280-1}
  {\path{doi:https://doi.org/10.1016/S0010-4655(02)00280-1}}.
\newline\urlprefix\url{https://www.sciencedirect.com/science/article/pii/S0010465502002801}

\end{thebibliography}

\appendix
\section{gFHR Surrogate Model and Pump Health Surrogate}\label{app:gFHRsurrogate}
This appendix section provides the supplemental information about the Virtual Asset submodules --- the gFHR Surrogate and the Pump Health Surrogate.
The Virtual Asset was developed by Lim et al.~\cite{Lim:25}, with the gFHR Surrogate serving as a reduced-complexity model of the physics-based SAM gFHR Model and the Pump Health Surrogate providing a degradation analysis on the HSCs, the pumps.

\begin{figure}
    \centering
    \includegraphics[width=0.8\linewidth]{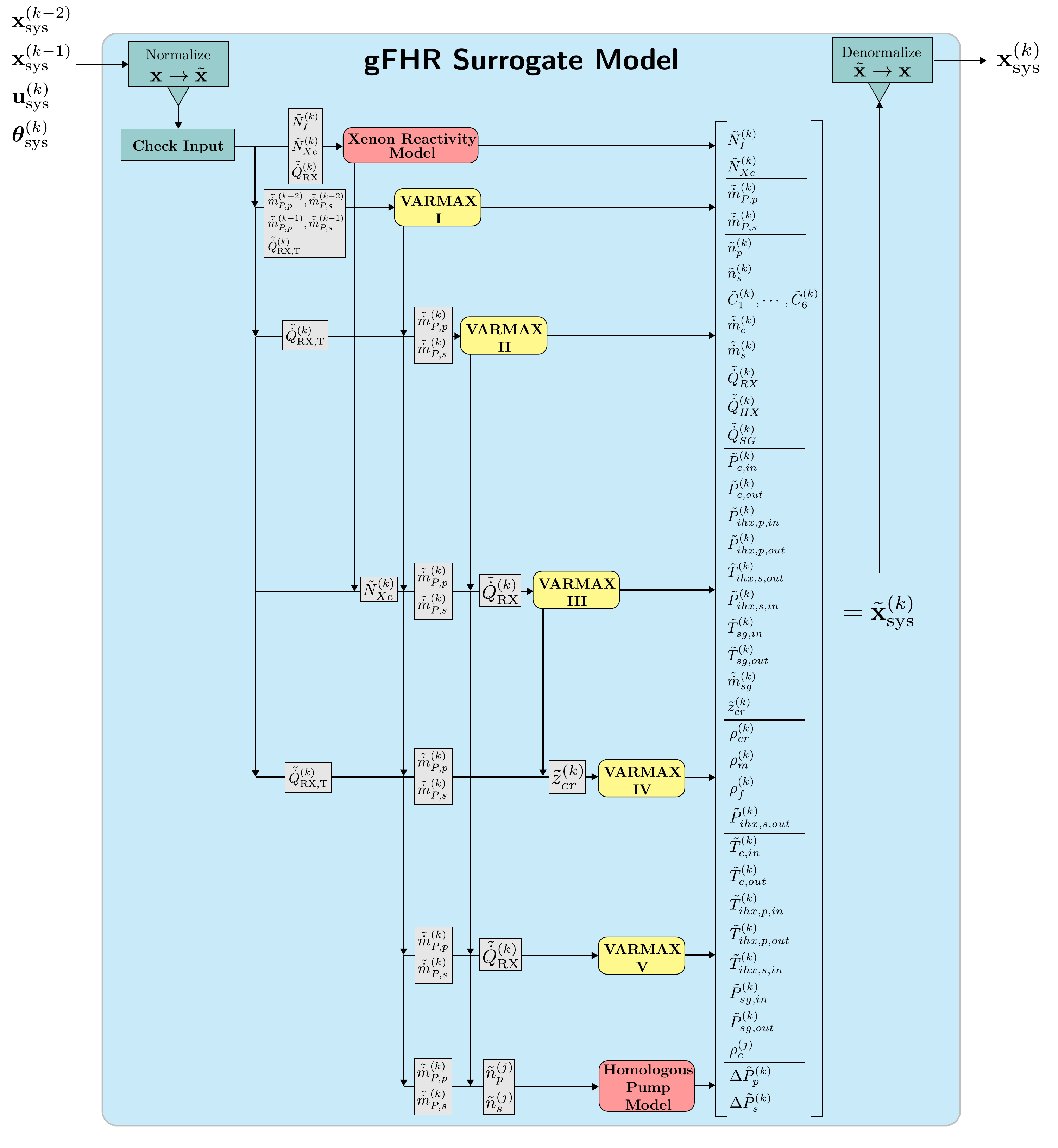}
    \caption{The gFHR Surrogate Model framework in the Virtual Asset}
    \label{fig:gFHRSurrogate}
\end{figure}

The gFHR Surrogate model, shown in Figure~\ref{fig:gFHRSurrogate}, has a hybrid structure made up of both physics-based models and a network of Vector AutoRegressive Moving-Average with eXogenous input (VARMAX) models. 
The xenon reactivity and homologous pump physics-derived models are integrated to enhance the surrogate's application specific accuracy, and the statistical VARMAX model enables the computational speed-up.
Further details about the VARMAX statistical model, including composition and trainable parameters, are detailed in~\ref{app:varmax}.

Similarly to the physics-based SAM gFHR model, the gFHR Surrogate model outputs a detailed system state response given a reactor target power distribution.
The gFHR Surrogate model $F_{\text{sys}}(\cdot)$ is mathematically represented as follows:
\begin{equation}\label{eq:surrogate}
    \mathbf{x}^{(k)} = F_{\text{sys}}\left(\mathbf{x}_{\text{sys}}^{(k-2)},\mathbf{x}_{\text{sys}}^{(k-1)},\mathbf{u}_{\text{sys}}^{(k)};\boldsymbol{\theta}_{\text{sys}}^{(k)}\right),
\end{equation}
where two time-delays of the system state $\mathbf{x}_{\text{sys}}\in\mathbb{R}^{n_{xs}}$, the surrogate input $\mathbf{u}_{\text{sys}}\in\mathbb{R}^{n_{us}}$, and the surrogate parameters $\boldsymbol{\theta}_{\text{sys}}$ are needed.
The gFHR Surrogate model input vector $\mathbf{u}_{\text{sys}}\in\mathbb{R}^{n_{us}}$ variables are listed in Table~\ref{tab:gFHR-SS-inputs} and the state vector $\mathbf{x}_{\text{sys}}\in\mathbb{R}^{n_{xs}}$ variables are listed in Table~\ref{tab:gFHR-SS-states}.
In this work, the surrogate parameters $\boldsymbol{\theta}_{\text{sys}}$ are the trainable VARMAX coefficients.

\begin{table}
    \centering
    \begin{tabular}{lll}
        \hline
        Variable & Description & Units \\
        \hline
        $t$ & Time & s \\
        $\dot{Q}_{RX,T}$ & Target Reactor Core Power & W \\ \hline
    \end{tabular}
    \caption{List of the gFHR Surrogate Model input vector $\mathbf{u}_{\text{sys}}\in\mathbb{R}^{2}$ variables with description and units.}
    \label{tab:gFHR-SS-inputs}
\end{table}

\begin{table}
    \centering
    \begin{tabular}{llll}
        \hline
        Variable & Description & Units & Type \\
        \hline
        $t$ & Time & s & P\\
        $N_{\text{I}}$ & Iodine-135 Concentration & $\Delta$k/k & C\\
        $N_{\text{Xe}}$ & Xenon-135 Concentration & $\Delta$k/k & C\\
        $T_{c,in}$ & Core Inlet Temperature & K &BC\\
        $T_{c,out}$ & Core Outlet Temperature & K &BC\\
        $P_{c,in}$ & Core Inlet Temperature & Pa &BC\\
        $P_{c,out}$ & Core Outlet Temperature & Pa &P\\
        $T_{ihx,p,in}$ & IHX Primary Inlet Temperature & K&BC\\
        $T_{ihx,p,out}$ & IHX Primary Outlet Temperature & K&BC\\
        $P_{ihx,p,in}$ & IHX Primary Inlet Temperature & Pa &P\\
        $P_{ihx,p,out}$ & IHX Primary Outlet Temperature & Pa &P\\
        $T_{ihx,s,in}$ & IHX Secondary Inlet Temperature & K&P\\
        $T_{ihx,s,out}$ & IHX Secondary Outlet Temperature & K&P\\
        $P_{ihx,s,in}$ & IHX Secondary Inlet Temperature & Pa &BC\\
        $P_{ihx,s,out}$ & IHX Secondary Outlet Temperature & Pa &P\\
        $T_{sg,in}$ & SG Inlet Temperature & K&P\\
        $T_{sg,out}$ & SG Outlet Temperature & K&BC\\
        $P_{sg,in}$ & SG Inlet Temperature & Pa &P\\
        $P_{sg,out}$ & SG Outlet Temperature & Pa &P\\
        $\dot{m}_c$ & Core Flow & kg/s & P\\
        $\dot{m}_s$ & Secondary Flow & kg/s & P\\
        $\dot{m}_{sg}$ & SG Flow & kg/s & P\\
        $\dot{Q}_{RX}$ & Reactor Core Power & W & P\\
        $\dot{Q}_{HX}$ & IHX Heat Extraction Power & W & P\\
        $\dot{Q}_{SG}$ & SG Heat Extraction & W & P\\
        $C_1,\hdots,C_6$ & Six-group Delay Neutron Precursor Concentrations &   & C\\
        $\rho_m$ & Moderator Reactivity & $\Delta$k/k & C\\
        $\rho_c$ & Coolant Reactivity & $\Delta$k/k & C\\
        $\rho_f$ & Fuel Reactivity & $\Delta$k/k & C\\
        $\rho_{cr}$ & External Reactivity Insertion from Control Rod& $\Delta$k/k & C\\
        $z_{cr}$ & Control Rod Position & m & CS\\
        $\dot{m}_{P,p}$ & Primary Pump Flow & kg/s & P\\
        $\dot{m}_{P,s}$ & Secondary Pump Flow & kg/s & P\\
        $n_{p}$ & Primary Pump Speed & RPM & P\\
        $n_{s}$ & Secondary Pump Speed & RPM & P\\
        $\Delta P_p$ & Primary Pump Head & Pa & CS\\ 
        $\Delta P_s$ & Secondary Pump Head & Pa & CS\\ \hline
    \end{tabular}
    \caption{List of gFHR Surrogate Model state vector $\mathbf{x}_{\text{sys}}\in\mathbb{R}^{42}$ variables with description, units, and types. The state types are (BC) Boundary Condition, (C) Computational, (CS) Control System, (H) Health, and (P) Physical.}
    \label{tab:gFHR-SS-states}
\end{table}

Using the state estimation provided by the gFHR Surrogate Model, the Pump Health Surrogate performs a physics-based health analysis of the pump based on the degradation model for a centrifugal pump in the report by Lee et al.~\cite{Lee:23}. 
The Pump Health Surrogate function $F_{\text{pump}}(\cdot)$ estimates the pump health states $\mathbf{x}_{\text{pump}}\in\mathbb{R}^{n_{x,\text{pump}}}$, which is mathematically defined as:
\begin{equation}
    \mathbf{x}_{\text{pump}}^{(k)} = F_{\text{pump}} \left(\mathbf{x}_{\text{sys}}^{(k)},\mathbf{x}_{\text{sys}}^{(k-1)},\mathbf{x}_{\text{pump}}^{(k-1)},\mathbf{u}_{\text{pump}}^{(k)}\right).
\end{equation}
The Pump Health Surrogate input vector $\mathbf{u}_{\text{pump}}$ variables are listed in Table~\ref{tab:gFHR-SS-Pump-inputs} and the state vector $\mathbf{x}_{\text{pump}}$ variables are listed in Table~\ref{tab:gFHR-SS-Pump-states}.
The models parameters $\boldsymbol{\theta}_{\text{pump}}(\cdot)$ are listed in Table~\ref{tab:pump-deg-params}, which define the degradation rate of each pump.

\begin{table}
    \centering
    \begin{tabular}{lll}
        \hline
        Variable & Description \\
        \hline
        $\mathbf{M}_{\text{pump},p}$ & Primary pump maintenance action\\
        $\boldsymbol{\theta}_{\text{pump},p}$ & Primary pump degradation parameters\\
        $\mathbf{M}_{\text{pump},s}$ & Secondary pump maintenance action\\
        $\boldsymbol{\theta}_{\text{pump},s}$ & Secondary pump degradation parameters\\\hline
    \end{tabular}
    \caption{List of Pump Health Surrogate input vector variables with description and units. The pump degradation parameters $\boldsymbol{\theta}_{\text{pump}}$ are listed in Table~\ref{tab:pump-deg-params}.}
    \label{tab:gFHR-SS-Pump-inputs}
\end{table}

\begin{table}
    \centering
    \begin{tabular}{llll}
        \hline
        Variable & Description & Units & Type \\
        \hline
        $t$ & Time & s & P\\
        $\dot{Q}_{P,D,p}$ & Required Primary Pump Power with Degradation & W & H\\
        $\dot{Q}_{P,p}$ & Required Primary Pump Power  & W & P\\
        $K_p$ & Primary Pump Degradation Loss Coefficient &  & H\\
        $\eta_p$ & Primary Pump Health Index Score &  & H\\
        $\dot{Q}_{P,D,s}$ & Required Secondary Pump Power with Degradation & W &H\\
        $\dot{Q}_{P,s}$ & Required Secondary Pump Power & W &P\\
        $K_s$ & Secondary Pump Degradation Loss Coefficient &  &H\\
        $\eta_s$ & Secondary Pump Health Index Score &  &H\\
        \hline
    \end{tabular}
    \caption{List of Pump Health Surrogate pump state vector $\mathbf{x}_{\text{pump}} \in\mathbb{R}^{9}$ variables with description and units. The state types are (BC) Boundary Condition, (C) Computational, (CS) Control System, (H) Health, and (P) Physical.}
    \label{tab:gFHR-SS-Pump-states}
\end{table}

\begin{table}
    \centering
    \begin{tabular}{lll}
        \hline
        Variable & Description & Unit\\ \hline
        $\lambda_D$ & Fractional pump head loss & \%\\
        $T_D$ & Time when $\lambda_D$ is reached & s\\
        $\sigma_D$ & Degradation uncertainty & -\\
        $\sigma_I$ & Time scaling effect on degradation & -\\
        $\alpha$ & Pump flow rate scaling effect on degradation & -\\
        $\alpha_{\dot{m}}$ & Pump flow rate change scaling effect on degradation & -\\
        $\varphi$ & Time-step of the pump health model & -\\ \hline
    \end{tabular}
    \caption{$\boldsymbol{\theta}_{\text{pump}}$: pump degradation parameters. These values are specified for both the primary and secondary pumps. [-] unit indicates that the parameter is dimensionless.}
    \label{tab:pump-deg-params}
\end{table}

\subsection{gFHR Surrogate Model for Shock Capturing Case}\label{app:gFHRSurrShock}
The original gFHR Surrogate model was trained in order to model plant dynamics, assuming that the system conditions remained the same. 
In case three of the digital twin demonstrations, a system shock is introduced through the steam generator, changing the boundary condition set on the heat exchanger secondary inlet temperature.
This section will cover the re-structuring and re-training procedure for the gFHR Surrogate in order to capture the specified system shock. 

\begin{figure}
    \centering
    \includegraphics[width=0.8\linewidth]{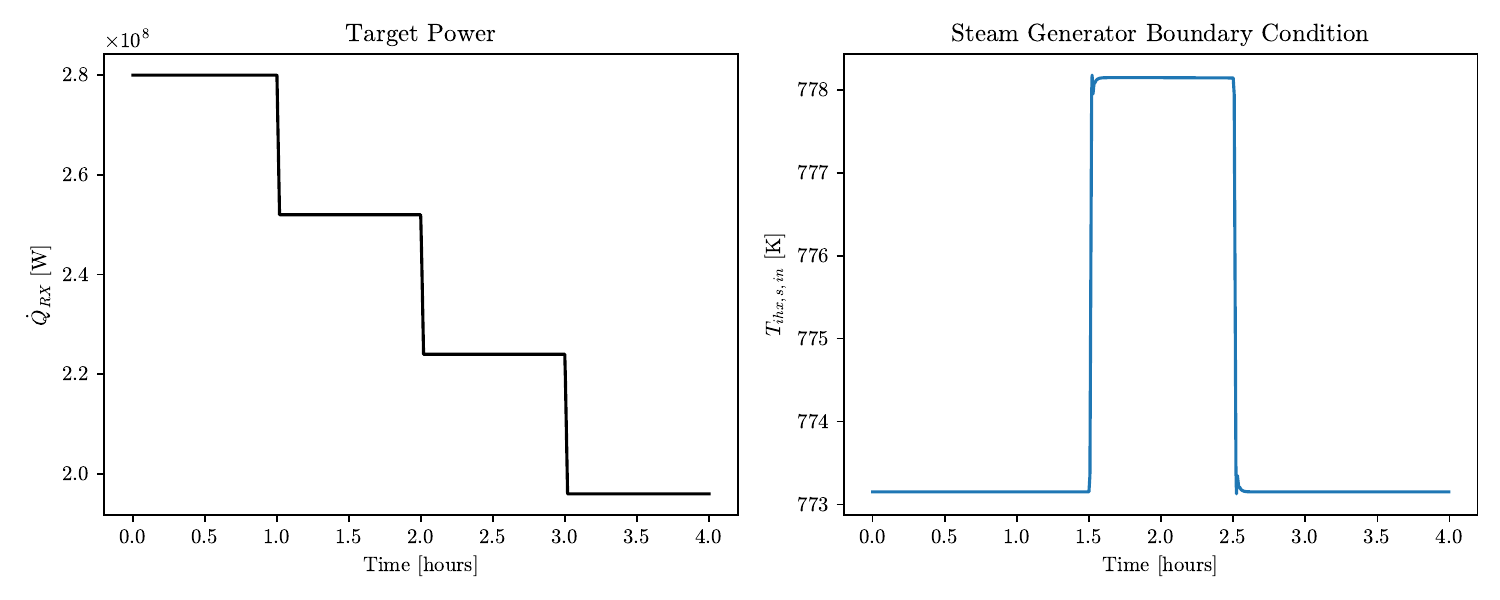}
    \caption{Toy three ramp load follow case with 5 degree Kelvin increase to the IHX secondary inlet temperature to model a shock in the steam generator.}
    \label{fig:toy-gFHRSurr}
\end{figure}

\begin{figure}
    \centering
    \includegraphics[width=0.9\linewidth]{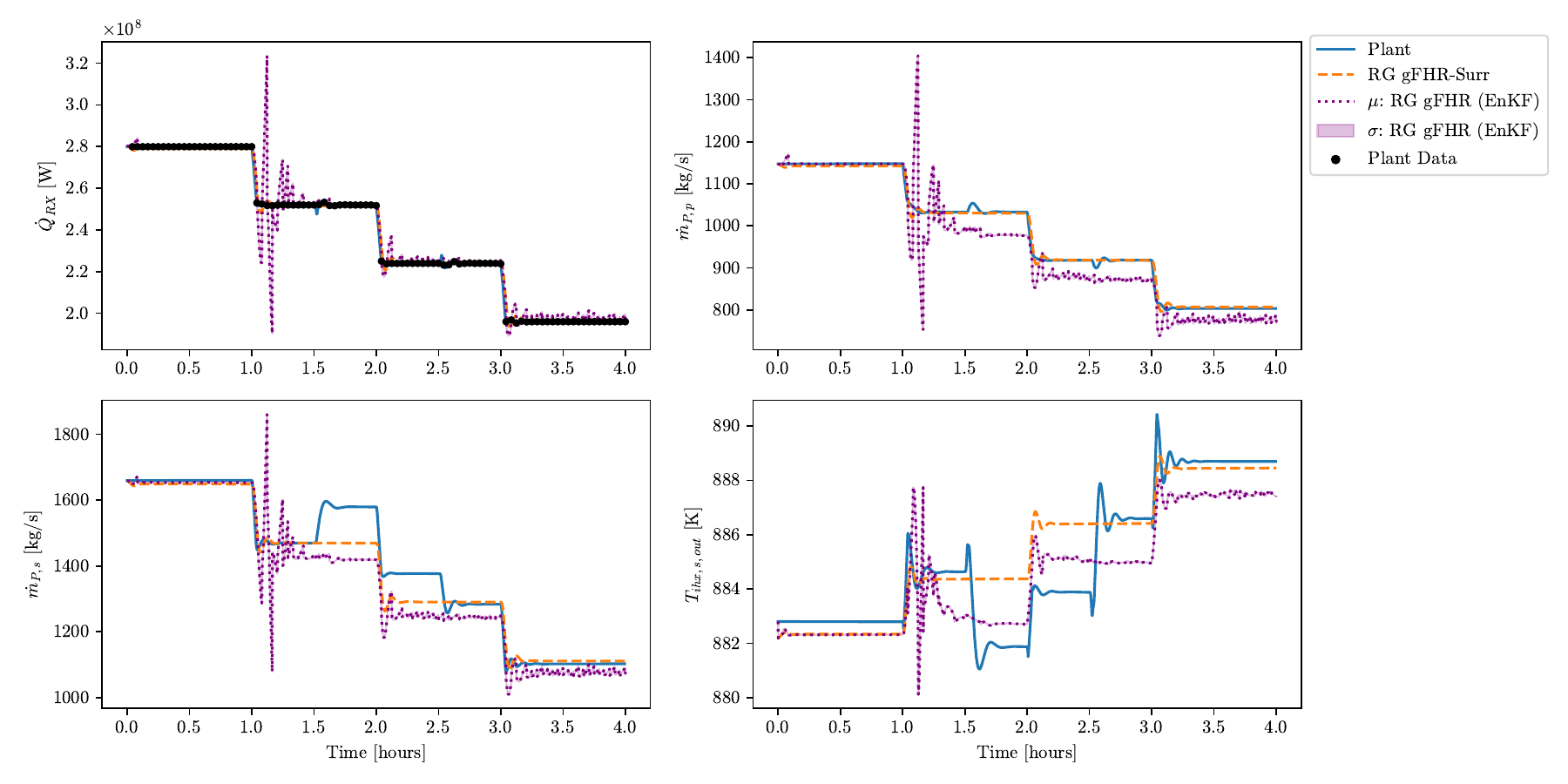}
    \caption{Toy three ramp load follow case with 5 degree Kelvin increase to the IHX secondary inlet temperature to model a shock in the steam generator. The originally developed gFHR Surrogate model is used as the dynamics model in a EnKF state-parameter estimation with observing the core energy. The estimation for the core energy and three defined constraints (primary mass flow rate, secondary mass flow rate and IHX secondary outlet temperature) are shown.}
    \label{fig:toy-states_gFHRSurr}
\end{figure}

A discussion of the original gFHR surrogate's short-comings for this specific shock case are first presented.
The surrogate is applied to a toy three-ramp load follows case with five Kelvin increase to the IHX secondary inlet temperature to model the steam generator shock; the case inputs are shown in Figure~\ref{fig:toy-gFHRSurr}. 
Applying the EnKF with the same set-up as the first digital twin demonstration shown in Section~\ref{subsec:demo1}, the estimation results for the core energy, primary and secondary mass flow rates, and IHX secondary outlet temperature are shown in Figure~\ref{fig:toy-states_gFHRSurr}.
Here, the core power is observed at a more frequent measurement frequency every $\beta=25$ time-steps.
From this plot, the EnKF fails to adapt the surrogate model to the system change because it is not assimilating the correct parameters to adjust the newly added observable state and the other system state prediction do not change because the VARMAX model was not trained with data that shows the shock behavior.
This is especially seen in the state where the shock is effecting the system, and in this case it is secondary pump mass flow and IHX secondary outlet temperatures. 
One can also observe that the estimation looks very noisy, which is an indication that the system is not computationally stable.
A higher observation frequency is necessary for this case because the surrogate strays away from the actual values in between observations, which is what is causing the noisy behavior.
Without a small measurement step, the ensemble sample spread would continue to grow and eventually cause the prediction to diverge. 
This gFHR Surrogate needs to be modified such that the expected shock changes to the state dynamics are incorporated and such that small noise perturbations do not cause estimation divergence. 

\begin{figure}
    \centering
    \includegraphics[width=0.85\linewidth]{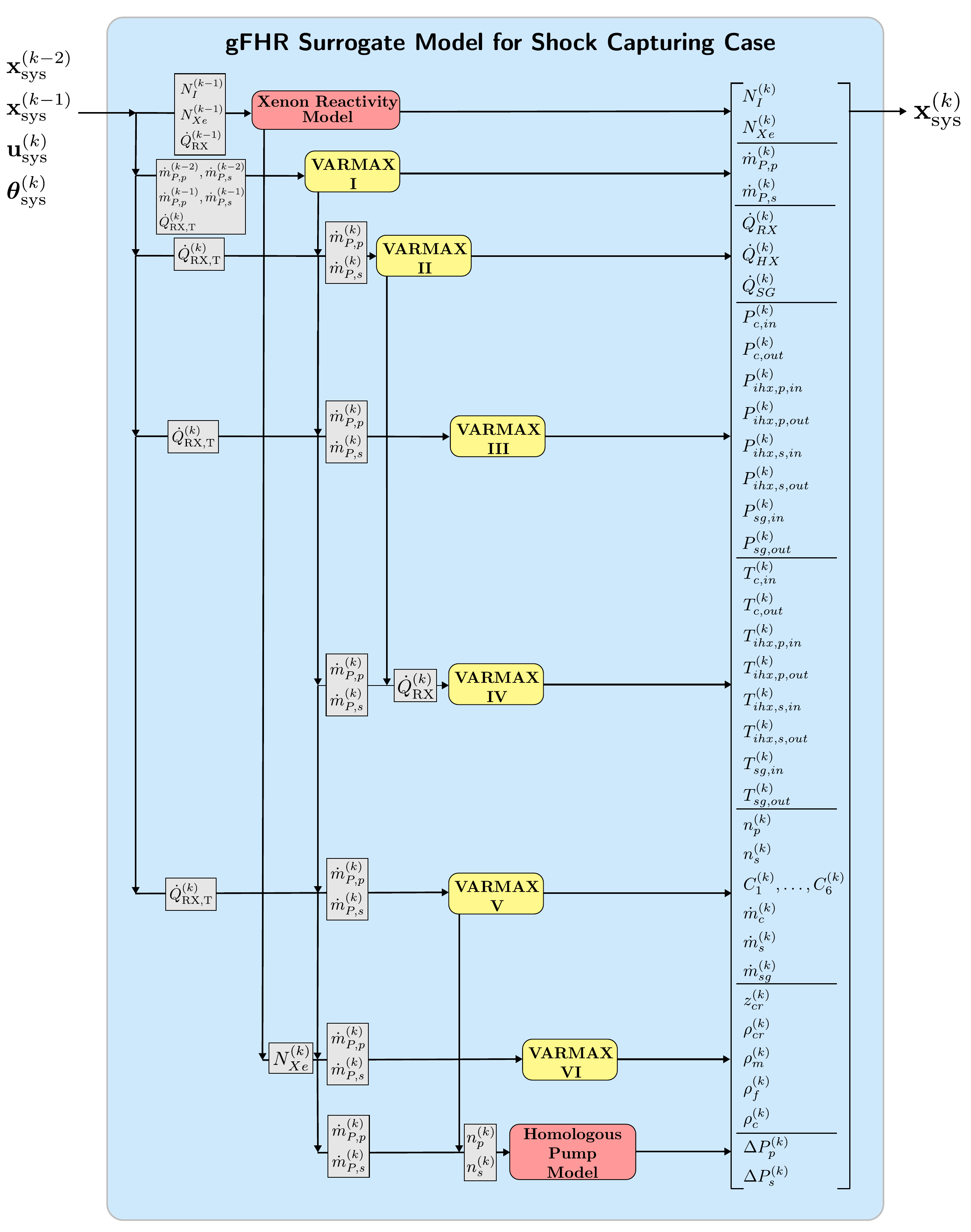}
    \caption{The modified gFHR Surrogate Model developed for the steam generator shock capturing case.}
    \label{fig:systemSurrogateSG}
\end{figure}

To address these issues, first a new gFHR surrogate model is built with data containing the shock behavior.
It is a design choice that the secondary pump mass flow rate $\dot{m}_{P,s}$ is added to the measurement model in order to detect the shock. 
The network of data-driven models is re-structured to a system of six VARMAX that group state based on their variable types; for instance, all pressure variables are group together.
Figure~\ref{fig:systemSurrogateSG} shows the modified gFHR surrogate model structure for shock capturing, where the measured secondary pump mass flow rate is predicted first and is the input to every following VARMAX model to integrate the information of data into the predictions.
The training data consists of 21 hours with power changes randomly occurring between 60\% and 100\% power, and random changes to the IHX secondary inlet temperature adding or subtracting up to eight degrees Kelvin. 
The first VARMAX model is not trained with the shock data because the only effected state, the secondary pump mass flow rate, can be corrected via measurements in the digital twin framework.
The remaining VARMAX models are trained with the shock training data such that is shock occurs, the digital twin assimilation will inform the remaining surrogate states via a measurement of the secondary pump mass flow rate. 

The new gFHR Surrogate model also does not normalize the state values in order to improve its robustness to perturbations it encounters in the filtering algorithm.
In the original surrogate model, the states are normalized by their total values at 100\% power, re-scaling the parameter space of every system state between 0 and 1 in the gFHR Surrogate predictions.
This causes issues in the EnKF algorithm when the digital twin needs to make major changes to the parameters because a small change in the parameter may translate to changing state prediction by several orders of magnitude. 
Therefore, the states are not normalized and motivated the design choice to create the sub VARMAX models grouped by similar state types. 

\begin{figure}
    \centering
    \includegraphics[width=0.9\linewidth]{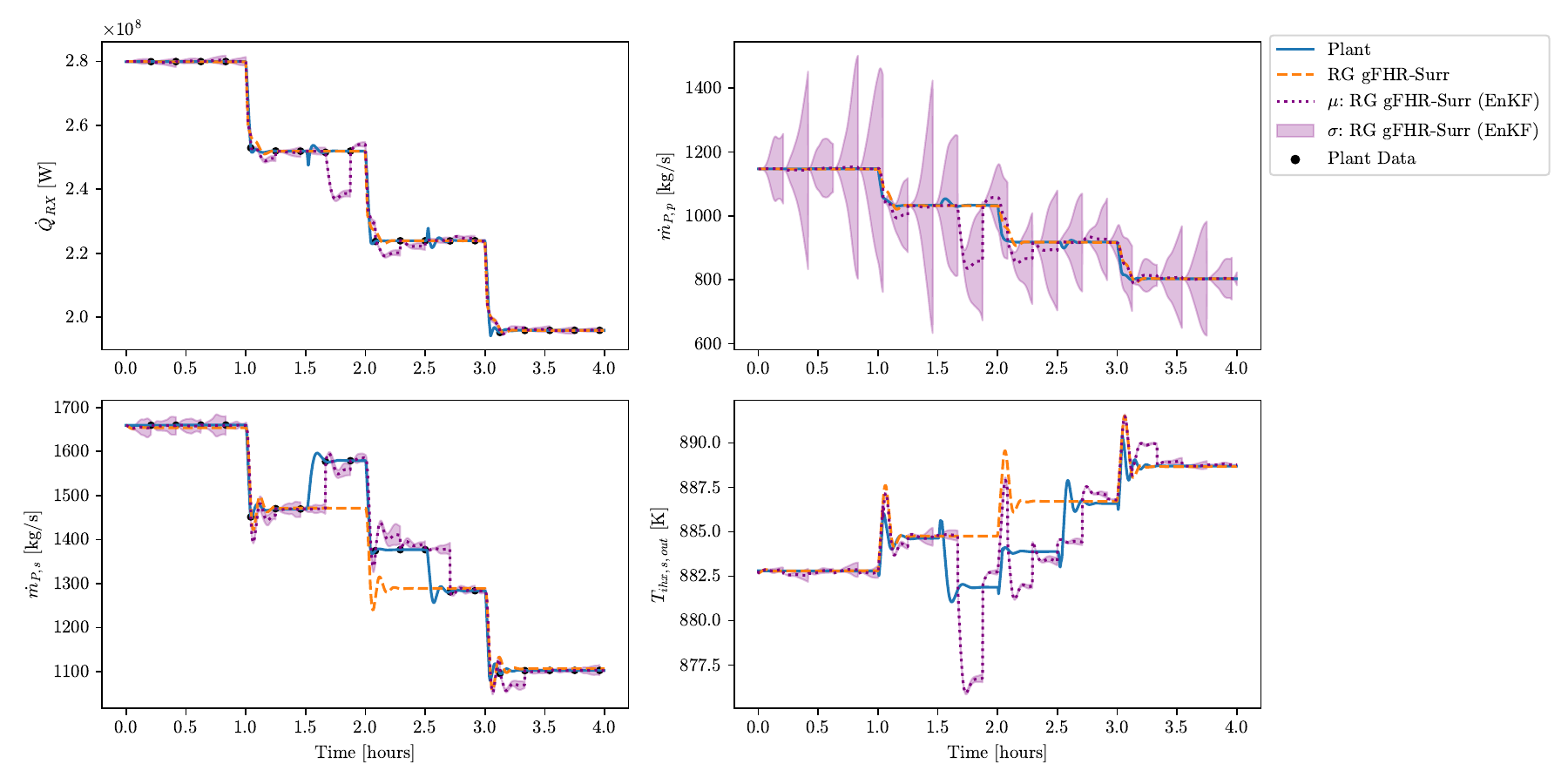}
    \caption{Toy three ramp load follow case with 5 degree Kelvin increase to the IHX secondary inlet temperature to model a shock in the steam generator. The modified gFHR Surrogate model is used as the dynamics model in a EnKF state-parameter estimation with observing the core energy and the secondary pump mass flow rate. The estimation for the core energy and three defined constraints (primary pump mass flow rate, secondary pump mass flow rate and IHX secondary outlet temperature) are shown.}
    \label{fig:toy-states_gFHRSurrMod}
\end{figure}

Since the secondary pump mass flow rate $\dot{m}_{P,s}$ is added as an observation, a new Sobol indices analysis is performed on the updated gFHR Surrogate to determine a new parameter vector $\boldsymbol{\theta}_A$, the results are shown in~\ref{app:sobol2}.
Applying the newly modified gFHR surrogate is tested on the same three ramp toy case, Figure~\ref{fig:toy-states_gFHRSurrMod} shows the updated surrogate.
With this updated model, the outputs can be observed less frequently without loss of computational stability; this is seen as the assimilation time-step is increased to $\beta = 150$. 
Additionally, the states with magnitude changes caused by the shock are captured with this new version of the surrogate.
The modified gFHR Surrogate is easily placed into the digital twin framework, since input and output functionality did not change.

\section{Vector Autoregressive Moving Average with Exogenous Term}\label{app:varmax}
The Vector AutoRegressive Moving-Average with eXogenous input (VARMAX) is the vectorized form of the AutoRegressive Moving-Average with eXogenous term (ARMAX). With the VARMAX model, the state value at the next time step $\mathbf{x}^{(k)}\in\mathbb{R}^{n_x}$ is determined by the linear combination of $p$ lagged state variables $\mathbf{x}^{k-1}, \mathbf{x}^{(k-2)},\ldots, \mathbf{x}^{(k-p)}$; the exogenous term $\mathbf{u}^{(k)}\in\mathbb{R}^{n_u}$; and a moving average model of order $q$. The trainable parameters are: the autoregressive coefficients $\mathbf{a}\in\mathbb{R}^{n_x}, \mathbf{A}_i\in\mathbb{R}^{(n_x\times n_x)}$ for $i\in[1,\hdots,p]$; the input coefficients $\mathbf{B}\in\mathbb{R}^{n_x \times n_u}$; and the moving average coefficients $\mathbf{M}_j \in\mathbb{R}^{n_x \times n_x}$ for $j=[1,\hdots,q]$. The total number of trainable parameters is $(1+n_u)n_x + (p+q)n_x^2$. If the $\mathbf{M}$ matrices are diagonal, the total number of trainable parameters is $(1+n_u+q)n_x + p n_x^2$. The VARMAX model is written as:
\begin{align}
    \mathbf{x}^{(k)} = V_{p,q}\left(\mathbf{x},\mathbf{u}\right) = \mathbf{a} + \mathbf{A}_{1} \mathbf{x}^{(k-1)} + \ldots + \mathbf{A}_p \mathbf{x}^{(k-p)} + \mathbf{B}\mathbf{u}^{(k)} + &\mathbf{M}_1 \boldsymbol\varepsilon^{(k-1)} + \ldots + \mathbf{M}_q \boldsymbol\varepsilon^{(k-q)}, \notag\\ 
    &\boldsymbol{\varepsilon}^{(0)} \sim \mathcal{N}(0,\boldsymbol{\Omega}),
\end{align}
where $\boldsymbol{\varepsilon}^{(k)}\in\mathbb{R}^{n_x}$ are independently identically distributed (i.i.d.) white noise error terms initialized from a normal distribution with zero bias and $\boldsymbol\Omega$ covariance matrix. It is noted that the VARMAX model has a fundamental identification issues, meaning that coefficient matrices $\mathbf{a},\mathbf{A}_i,\mathbf{M}_j$ are not unique and there may be multiple value combinations that equate the same results for a give time-series~\cite{Lutkepol:07}.

The total number of trainable parameters in the VARMAX model for a non-diagonal $\boldsymbol\Omega$, the covariance matrix for the white-noise terms, is:
\begin{equation}
    \text{Number of Trainable Parameters (non-diagonal $\boldsymbol{\Omega}$)} =\left(p+q+\frac{1}{2}\right)n_x^2 + \left(\frac{3}{2}+n_u\right)n_x,
\end{equation}
and the total number of trainable parameters in the VARMAX model for a diagonal $\boldsymbol\Omega$, the covariance matrix for the white-noise terms, is:
\begin{equation}
    \text{Number of Trainable Parameters (diagonal $\boldsymbol{\Omega}$)} =(p+q)n_x^2 + (2+n_u)n_x. \label{eq:num-train-param-diagonal}
\end{equation} 

\section{Sobol Indices for Global Parameter Sensitivity Analysis}\label{app:sobol}

A digital twin framework is developed such that the virtual model is modifiable by allowing the trainable parameters to be adapted online through measured observations. 
For this application, the low-complexity surrogate model contains more than 500 trainable parameters that need to be analyzed to methodically selected.
The parameters are selected through a global parameter sensitivity analysis using Sobol indices~\cite{sobol:01}. The Sobol indices provide a variance-based measure to quantify the effects of inputs on a function output by fractionally comparing the variance effects of one input compared to the total variance. Given the model $Y = f(\mathbf{X})$, where $\mathbf{X}$ is a vector of uncertain model inputs $\mathbf{X} = [X_1, \hdots, X_d]$, the first order index $S_i$ and total order index $S_{T_i}$ of an input variable $X_i$ are computed by:
\begin{align}
    S_i &= \frac{V_i}{\mathbb{V}\text{ar}[Y]} \label{eq:sobol_first}\\
    S_{T_i} &= \frac{\mathbb{E}_{X_{\sim i}}\left[\mathbb{V}\text{ar}_{X_i}[Y\mid X_{\sim i}]\right]}{\mathbb{V}\text{ar}[Y]},\label{eq:sobol_total}
\end{align}
where $V_i = \mathbb{V}\text{ar}_{X_i}[\mathbb{E}_{X\sim i}[Y\mid X_i]]$.
The expectations and variances in Equations~\ref{eq:sobol_first} and~\ref{eq:sobol_total} respectively are estimated by generating many samples of the forward model while varying the inputs in a defined parameter space. 

A Sobol analysis is conducted for two cases. 
The first is considering only one output, the core energy. The parameter identified for this case are used in the first two digital twin demonstration, the long-term and transient capturing. 
The second analysis is performed on two outputs, the core energy and the secondary pump mass flow rate, to be used in the third digital twin demonstration featuring the system shock capturing.
The remainder of this appendix section will discuss the results of the Sobol analysis and the resulting augmented parameter vectors for the EnKF state-parameter estimation. 
The Sobol analysis is performed with the \texttt{SALib}~\cite{Iwanaga2022,Herman2017} python package using the Saltelli sequence~\cite{saltelli:02} to generate $N(2D+2)$ samples, where $N$ is the number of samples and $D$ is the number of inputs.
For this work, $N=2^{13}$.

\subsection{One output: core energy parameter analysis}\label{app:sobol1}
The gFHR Surrogate framework shown in Figure \ref{fig:gFHRSurrogate} is analyzed to select the VARMAX trainable parameters that have the greatest impact on the core energy prediction. 
From the figure, it shows that the core energy is predicted with the VARMAX II model, which also uses former predictions from the VARMAX I model.
Since only the VARMAX I and VARMAX II models are used for the core energy prediction, only the parameters in these two models are considered in this Sobol indices study; thus the total number of parameters considered in this study is $D=49$.

\begin{figure}
    \centering
    \includegraphics[width=\textwidth]{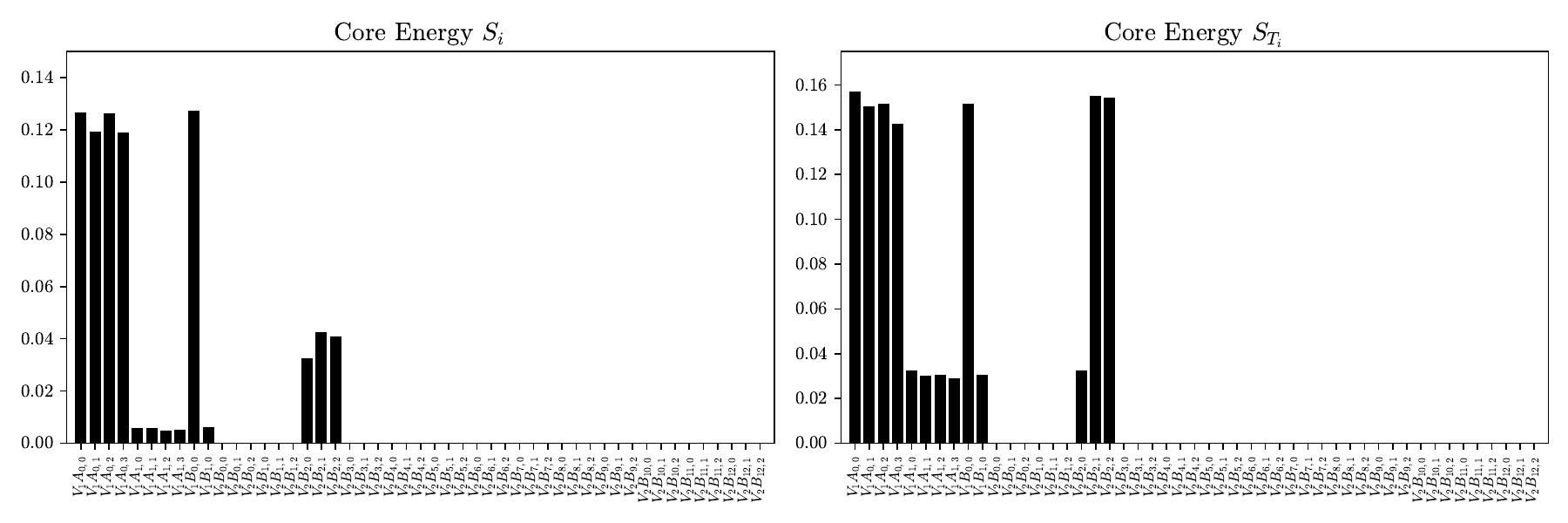}
    \caption{Sobol first order and total order index analysis of the gFHR surrogate model when predicting the core energy. Only the parameters of VARMAX I and VARMAX II models are considered, since the core energy is predicted by the VARMAX II model.}
    \label{fig:sobol}
\end{figure}

A bounds space for each parameter is defined to be $X_i \pm 0.5$, where $X_i$ are the originally trained values; and the function $f(\cdot)$ is the gFHR Surrogate model taking one time-step at 100\% power.
The results are shown in Figure~\ref{fig:sobol}, where eight parameters are clearly identified to have the greatest impact on the core power prediction.
These eight parameters with $S_i, S_{T_i} > 0.10$ are augmented to the EnKF assimilation parameter vector $\boldsymbol{\theta}_{\text{A}}$:
\begin{equation}
    \boldsymbol{\theta}^{(k)}_A \triangleq \begin{bmatrix} \mathbf{A}_{\text{VI},0,0}^{(k)} & \mathbf{A}_{\text{VI},0,1}^{(k)} & \mathbf{A}_{\text{VI},0,2}^{(k)} & \mathbf{A}_{\text{VI},0,3}^{(k)} & \mathbf{B}_{\text{VI},0,0}^{(k)} & \mathbf{B}_{\text{VII},2,0}^{(k)} & \mathbf{A}_{\text{VII},2,1}^{(k)} & \mathbf{A}_{\text{VII},2,2}^{(k)}\end{bmatrix}^T.
\end{equation}

\subsection{Two output: core energy and secondary pump mass flow rate analysis}\label{app:sobol2}

In the third digital twin demonstration (Section~\ref{subsec:demo3}), the gFHR Surrogate model was reconstructed in order to capture a system shock caused by a change in the steam generator boundary condition. 
To capture the shock, the secondary pump mass flow is added as an observable measurement.
A sobol analysis is performed in order to determine the parameters to be added to the EnKF state-parameter estimation with the modified gFHR Surrogate model and the additional output.
In the new surrogate framework, shown in Figure~\ref{fig:systemSurrogateSG}, the secondary pump mass flow rate is predicted from the VARMAX~I and the core energy is predicted from the VARMAX~II model, therefore only the $D=28$ trainable parameters in these sub-models are considered. 

\begin{figure}
    \centering
    \includegraphics[width=\textwidth]{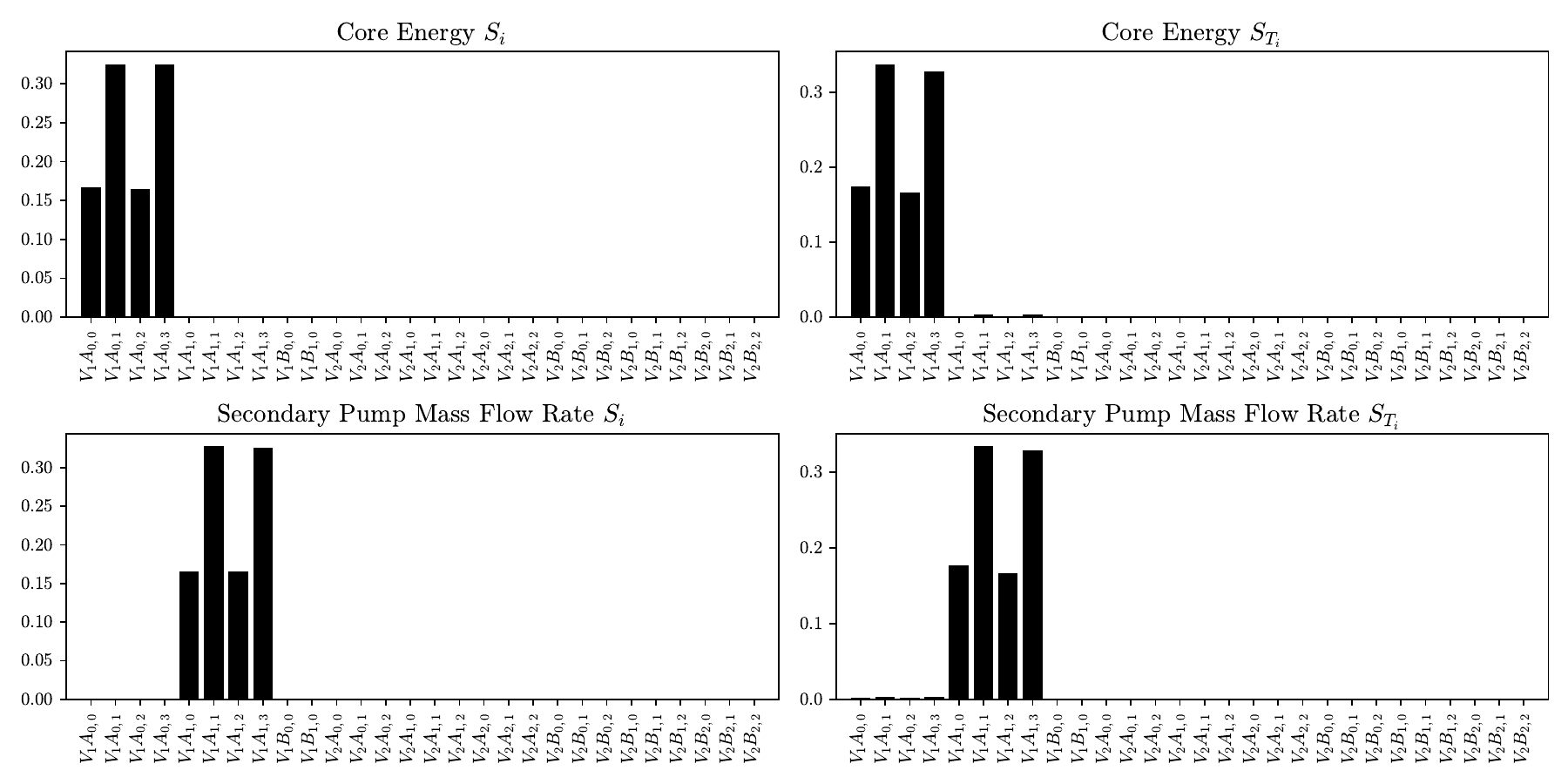}
    \caption{Sobol first order and total order index analysis of the modified gFHR Surrogate model for the shock capturing case when predicting the core energy and the secondary pump mass flow rate. Only the parameters of VARMAX~I and VARMAX~II models are considered, since the secondary pump mass flow rate is prediction by the VARMAX~I model and the core energy is predicted by the VARMAX II model.}
    \label{fig:sobol2}
\end{figure}

The bounds space for each parameter is $X_i \pm 0.5$, where $X_i$ is the parameter's original value. 
Each sample is generated propagating the gFHR modified surrogate one time-step at 100\% power, and a Sobol indices study is done individually for each of the two outputs.
The results are shown in Figure~\ref{fig:sobol2} for the core energy and the secondary pump mass flow rate and parameters with $S_i, S_{T_i} > 0.20$ are appended to the augmented parameter vector.
Additionally, the two coefficients in the VARMAX II input matrix $\mathbf{B}$ are added, since they are the parameters that determine the core prediction energy given the inputs. 
The final parameter vector for this modified gFHR Surrogate model is:
\begin{equation}
    \boldsymbol{\theta}^{(k)}_A \triangleq \begin{bmatrix} \mathbf{A}_{\text{VI},0,1}^{(k)} & \mathbf{A}_{\text{VI},0,3}^{(k)} & \mathbf{A}_{\text{VI},1,1}^{(k)} & \mathbf{A}_{\text{VI},1,3}^{(k)} & \mathbf{B}_{\text{VII},0,0}^{(k)} & \mathbf{B}_{\text{VII},1,0}^{(k)} \end{bmatrix}^T.
\end{equation}

\end{document}